\documentclass[journal,twoside]{IEEEtran}
\usepackage{cite}             % improves presentation of citations
\usepackage[utf8]{inputenc}   % allow utf-8 input
\usepackage[T1]{fontenc}      % use 8-bit T1 fonts
\usepackage{url}              % simple URL typesetting   
\usepackage[pdftex]{graphicx} % \graphicspath{ {./images/} }
\usepackage[cmex10]{amsmath}  
\usepackage{amsfonts,amssymb,mathtools,bbm,verbatim}  
\usepackage{amsthm}
\usepackage{array}
\usepackage{mleftright}       % fix to wrong spacing of \left-,
\mleftright                   % \middle- \right-commands 
\usepackage{comment}
\usepackage{hyperref}
\hypersetup{colorlinks = true, linkcolor = blue, urlcolor = blue, citecolor = blue}  
\usepackage[all]{hypcap}      % for going to the top of an image
\usepackage[capitalize]{cleveref}
\usepackage{cuted}

\allowdisplaybreaks           % to allow display breaks in align environment
\crefformat{equation}{(#2#1#3)} 
\usepackage{flushend}

\def\X{{\mathcal{X}}}
\def\Y{{\mathcal{Y}}}
\def\Z{{\mathcal{Z}}}

\def\P{{\mathbb{P}}}
\def\E{{\mathbb{E}}}

\def\R{{\mathbb{R}}}
\def\N{{\mathbb{N}}}

\def\1{{\mathbf{1}}}
\def\0{{\mathbf{0}}}
\def\I{{\mathbbm{1}}}

\def \A {{A}}
\def \calP {{\mathcal{P}}}

\def \calV {{\mathcal{V}}}
\def \calI {{\mathcal{I}}}
\def \calT {{\mathcal{T}}}
\def \calA {{\mathcal{A}}}
\def \calB {{\mathcal{B}}}
\def \calZ {{\mathcal{Z}}}
\def \calZt {\tilde{\mathcal{Z}}}

\DeclareMathOperator\diag{{diag}}
\DeclareMathOperator*{\argmax}{arg\,max}
\DeclareMathOperator*{\argmin}{arg\,min}

\renewcommand{\complement}{\mathrm{c}}

\newcommand{\T}{\mathrm{T}} % transpose
\newcommand{\pa}{\mathrm{pa}}

\newtheorem{theorem}{Theorem}
\newtheorem{lemma}{Lemma}
\newtheorem{proposition}{Proposition}
\newtheorem{corollary}{Corollary}
\newtheorem{definition}{Definition}

\begin{document}

\bstctlcite{IEEEexample:BSTcontrol} % to ensure that adjacent author names that are the same are not dashed in the bibliography

\title{Doeblin Coefficients and Related Measures}
\author{Anuran~Makur~and~Japneet~Singh%
\thanks{The author ordering is alphabetical. An earlier version of this paper was presented in part at the 2023 IEEE International Symposium on Information Theory \cite{MakurSingh2023a}.}%
\thanks{A. Makur is with the Department of Computer Science and the Elmore Family School of Electrical and Computer Engineering, Purdue University, West Lafayette, IN 47907, USA (e-mail: amakur@purdue.edu).}%
\thanks{J. Singh is with the Elmore Family School of Electrical and Computer Engineering, Purdue University, West Lafayette, IN 47907, USA (e-mail: sing1041@purdue.edu).}%
}

\maketitle

\begin{abstract}
Doeblin coefficients are a classical tool for analyzing the ergodicity and exponential convergence rates of Markov chains. Propelled by recent works on contraction coefficients of strong data processing inequalities, we investigate whether Doeblin coefficients also exhibit some of the notable properties of canonical contraction coefficients. In this paper, we present several new structural and geometric properties of Doeblin coefficients. Specifically, we show that Doeblin coefficients form a multi-way divergence, exhibit tensorization, and possess an extremal trace characterization. We then show that they also have extremal coupling and simultaneously maximal coupling characterizations. By leveraging these characterizations, we demonstrate that Doeblin coefficients act as a nice generalization of the well-known total variation (TV) distance to a multi-way divergence, enabling us to measure the ``distance'' between multiple distributions rather than just two. We then prove that Doeblin coefficients exhibit contraction properties over Bayesian networks similar to other canonical contraction coefficients. We additionally derive some other results and discuss an application of Doeblin coefficients to distribution fusion. Finally, in a complementary vein, we introduce and discuss three new quantities: \textit{max-Doeblin coefficient}, \textit{max-DeGroot distance}, and \textit{min-DeGroot distance}. The max-Doeblin coefficient shares a connection with the concept of maximal leakage in information security; we explore its properties and provide a coupling characterization. On the other hand, the max-DeGroot and min-DeGroot measures extend the concept of DeGroot distance to multiple distributions. 
\end{abstract}


\section{Introduction}
Recently, there has been a flurry of research activity on \emph{contraction coefficients}, which mathematically capture the degree to which information contracts in channels or Markov kernels. Much of this literature in information theory has focused on \emph{strong data processing inequalities} (SDPIs) \cite{AhlswedeGacs1976, Cohenetal1993, CohenKempermanZbaganu1998}, which insert contraction coefficients into classical data processing inequalities for $f$-divergences \cite{Morimoto1963, AliSilvey1966, Csiszar1967, Csiszar1972} in order to strengthen them. Various aspects of SDPIs and contraction coefficients, such as bounds between different definitions, their behavior over Bayesian networks, relations to other statistical and probabilistic notions (such as hypercontractivity, maximal correlation, and logarithmic Sobolev inequalities), etc. have been studied extensively, cf. \cite{Anantharametal2013, Anantharametal2014, MakurZheng2015, PolyanskiyWu2016, Raginsky2016, Polyanskiy2017, MakurPolyanskiy2017, Huangetal2017, MakurZheng2017, Calmonetal2017, CalmonPolyanskiyWu2018, MakurPolyanskiy2018, Makur2019, MakurZheng2020, MakurWornellZheng2020, Huangetal2023} and the references therein. The resulting ideas have found applications in diverse fields such as non-interactive simulation of distributions \cite{KamathAnantharam2012}, distributed estimation \cite{Xu2015}, differential privacy \cite{Duchi2013, Asoodehetal2016, Asoodeh2021}, and secret key generation \cite{Liu2015}, among others. Rather interestingly, it was shown in \cite{Polyanskiy2017} that the contraction coefficient for Kullback-Leibler (KL) divergence of a given channel is completely characterized by the extremal erasure probability so that an erasure channel dominates the given channel in the well-known \emph{less noisy} preorder sense \cite{korner1977}. This observation was generalized for a broader class of $f$-divergences in \cite{MakurZheng2020}, and to broader classes of dominating channels such as symmetric channels in \cite{MakurPolyanskiy2018}. 
These observations portrayed that the less noisy preorder could be used to define canonical contraction coefficients. In this paper, we study the contraction coefficients generated by the above recipe when the well-known (output) \emph{degradation} preorder over channels \cite{Bergmans1973, Cover1972} is used instead of the less noisy preorder. These coefficients, which we define next, were originally introduced in the analysis of Markov chains and are known as \emph{Doeblin coefficients} \cite{Doeblin1937}.

 \begin{definition}[Doeblin Coefficient]
    \label{Definition of Doeblin}
        For any channel $W \in \R^{n\times m}_{\mathsf{sto}}$, its Doeblin coefficient $\tau(W)$ is defined as
    \begin{equation*}
        \tau(W) \triangleq \sum_{j =1}^m \min_{i \in \{1,\dots,n \}} W_{ij},
    \end{equation*} 
    where $\R^{n\times m}_{\mathsf{sto}}$ denotes the set of all $n \times m$ row stochastic matrices.
\end{definition}

To see the connection between \cref{Definition of Doeblin} and degradation, recall that a channel (or row stochastic matrix) $W \in \R^{r\times m}_{\mathsf{sto}}$ is said to be a \emph{degraded} version of another channel $V \in \R^{r \times n}_{\mathsf{sto}}$ (with common input alphabet), denoted $V \succeq_{\textsf{\tiny deg}} W$, if there exists a channel $U\in \R^{n\times m}_{\mathsf{sto}}$ such that $W = V U$. It has been established that the Doeblin coefficient of a given channel $W \in \R^{r\times m}_{\mathsf{sto}}$ is characterized by the extremal erasure probability such that $W$ is a degraded version of an erasure channel \cite{Chestnut2010, MakurPolyanskiy2018, GohariGunluKramer2020}: 
\begin{equation}
\label{Eq: IT char}
\tau(W) = \sup\{\epsilon \in [0,1] : \text{E}_{\epsilon} \succeq_{\textsf{\tiny deg}} W\} ,
\end{equation}
where $\text{E}_{\epsilon} \in \R^{r \times (r+1)}$ is an $r$-ary erasure channel which erases its input and produces an erasure symbol with probability $\epsilon$ and keeps the input unchanged with probability $1-\epsilon$. 

Our broad objective in this work (also see \cite{MakurSingh2023a}) is to develop several structural and geometric results on Doeblin coefficients and illustrate that Doeblin coefficients share most of the ``nice'' properties of canonical contraction coefficients of SDPIs, such as their behavior over Bayesian networks. The next subsection briefly discusses further background literature pertaining specifically to Doeblin coefficients (rather than contraction coefficients more broadly), and the subsequent subsection delineates our main contributions in this work.

\subsection{Related Literature on Doeblin Coefficients}

Historically, Doeblin introduced two important techniques to analyze the ergodicity and exponential convergence of Markov chains in \cite{Doeblin1937} and \cite{Doeblin1938}. In \cite{Doeblin1938}, he developed coupling as a technique for proving uniform geometric rates of convergence of Markov chains to their stationary distributions in terms of \emph{total variation (TV) distance} (see \cite{Kimbleton1969, BhattacharyaWaymire2001, Chestnut2010} for more details). In \cite{Doeblin1937}, Doeblin characterized conditions for weak ergodicity of possibly inhomogeneous Markov chains using the notion of Doeblin minorization, which is closely related to Doeblin coefficients. Indeed, it can be shown that for a given channel (or Markov kernel) $W \in \R_{\mathsf{sto}}^{r \times m}$, its Doeblin coefficient is the largest constant $\alpha \in [0,1]$ such that the \emph{Doeblin minorization} condition holds \cite{Doeblin1937}, i.e., there exists a probability distribution $\mu = (\mu_1,\dots,\mu_m)$ such that
\begin{equation}
W_{ij} \geq \alpha \mu_j ,
\end{equation}
for all $i \in \{1,\dots,r\}$ and $j \in \{1,\dots,m\}$. Moreover, a Markov chain $W$ (when $r = m$) with a large Doeblin coefficient is said to be ``well-minorized,'' because it can be shown that the chain converges quickly to its invariant distribution. 

The Doeblin minorization condition has also been studied in the context of information theory, where it was used to derive universal upper bounds on contraction coefficients of SDPIs for any $f$-divergence \cite[Remark 3.2]{Raginsky2016}, \cite[Section I-D]{MakurPolyanskiy2018}. In fact, Doeblin minorization is known to be equivalent to degradation by erasure channels (see \cite[Theorem 3.1]{BhattacharyaWaymire2001}, \cite[Lemma 5]{GohariGunluKramer2020}, and \cite[Section IV-D]{Makur2020} for more details). So, the characterization of Doeblin coefficients as extremal minorization constants is essentially equivalent to the information-theoretic characterization in \cref{Eq: IT char}. In the theory of Harris chains, this idea can be viewed as a specialization of the \emph{regeneration} or Nummelin splitting technique  \cite{AthreyaNey1978, Nummelin1978}. (The regeneration technique involves dividing the state space of a Markov chain into disjoint subsets such that the chain can be split into independent and identically distributed (i.i.d.) blocks within each subset. This allows for the application of the Doeblin minorization condition on each subset, which ensures the convergence of the chain to a unique stationary distribution.) 
As noted earlier, due to this equivalence between minorization and degradation by erasure channels, Doeblin coefficients can be construed as extensions of the broader theory of canonical contraction coefficients \cite{CohenKempermanZbaganu1998}, \cite[Definition 5.1]{Cohenetal1993}, which are characterized using domination by erasure channels under the less noisy preorder. 

Doeblin coefficients and minorization have several applications in the analysis of applied probability problems. 
For instance, \cite{chen2022change} developed a change detection algorithm for Markov kernels and used the Doeblin minorization condition to derive an upper bound on the mean delay. Similarly, \cite{Vrettos2020} used the Doeblin minorization condition to provide finite-time upper bounds for the regret incurred in a multi-armed bandit problem involving multiple plays and Markovian rewards in the rested bandits setting. 
Furthermore, Doeblin coefficients are classically used to construct Markov chain Monte Carlo (MCMC) methods \cite{rosenthal1995} and to design algorithms for decision-making with inhomogeneous Markov decision processes \cite{alden1992}. 
The Doeblin coefficient is also closely related to other important concepts in probability theory, such as mixing times \cite{doeblin1940elements}. 
As another application, in structured prediction, Doeblin chains \cite{doeblin1940elements} are utilized to learn fast-mixing models for approximate inference \cite{steinhardt2015}. Doeblin-type assumptions have also been used in the development of entropy rate estimators for stationary ergodic processes, e.g., \cite{kontoyiannis1994pre} extended the relation between the entropy rate and asymptotic lengths of shortest new prefixes to any finite valued stationary ergodic process that satisfies a Doeblin-type condition, and \cite{kontoyiannis1998nonparametric} established the pointwise consistency of three different estimators of entropy rate for a stationary ergodic process satisfying a Doeblin-type mixing condition.      

Finally, in a complementary vein, while various notions of divergence between two distributions, such as $f$-divergences, are widely utilized in information theory, machine learning, and statistics, an intriguing line of work has defined new notions of \emph{multi-way divergences} and metrics over probability distributions. Multi-way divergences measure ``distances'' between two \emph{or more} distributions. We refer readers to \cite{Williamson2022, Gushchin, Ginebra2007} and the references therein for a detailed overview of this line of research. Interestingly, we will show in this work that Doeblin coefficients possess properties which make them suitable for measuring divergence between multiple distributions as well.

\subsection{Main Contributions}
    The contributions of this paper are summarized below.
    \begin{enumerate}
        \item We will explore various structural and geometric properties of Doeblin coefficients. Firstly, we will prove various new properties and review several known properties of Doeblin coefficients for the readers' convenience in \cref{Properties of Doeblin coefficient}. For example, we will demonstrate that the Doeblin coefficient possesses desirable properties that make it suitable for defining a notion of ``distance" between multiple distributions. In particular, we will show that the complement of the Doeblin coefficient is a multi-way metric, enabling us to measure the ``distance'' between multiple distributions.  
        \item Next, we will introduce a maximal coupling characterization for the Doeblin coefficient in \cref{Maximal Coupling}, extending the known coupling characterization of TV distance to the case of multiple distributions. Combining this coupling characterization with properties such as multi-way metric will demonstrate how the Doeblin coefficient can act as a nice generalization of TV distance for multiple distributions. 
        \item Then, we will introduce a new coefficient dubbed the \emph{max-Doeblin coefficient} and explore its properties and provide its corresponding coupling characterization in \cref{Properties of max-Doeblin coefficient,thm:MaxDoeblinCoupling} and ensuing propositions. Notably, the max-Doeblin coefficient acts as a multi-way metric and provides another generalization of TV distance to multiple distributions. Moreover, its analogous coupling characterization to the Doeblin coefficient holds under a specific condition described later in the paper. Interestingly, in the case of two distributions, both the generalizations obtained from max-Doeblin and Doeblin coefficients are equivalent to the conventional TV distance.
        \item We will also show that the well-known DeGroot distance or Bayes statistical information \cite{Raginsky2016}, which is closely related to TV distance, can be similarly generalized to max-DeGroot and min-DeGroot counterparts. 
        \item Next, we will study the contraction properties of Doeblin coefficients over Bayesian networks (or directed graphical models). To do this, we will first develop a simultaneously maximal coupling result for the Doeblin coefficient in \cref{Simultaneous Coupling}. This result will then be used to develop bounds that illustrate the contraction properties of Doeblin coefficients over Bayesian networks in \cref{Doeblins Coefficient in Networks}. This will be further followed by other related results, such as \cref{Samorodnitskys SDPI in terms of Doeblin Coefficient}.
        \item Finally, we will briefly illustrate an application where Doeblin coefficients will be used to recover a fusion or aggregation rule for probability mass functions (PMFs). Specifically, we will show that Doeblin coefficients allow us to develop an {optimization-based approach} for the aggregation of PMFs.  
    \end{enumerate}

\subsection{Outline}
    The paper is organized as follows. In \cref{Sec: Main Results and Discussions}, we present the main results of our study. We begin by introducing some new properties of the Doeblin coefficient in \cref{Sec: Properties of Doeblin Coefficients}. In \cref{Coupling Characterization}, we provide a maximal coupling characterization of the Doeblin coefficient. 
    Then, we introduce the max-Doeblin coefficient in \cref{sec: Max-Doeblin Coefficient and its Minimal Coupling Characterization} and explore its properties and minimal coupling characterization in \cref{sec: Max-Doeblin Coefficient,sec: Minimal Coupling Characterization of max-Doeblin Coefficient}, respectively. In \cref{sec: DeGroot Distance}, we discuss a similar extension of the well-known DeGroot distance to max-DeGroot and min-DeGroot.  
    In \cref{Sec: Simultaneously Maximal Coupling}, we demonstrate a simultaneously maximal coupling result for the Doeblin coefficient. Moreover, in \cref{Sec: Contraction over Bayesian Networks}, we show that the Doeblin coefficient exhibits contraction properties over Bayesian networks, similar to other canonical contraction coefficients. Next, we apply the Doeblin coefficient to the problem of PMF fusion in \cref{Sec: Application to PMF Fusion}, where we propose a fusion rule.  In \cref{sec: Proofs,sec: proofs2,Proof of Samorodnitskys SDPI in terms of Doeblin Coefficient}, we present the proofs of various results discussed above. Finally, in \cref{Sec: Conclusion}, we reiterate our results and provide some directions for future research. 

\subsection{Notation}
    We briefly collect some notation that is used throughout this work. For $n \in \N \triangleq \{1,2,3,\dots\}$, let $[n]$ denote the set $\{1,\dots,n\}$. Let $\1$ denote a column vector with all entries equal to $1$ of appropriate dimension. Let $\calP_n$ denote the $(n-1)$-dimensional probability simplex of row vectors in $\R^n$, i.e., the set $\{P \in \R^n: P \geq 0 \text{ entry-wise and } P\1 = 1 \}$. For any two PMFs $P_1,P_2 \in \calP_n$, we define $\|P_1 - P_2\|_{\mathsf{TV}} \triangleq \frac{1}{2}\|P_1 - P_2\|_1$ as the TV distance between $P_1$ and $P_2$, where $\|\cdot\|_1$ denotes the $\ell^1$-norm. 
    For an arbitrary input alphabet $\X \triangleq \{x_1,\ldots,x_n\}$ and arbitrary output alphabet $\Y = \{y_1,\ldots,y_m\}$ with $|\X| = n $ and $|\Y| = m \in \N$ (without loss of generality), let $X$ and $Y$ be random variables taking values in $\X$ and $\Y$, respectively, and let $\R^{n\times m}_{\mathsf{sto}}$ denote the set of all $n \times m$ row stochastic matrices in $\R^{n\times m}$, which are conditional distributions of $Y$ given $X$. For any matrix $W \in \R^{n\times m}_{\mathsf{sto}}$ such that $W = \big[P_1^{\T} \, P_2^{\T} \, \dots \, P_n^{\T}\big]^{\T}$ is formed by stacking the row vectors $P_1,P_2, \dots ,P_n \in \calP_m$, we will interchangeably use the notation $\tau(P_1,P_2, \dots ,P_n)$ to denote the Doeblin coefficient $\tau(W)$. Let $\mathsf{Tr}(A)$ denote the trace of any matrix $A \in \R^{n\times n}$ and let $I$ denote an identity matrix of appropriate size. Finally, for any vector $x \in \R^n$, $\diag(x) \in \R^{n \times n},$ is the diagonal matrix with $x$ along its principal diagonal and $(a)_+$ represents $\max\{a,0\}$ for $a \in \R$. Throughout this paper, we adopt the convention that any expression in the form $0/0$ is treated as zero.
   
\section{Main Results and Discussion}\label{Sec: Main Results and Discussions}
    In this section, we will present our main results on the properties and various characterizations of Doeblin coefficients as well as related quantities. 

\subsection{Doeblin Coefficient and Its Maximal Coupling Characterization} \label{sec:Doeblin Coefficient and its Maximal Coupling Characterization}
    We begin by presenting some properties of Doeblin coefficients and then move on to introducing its maximal coupling representation.
\subsubsection{Properties of Doeblin Coefficients} \label{Sec: Properties of Doeblin Coefficients}
    The ensuing theorem summarizes various properties of Doeblin coefficients.
    \begin{theorem}[Properties of Doeblin Coefficients] \label{Properties of Doeblin coefficient} Let $W = \big[P_1^{\T} \, \dots \, P_n^{\T}\big]^{\T} \in \R^{n \times m}_{\mathsf{sto}}$ be a channel formed by stacking the PMFs $P_{1}, \ldots, P_{n} \in \calP_m$. Then, the Doeblin coefficient $\tau(W)$ of $W$ satisfies the following properties:
    \begin{enumerate}
    \item  \emph{(Normalization)} We have $0 \leq \tau(W) \leq 1$, where $\tau(W)=1$ if and only if $P_{1}=P_{2}=\dots=P_{n}$ (entry-wise), and $\tau(W)=0$ if and only if at least one of $P_{1}(y), \dots, P_{n}(y)$ is zero for all $y \in \Y$.
    \item  \emph{($n$-way Metric)} The complement of the Doeblin coefficient $\gamma(P_1,\dots,P_n) \triangleq 1-\tau(P_1, \dots,P_n) \in [0,1]$ is an $n$-way metric \cite{Warrens2010} i.e., it satisfies the following properties:
        \begin{itemize}
                \item[a)] \emph{(Total Symmetry)} 
                For any permutation $\pi$ of $[n]$, we have
                \[ \gamma ( P_{\pi(1)}, \ldots , P_{\pi(n)})= \gamma (P_{1},\ldots , P_{n}). \]
                
                \item[b)] \emph{(Positive Definiteness)} $\gamma(P_{1}, \dots, P_{n})=0 $ if and only if $P_1  = \dots =P_n$.
                
                \item[c)] \emph{(Polyhedron Inequality)} For any $P_{n+1} \in \calP_m$, $\gamma(P_1,\dots,P_n)$ satisfies the inequality
                    \begin{equation*}
                    \begin{aligned}
                        (n-1)  \gamma(& P_{1}, \dots, P_{n} ) \leq \\  
                        & \sum_{i=1}^{n} \gamma(P_{1}, \dots, P_{i-1}, P_{i+1}, \dots, P_{n+1}).
                    \end{aligned}
            \end{equation*}
        \end{itemize}
    \item  \emph{(Concavity)} The map $W \mapsto \tau(W)$ is concave.

    \item \emph{(Sub-multiplicativity \cite{Chestnut2010, Lladser2014})} For channels $V \in \R^{k \times n}_{\mathsf{sto}} $ and $W \in \R^{n \times m}_{\mathsf{sto}}$, the complement of the Doeblin coefficient is sub-multiplicative:
         \[ 1- \tau(VW) \leq (1-\tau(V) )(1-\tau(W)).\]
    \item  \emph{(Tensorization)} For channels $W \in \R^{n \times m}_{\mathsf{sto}} $ and $V \in \R^{l \times k}_{\mathsf{sto}}$, we have
        \[\tau(W\otimes V) =\tau(W) \tau(V), \]
    where $\otimes$ denotes the Kronecker product of matrices.
    \item  \emph{(Upper Bound \cite{Raginsky2016})}  $\tau(W)$ satisfies the following bound:  
         \[\tau(W) \leq 1 - \eta_{\mathsf{T V}}(W) ,\]
          where
          $\eta_{\mathsf{TV}}(W) = \max_{i,j \in [n]}\|P_i - P_j\|_{\mathsf{TV}}$ denotes the \emph{Dobrushin contraction coefficient} for TV distance.
    \item \emph{(Minimum Trace Characterization)} $\tau(W)$ is the solution to the following optimization problem:
         $$\tau(W) = \min_{P \in \R^{m\times n}_{\mathsf{sto}} } \mathsf{Tr}(P W),$$
         where the minimum is over all possible $m \times n$ row stochastic matrices $P$. 
    \item  \emph{(Optimal Estimator)} 
    Consider a hidden random variable $X$ that is uniformly distributed on $\X$, i.e., $X \sim \textup{unif}(\X)$, and the fixed channel (or observation model) $W = P_{Y|X} \in \R^{n\times m}_{\mathsf{sto}}$. Let $\hat{X} \in \X$ denote any (possibly randomized) estimator of $X$ based on the observation $Y$, which is defined by a kernel $P_{\hat{X}|Y} \in \R^{m\times n}_{\mathsf{sto}}$ such that $X \rightarrow Y \rightarrow \hat{X}$ forms a Markov chain. Then, we have
    $$\frac{\tau(P_{Y|X})}{n} = \min_{P_{\hat{X}|Y} \in \R^{m\times n}_{\mathsf{sto}}} \P( \hat{X} = X),$$
    where the minimum is computed over all estimators $\hat{X}$, or equivalently, over all kernels $P_{\hat{X}|Y} \in \R^{m\times n}_{\mathsf{sto}}$, and $\P(\cdot)$ denotes the probability law of $(X,Y,\hat{X})$. 
    \end{enumerate}    
    \end{theorem}
    
    The proof is provided in \cref{Proof of Properties of Doeblin}. Properties 1 and 3 in \cref{Properties of Doeblin coefficient} are straightforward to prove (and property 1 is of course well-known). Property 4 was already proved in \cite{Chestnut2010, Lladser2014}, but we provide a new proof for it. 
    The upper bound in property 6 is also well-known, cf. \cite[Remark III.2]{Raginsky2016}. 
    The rest of the properties are new to our knowledge. It is worth reiterating that the complement of the Doeblin coefficient $\gamma(\cdot)$ acts as a multi-way metric, which allows us to simultaneously measure ``distance'' between multiple distributions. In \cref{Coupling Characterization}, we will establish that $\tau(\cdot)$ admits a maximal coupling characterization similar to the TV distance. 
    Furthermore, the relevance of property 8 becomes evident in settings like cryptographic systems or the old game ``Minesweeper'', where the primary objective is reducing the likelihood of selecting specific actions. For example, in Minesweeper, the aim is to uncover cells while evading mines, or in cryptographic systems, the goal is to avoid unauthorized decryption attempts.
    
\subsubsection{Maximal Coupling Characterization for Doeblin Coefficient} \label{Coupling Characterization}
    In this subsection, we develop an extremal-coupling-based characterization of Doeblin coefficients that generalizes the known maximal coupling characterization for TV distance to the setting of multiple distributions.  

    Recall that for random variables $X$ and $Y$ taking values in some common alphabet $\Y$ with probability distributions $P$ and $Q$, respectively, a \emph{coupling} of $P$ and $Q$ is a joint distribution $P_{X, Y}$ on the product space $\Y \times \Y$ such that the corresponding marginal distributions of $X$ and $Y$ are equal to $P$ and $Q$, respectively. A \emph{maximal coupling} of two distributions $P$ and $Q$ on $\Y$ is a coupling that maximizes the probability that the two random variables are equal. It is known that the TV distance between $P$ and $Q$ can be characterized by Dobrushin's maximal coupling of $P$ and $Q$ \cite[Proposition 4.7]{LevinPeresWilmer2009}: 
    \begin{equation}
        1 - \|P-Q\|_{\mathsf{TV}} = \max_{ \substack{ P_{X,Y}:\\ \, P_X = P, \, P_{Y} = Q}} \P(X = Y),  \label{Eq:TVdistance}  
    \end{equation}
    where the maximum is over all couplings of $P$ and $Q$, and $\P(\cdot)$ is the probability law corresponding to the coupling $P_{X,Y}$.
    Moreover, under the maximal coupling, for any $y \in \Y$,
    \begin{equation}
           \P(X = Y= y) = \min\{P(y),Q(y)\}. \label{Eq:TV Coupling Distribution} 
    \end{equation}   
    We will now generalize this result for the case where we have $n$ distributions $P_1, P_2,\dots, P_n \in \calP_m$. A coupling of random variables $Y_1,\dots,Y_n \in \Y$ distributed as $P_1,\dots,P_n$, respectively, is defined as a joint distribution $P_{Y_1,\dots,Y_n}$ that preserves the marginals. Specifically, under the maximal coupling of $P_1, P_2, \dots, P_n$, we will show that the measure of the ``diagonal set'' of $\Y^n$ equals $\tau(P_1, \dots, P_n)$.
    \begin{theorem}[Maximal Coupling for Doeblin Coefficient] \label{Maximal Coupling}
        For any random variables $Y_1, \dots, Y_n \in \Y$ distributed according to the PMFs $P_{1}, \dots, P_{n} \in \calP_m$, respectively, let $W \in \R^{n\times m}_{\mathsf{sto}}$ be a channel defined by $W = \big[P_1^{\T} \, P_2^{\T} \, \dots \, P_n^{\T}\big]^{\T}$. Then, we have
        \begin{equation*}
            \tau(W) = 1 -\gamma(W) = \max_{\substack{ P_{Y_1, \dots, Y_n} : \\ P_{Y_1 } = P_1, \, \dots \, , \, P_{Y_n} =P_n} } \!\P( Y_1 =  \dots = Y_n), 
        \end{equation*}
    where the maximum is over all couplings of $P_{1}, \dots, P_{n}$, and $\P(\cdot)$ denotes the probability law corresponding to a coupling. Moreover, under the maximal coupling, for any $y \in \Y$, 
    \begin{equation*}
       \P( Y_1 =  \dots = Y_n = y )  = \min \{ P_1(y), \dots, P_n(y) \}.    
    \end{equation*}
    \end{theorem}
    
    The proof is provided in \cref{Proof of Maximal Coupling}. \cref{Maximal Coupling} illustrates that $\gamma(\cdot)$ can be perceived as a generalization of TV distance to multiple distributions.  Moreover, by property 2 in \cref{Properties of Doeblin coefficient}, $\gamma(\cdot)$ is an $n$-way metric just like TV distance is a metric between two distributions. Notably, when considering the special case of $n=2$ distributions, $\gamma(\cdot)$ reduces to the familiar TV distance. 
    \subsection{Max-Doeblin Coefficient and Its Minimal Coupling Characterization} \label{sec: Max-Doeblin Coefficient and its Minimal Coupling Characterization}
    The affinity characterization of TV distance between distributions $P, Q  \in \calP_m$ allows us to express it as
    \begin{equation}
         \| P -Q \|_\mathsf{TV} = 1 - \sum_{y \in \Y} \min\{P(y), Q(y)\}.
    \end{equation}
    As discussed in the previous section, the Doeblin coefficient allows the generalization of this characterization to $n$ distributions. Additionally, we can also express the TV distance between two probability distributions in the ``max-form'' as shown in the following equation:
    \begin{equation}
        \|P-Q\|_{\mathsf{TV}} = \sum_{y\in \Y} \max\{P(y),Q(y)\} - 1.
    \end{equation}
    Motivated by this formulation, we introduce a new coefficient called the \emph{max-Doeblin coefficient} as an alternative approach to generalize the TV distance using the maximum of distributions. The max-Doeblin coefficient is defined as follows.
    \begin{definition}[Max-Doeblin Coefficient]
        For a stochastic matrix $W \in \R^{n \times m}_{\mathsf{sto}}$, we define the max-Doeblin coefficient $\tau_{\max}(W)$ as 
    \begin{equation}
     \tau_{\max}(W) \triangleq \sum_{j=1}^m \max_{i\in [n]} W_{ij}.
     \end{equation}     
    \end{definition}
    We note that the logarithm of the {max-Doeblin} coefficient is also known as maximal leakage, and has been studied in both the computer security and information theory literature. Maximal leakage \cite{Geoffrey2009, Issa2020, Alvim2014AdditiveAM, BRAUN200975}, denoted as $\mathcal{L}(X \rightarrow Y)$, quantifies the amount of information leakage from a random variable $X$ to another random variable $Y$. 
    In particular, maximal leakage from $X$ to $Y$ is defined as
    \begin{equation}
    \mathcal{L}(X \rightarrow Y) \triangleq \log \bigg(\sum_{y \in \Y} \max_{x: P_X(x)>0} P_{Y|X}(y|x) \bigg).
    \end{equation}
    Moreover, $\mathcal{L}(X \rightarrow Y)$ is known to be equal to Sibson's mutual information between $X$ and $Y$ of order infinity \cite{sibson1969information, verdu2015alpha, esposito2022sibson}. 
    Hence, our ensuing results on the max-Doeblin coefficient also carry over to these concepts.

    Note that the max-Doeblin coefficient $\tau_{\max}(W)$ satisfies the inequality $1 \leq \tau_{\max}(W) \leq n$. To obtain a normalized version of  $\tau_{\max}(W)$ taking values between $0$ and $1$, we introduce a new quantity $\gamma_{\max}(W)$ defined as follows:
    \begin{equation}
        \gamma_{\max}(W) \triangleq \frac{\tau_{\max}(W) -1}{n-1}. \label{Eq:maxTV}
    \end{equation}
    We will show that $\gamma_{\max}(W)$ is another suitable generalization of the TV distance to multiple distributions. Similar to the Doeblin coefficient $\tau(\cdot)$, we will interchangeably use the notation $\tau_{\max}(W) = \tau_{\max}(P_1,\ldots,P_n)$ for PMFs $P_{1}, P_{2}, \ldots, P_{n} \in \calP_m$ and channel $W = \big[P_1^{\T} \, \dots \, P_n^{\T}\big]^{\T} \in \R^{n \times m}_{\mathsf{sto}}$, and we will use analogous notation for $\gamma_{\max}(\cdot)$.  The ensuing subsections demonstrate that $\tau_{\max}(\cdot)$ and $\gamma_{\max}(\cdot)$ share several analogous properties to $\tau(\cdot)$ and $\gamma(\cdot)$ (discussed in \cref{Properties of Doeblin coefficient,Maximal Coupling}). 

    \subsubsection{Properties of Max-Doeblin Coefficient}\label{sec: Max-Doeblin Coefficient}
    In the following theorem, we will summarize various properties of max-Doeblin coefficients.
    \begin{theorem}[Properties of Max-Doeblin Coefficient] \label{Properties of max-Doeblin coefficient} Let $W = \big[P_1^{\T} \, \dots \, P_n^{\T}\big]^{\T} \in \R^{n \times m}_{\mathsf{sto}}$ be a channel formed by stacking the PMFs $P_{1}, \ldots, P_{n} \in \calP_m$. Then, the quantity $\gamma_{\max}(W)$ satisfies the following properties:
    \begin{enumerate}
    \item  \emph{(Normalization)} We have $0 \leq \gamma_{\max}(W) \leq 1$, where $\gamma_{\max}(W)=0$ if and only if $P_{1}=P_{2}=\dots=P_{n}$ (entry-wise), and $\gamma_{\max}(W)=1$ if and only if $P_{1}, \dots, P_{n}$ are pairwise disjoint. 
    \item  \emph{($n$-way Metric)} $\gamma_{\max}(P_1,\dots,P_n)$ is an $n$-way metric.
    \item  \emph{(Convexity {\cite{Issa2020}})} The map $W \mapsto \tau_{\max}(W)$ is convex.
    \item \emph{{(Monotonicity {\cite{Issa2020}})}} For channels $V \in \R^{k \times n}_{\mathsf{sto}} $ and $W \in \R^{n \times m}_{\mathsf{sto}}$,  the max-Doeblin coefficient of composite channel $VW$ is bounded as:
         \[ \tau_{\max}(VW) \leq \min\{\tau_{\max}(V), \tau_{\max}(W)\}.\]
    \item  \emph{(Tensorization {\cite{Issa2020}})} For channels $W \in \R^{n \times m}_{\mathsf{sto}} $ and $V \in \R^{l \times k}_{\mathsf{sto}}$, we have 
         \[\tau_{\max}(W\otimes V) =\tau_{\max}(W) \tau_{\max}(V). \]
    \item \emph{(Lower Bound {\cite{Issa2020}})} $\tau_{\max}(P_{Y|X})$ is lower bounded in terms of the mutual information between the random variables $(X, Y)$, where $(X, Y) \sim (P_X, P_{Y|X})$ and $P_X(x) > 0$ for all $x \in \mathcal{X}$:
    \begin{equation*}
        \tau_{\max}(P_{Y|X}) \geq \exp(I(X;Y)).
    \end{equation*}
    \item \emph{(Maximum Trace Characterization)} $\tau_{\max}(W)$ is the solution to the following optimization problem:
         $$\tau_{\max}(W) = \max_{P \in \R^{m\times n}_{\mathsf{sto}} } \mathsf{Tr}(P W),$$
         where the maximum is over all possible $m \times n$ row stochastic matrices $P$. 
    \item  \emph{(Optimal Estimator)} 
    Consider a hidden random variable $X$ that is uniformly distributed on $\X$, i.e., $X \sim \textup{unif}(\X)$, and the fixed channel (or observation model) $W = P_{Y|X} \in \R^{n\times m}_{\mathsf{sto}}$. Let $\hat{X} \in \X$ denote any (possibly randomized) estimator of $X$ based on the observation $Y$, which is defined by a kernel $P_{\hat{X}|Y} \in \R^{m\times n}_{\mathsf{sto}}$ such that $X \rightarrow Y \rightarrow \hat{X}$ forms a Markov chain. Then, we have
    $$\frac{\tau_{\max}(P_{Y|X})}{n} = \max_{P_{\hat{X}|Y} \in \R^{m\times n}_{\mathsf{sto}}} \P( \hat{X} = X),$$
    where the minimum is computed over all estimators $\hat{X}$, or equivalently, over all kernels $P_{\hat{X}|Y} \in \R^{m\times n}_{\mathsf{sto}}$, and $\P(\cdot)$ denotes the probability law of $(X,Y,\hat{X})$. 
    \end{enumerate}
    \end{theorem}
    The proof of \cref{Properties of max-Doeblin coefficient} is provided in \cref{Proof of Properties of Max-Doeblin coefficient}. Much of the proof follows from arguments that draw parallels to the corresponding properties of the Doeblin coefficient in \cref{Properties of Doeblin coefficient}. 
    These properties, particularly the $n$-way metric property, convexity, etc., strengthen the assertion that $\gamma_{\max}(\cdot)$ serves as another appropriate generalization of the TV distance to multiple distributions. While some of these properties were previously established in \cite[Corollary 1]{Issa2020}, we present the proofs in \cref{Proof of Properties of Max-Doeblin coefficient} for sake of completeness. Moreover, it is interesting to note that $\tau_{\max}$ also exhibits a minimal coupling characterization {(under a specific condition)} as discussed in the next subsection.

    \subsubsection{Minimal Coupling Characterization of Max-Doeblin Coefficient} \label{sec: Minimal Coupling Characterization of max-Doeblin Coefficient}
    In this subsection, we establish a minimal coupling characterization for the max-Doeblin coefficient, which is analogous to the maximal coupling characterization of the standard Doeblin coefficient.
    For any stochastic matrix $W \in \R_{\mathsf{sto}}^{n \times m}$, we first begin by defining $\tau_{\max_{2}}(W)$ as 
    \begin{equation}
    \tau_{\max\nolimits_{2}}(W) \triangleq \sum_{j=1} \max\nolimits_{2}\{W_{1j}, \dots, W_{nj}\}, \label{definition of max2}
    \end{equation} 
    where $\max\nolimits_2\{x_1,\dots, x_n\}$ represents the second largest value among $x_1,\dots,x_n$. In instances where multiple values are equal to the maximum value, $\max\nolimits_2\{\cdot\}$ would yield the same value as $\max\{\cdot\}$, which is the maximum value itself.
    The following theorem provides a comprehensive minimal coupling characterization of the max-Doeblin coefficient when $\tau_{\max_2}(W) \leq 1$.

    \begin{theorem}[Minimal Coupling for Max-Doeblin Coefficient] \label{thm:MaxDoeblinCoupling}
        For any random variables $Y_1, \dots, Y_n \in \Y$ distributed according to the PMFs $P_{1}, \dots, P_{n} \in \calP_m$, respectively, let $W \in \R^{n\times m}_{\mathsf{sto}}$ be a channel defined by $W = \big[P_1^{\T} \, P_2^{\T} \, \dots \, P_n^{\T}\big]^{\T}$.  
        If $\tau_{\max_2}(W) \leq 1 $, then we have 
        \begin{equation*}
            \tau_{\max}(W) = \min_{\substack{ P_{Y_1, \dots, Y_n} : \\ P_{Y_1 } = P_1, \, \dots \, , \, P_{Y_n} =P_n} } \! \sum_{y\in \Y} \P( \cup_{i=1}^n \{Y_i =y\}), \label{Eq:MaxDoeblinMinimalCoupling}
        \end{equation*}
    where the minimum is over all couplings of $P_{1}, \dots, P_{n}$, and $\P(\cdot)$ denotes the probability law corresponding to a coupling.
    \end{theorem}
    The proof of \cref{thm:MaxDoeblinCoupling} is provided in \cref{sec:Truelabel}. 
    For $n=2$, it is worth noting that the condition $\tau_{\max_2}(P_1,P_2) = \sum_{y \in \Y} \max\nolimits_2\{P_1(y),P_2(y)\}\leq 1$ is always satisfied, as it is equivalent to $\sum_{y\in \Y} \min\{P_1(y),P_2(y)\} \leq 1$, which is always true.
    Moreover, when $n=2$, our minimal coupling construction for the max-Doeblin coefficient reduces to Dobrushin's maximal coupling characterization of TV distance \cite[Proposition 4.7]{LevinPeresWilmer2009}. However, for $n \geq 3$, the condition $\tau_{\max_2}(W) \leq 1$ may not be true in general. For example, when $W = W_\delta$, where $W_\delta$ is a $q$-ary symmetric channel with cross-over probability $\delta$ \cite[Equation (10)]{MakurPolyanskiy2018}, the condition $\tau_{\max_2}(W_\delta) \leq 1 $ is satisfied only when $\delta  \leq 1 - 1/q$. (On the other hand, it turns out that an erasure channel $\text{E}_\epsilon$ with erasure probability $\epsilon$ always satisfies the condition $\tau_{\max_2}(\text{E}_\epsilon) \leq 1$.) 
    
     We will first provide our coupling construction for $n=3$ in \cref{sec: Proof of MaxDoeblinCoupling}. Then, we will generalize this construction to any general $n$ in \cref{Proof of MaxDoeblinCoupling for general n}. The generalization of the coupling construction utilizes the maximum-minimums identity and the inclusion-exclusion principle from probability theory. It is also worth mentioning that our coupling construction (in \cref{Proof of MaxDoeblinCoupling for general n}) becomes invalid when $\tau_{\max_2}(W) > 1$. This observation raises the question of whether the condition $\tau_{\max_2}(W) \leq 1$ is necessary, or is it merely an outcome of the construction technique employed.
    
    In the following proposition, we demonstrate that for $n=3$, the condition $\tau_{\max_2}(W) \leq 1$ is in fact needed. 
    
    \begin{proposition}[Minimal Coupling for Max-Doeblin Coefficient when $n=3$] \label{New Minimal Coupling for max-Doeblin Coefficient}
    For any random variables $Y_1, Y_2, Y_3 \in \Y$ distributed according to the PMFs $P_{1}, P_2, P_{3} \in \calP_m$, respectively, let $W \in \R^{3\times m}_{\mathsf{sto}}$ be a channel defined by $W = \big[P_1^{\T} \, P_2^{\T} \, P_3^{\T}\big]^{\T}$. 
    Then, we have 
        \begin{equation*}
        \begin{aligned}
            \tau_{\max}(W) + &  (\tau_{\max_2}(W)-1)_+  \\ 
            & = \min_{\substack{ P_{Y_1, Y_2, Y_3} : \\ P_{Y_1 } = P_1\,, \dots\, , P_{Y_3} =P_3} } \! \sum_{y\in \Y} \P( \cup_{i=1}^3 \{Y_i =y\}). 
        \end{aligned}
        \end{equation*}
    \end{proposition}
    The proof is provided in \cref{Proof of New Minimal Coupling for max-Doeblin Coefficient}. The proof utilizes two distinct coupling constructions, each applicable to different regimes based on the value of $\tau_{\max_2}(W)$. Specifically, the coupling construction for $\tau_{\max_2}(W) \leq 1$ is the same as that described in \cref{sec: Proof of MaxDoeblinCoupling}. However, a slightly different coupling construction is used when $\tau_{\max_2}(W) > 1 $. (Also, it is worth mentioning that when $\tau_{\max_2}(W) = 1$, then both the constructions coincide.) 

    We remark that in addition to the maximal leakage motivation discussed earlier, the notion of max-Doeblin coefficient can also be interpreted using the following scenario. Let $Y$ be an unknown random variable taking values uniformly from a finite set $\Y$. Let ${P_1,\dots, P_n \in \calP_m}$ be a collection of PMFs on $\Y$ representing the beliefs of $n$ experts or agents with estimates $Y_1,  \dots, Y_n$ that are independent of $Y$, respectively, about the value of the unknown variable $Y$. Under any joint probability law $\P$, the probability that at least one agent is correct is given by
    \begin{equation}
    \begin{aligned}
        \P(\text{at least one agent is correct}) = \frac{1}{m} \sum_{j=1}^m \P(\cup_{i=1}^n \{Y_i =y_j\}).
    \end{aligned}
    \end{equation}
    By \cref{thm:MaxDoeblinCoupling}, the minimum probability that at least one of the agents correctly predicts the unknown variable $Y$ over all possible couplings of their beliefs (under the condition that $\tau_{\max_2} (P_1,\dots,P_n) \leq 1$) is given by  
        \begin{equation*}
            \frac{\tau_{\max}(W)}{m} = \min_{\substack{ P_{Y_1, \dots, Y_n} : \\ P_{Y_1 } = P_1, \, \dots \, , \, P_{Y_n} =P_n} } \!  \frac{1}{m} \sum_{j=1}^m\P( \cup_{i=1}^n \{Y_i =y_j\}). 
        \end{equation*}
    This provides another interesting interpretation of the max-Doeblin coefficient.
    
    Finally, a result that follows immediately from the coupling construction in \cref{thm:MaxDoeblinCoupling} is that we can establish a coupling for $n$ random variables $Y_1, \ldots, Y_n$ with PMFs $P_1, \ldots, P_n$ that simultaneously achieves $ \sum_{y \in \Y} \P(\cap_{i=1}^n \{Y_i =y \}) = \tau(P_1,\ldots,P_n)$ and  $ \sum_{y \in \Y} \P(\cup_{i=1}^n \{Y_i =y \}) = \tau_{\max}(P_1,\ldots,P_n)$ when $\tau_{\max_2}(P_1, \dots, P_n) \leq 1$. This result is stated in the next proposition. 
    \begin{proposition}[Simultaneous Coupling for Doeblin and Max-Doeblin Coefficients] \label{Simultaneous Coupling for Doeblin and max-Doeblin}
        For any random variables $Y_1,\ldots,Y_n$ with PMFs $P_{1},\dots, P_{n} \in \calP_m$, respectively, if $ \tau_{\max_2}(P_1,\dots,P_n) \leq 1$, then there exists an extremal coupling $\P$ of $P_{1},\dots, P_{n}$ such that
        \begin{equation}
        \begin{aligned}
            \sum_{y \in \Y}\P(\cap_{i=1}^n \{Y_i =y \})  &= \tau(P_1,\ldots,P_n),\\
            \sum_{y \in \Y}\P(\cup_{i=1}^n \{Y_i =y \})  &= \tau_{\max\nolimits}(P_1,\ldots,P_n).
        \end{aligned}
        \end{equation}
    \end{proposition}
    The proof is presented in \cref{Proof of Simultaneous Coupling for Doeblin and max-Doeblin}.

    {It is worth mentioning at this stage that many of the results in this work can be extended beyond the finite alphabet setting considered. For example, \cref{Properties of Doeblin coefficient,Maximal Coupling} could be extended to settings with output alphabet sets $\Y$ that are Euclidean spaces. However, extending the scope of the results presented in this work, e.g., \cref{Maximal Coupling,thm:MaxDoeblinCoupling}, to general measurable input and output spaces poses certain challenges, especially in the context of the coupling characterization of the max-Doeblin coefficient. In fact, even defining a maximal coupling for Doeblin coefficients requires a careful measure-theoretic treatment if the input alphabet set is uncountable. So, we leave the development of such results over general measurable spaces for future work in the area.} 

\subsection{DeGroot Distance} \label{sec: DeGroot Distance}
    In this subsection, we extend the concept of DeGroot distance to the case of multiple probability distributions, resulting in two new information measures: the min-DeGroot and max-DeGroot distances (akin to the standard Doeblin and max-Doeblin coefficients). {Consider a Markov chain $X \rightarrow Y \rightarrow \hat{X}$, where $X$ is a random variable following a Bernoulli distribution with parameter $\lambda \in [0,1]$. The conditional probability distribution $P_{Y|X}$ is defined via the PMFs $P_1$ when $X = 0$ and $P_2$ when $X = 1$. The goal here is to develop an estimator $\hat{X}$ for $X$ based on $Y$ that minimizes the risk for the loss function $\I_{\hat{X} \neq  X}$, where $\I$ denotes the indicator variable. The \emph{DeGroot distance} \cite{DeGroot1962}, also known as Bayes statistical information, quantifies the reduction in Bayes risk when estimating $X$ upon observing $Y$, as compared to the scenario when $Y$ is not observed. Recall that DeGroot distance $D_{\lambda}(P||Q)$ between two distributions $P$ and $Q$ is given by \cite{Raginsky2016}: 
    \begin{equation}
         D_{\lambda}(P||Q) = \|\lambda P -  (1-\lambda) Q\|_\mathsf{TV} - \frac{1}{2} |1 - 2\lambda|      
    \end{equation}  
    Since TV distance allows us to define DeGroot distance, we extend this concept further. Building upon the $n$-way generalization of TV distance, we now generalize the DeGroot distance to the scenario where we have multiple probability distributions (i.e., $X$ is a general categorical or discrete random variable). This yields two distinct notions of statistical information: the min-DeGroot and max-DeGroot distances.
    
    \subsubsection{Min-DeGroot Distance} \label{Min-DeGroot distance}   
        Given any prior PMF $\lambda \in \calP_n$, consider a hidden random variable $X \in \X$ and an observed random variable $Y \in \Y$ such that $|\X| = n$, $|\Y| = m$, and  
            \begin{equation}
                 X \sim \lambda \text{ and } P_{Y | X=x_i}=P_{i} \label{Distribution}
            \end{equation}  
        for $i \in [n]$, where $P_{1}, \dots , P_{n} \in \calP_m$ determine the channel (or observation model) $P_{Y|X}$, and we let $\X = \{x_1,\dots,x_n\}$ without loss of generality. Let $\hat{X} \in \X$ denote any (possibly randomized) estimator of $X$ based on $Y$, which is defined by a kernel $P_{\hat{X}|Y} \in \R^{m\times n}_{\mathsf{sto}}$ such that $X \rightarrow Y \rightarrow \hat{X}$ forms a Markov chain. 
        Suppose the goal is to minimize the \emph{risk} for the loss function $l(X, \hat{X}) =  \I_{\hat{X}=X}$. When we have no observations related to $X$, the Bayes optimal estimator is to choose the least likely $x \in \X$, i.e., $\hat{X}^* = x_{i^*}$ with $i^* =  \argmin_{i \in [n]} \lambda_{i}$, and the corresponding Bayes risk $R^{*}_\lambda$ is 
            \begin{equation}
                R^{*}_\lambda = \min\{ \lambda_{1}, \dots , \lambda_{n}\}.
             \end{equation} 
             
        On the other hand, when $Y$ is observed, the Bayes optimal estimator $\hat{X}^*$ is given by the solution to the following problem:
        \begin{equation}
        P_{\hat{X}^*|Y} = \argmin_{P_{\hat{X}|Y} \in \R^{m\times n}_{\mathsf{sto}} } \P(\hat{X}=X),
        \end{equation}
        where the minimum is over all possible estimators. In statistical decision theory, is it well-known that for an observation model $P_{Y|X} \in \R^{n\times m}_{\mathsf{sto}}$, prior PMF $P_X \in \calP_n$, loss function $l: \X \times \X \rightarrow \R$, and randomized estimator $P_{\hat{X}|Y} \in \R^{m\times n}_{\mathsf{sto}}$, the risk function is defined as
                \begin{align}
                    R_{P_X}( & P_{Y |X}, l, P_{\hat{X}|Y} )  \triangleq \E[ l(X, \hat{X})] \nonumber \\
                    &  = \sum_{x \in \X}\sum_{y\in \Y}\sum_{\hat{x}\in \X} l(x, \hat{x}) P_X(x) P_{Y |X} (y|x) P_{\hat{X}|Y}(\hat{x}|y) \nonumber \\
                   &  = \mathsf{Tr}(L^{\T} \diag(P_X) P_{Y|X} P_{\hat{X}|Y }),
                    \label{Eq: Trace risk}
                \end{align}
        where $L \in \R^{n \times n}$ is the matrix representation of the loss function $l : \X \times \X \rightarrow \R$, i.e., the $(i,j)$th entry of $L$ is $l(x_i, \hat{x}_j)$ for $ i, j \in [n]$ and $x_i, \hat{x}_j \in \X$.  By our earlier choice of the loss function, we have $L = I$. Moreover, we also have $P_{X} = \lambda$. Letting $\Lambda = \diag(\lambda)$ be the diagonal matrix with $\lambda$ along its principal diagonal, we see that the Bayes risk is 
                \begin{align}
                   R^*_{\lambda}( &P_{Y|X})  = \nonumber \min_{P_{\hat{X}|Y} \in \R^{m\times n}_{\mathsf{sto}}}  \P(\hat{X}  = X)\\
                   &    = \nonumber \min_{P_{\hat{X}|Y} \in \R^{m\times n}_{\mathsf{sto}}} \mathsf{Tr}( \Lambda P_{Y \mid X} P_{\hat{X} \mid Y}) \\
                   & =  \sum_{y\in \Y} \min\{\lambda_1 P_{Y|X}(y|x_1), \dots ,\lambda_n P_{Y|X}(y|x_n) \}.
                \end{align}            
        Finally, we can define the \emph{min-DeGroot distance} as $\tilde{\tau}_{\min}(\lambda, P_{Y|X}) \triangleq R^*_{\lambda} - R^*_{\lambda}(P_{Y|X}) $, which gives 
            \begin{equation}
                \begin{aligned}
                     \tilde{\tau}_{\min }(&\lambda, P_{Y|X})   = \min\{\lambda_1, \dots, \lambda_n\}  \\ 
                     & - \sum_{y\in \Y} \min\{\lambda_1 P_{Y|X}(y|x_1), \dots ,\lambda_n P_{Y|X}(y|x_n) \}.
                \end{aligned}
            \end{equation}
        Note that $\tilde{\tau}_{\min }(P_{Y|X}) \geq 0$ can be interpreted as a measure of the statistical information obtained about $X$ after observing $Y$. 
    \subsubsection{Max-DeGroot Distance} 
        Consider the same setting as \cref{Min-DeGroot distance}, where we are given a pair of random variables $(X, Y)$ distributed as in \eqref{Distribution}. This time, fix the loss function as $l(X,\hat{X}) = \I_{\hat{X} \neq X}$. Then, the Bayes risk $R^*_\lambda $ in estimating $X$ when we do not observe $Y$ is  
            \begin{equation}
                R^*_\lambda = 1 - \max\{\lambda_1, \dots ,\lambda_n\}.
            \end{equation}
        On the other hand, when $Y$ is observed, since $L = \1\1^{\T} - I$ and $\diag(P_X) = \Lambda $ in the context of \eqref{Eq: Trace risk}, the Bayes risk $R^*_\lambda(P_{Y|X})$ is given by 
                \begin{align}
                     & R^*_\lambda (P_{Y|X})   = \min_{P_{\hat{X}|Y} \in \R^{m\times n}_{\mathsf{sto}}} \mathsf{Tr}( (\1\1^\T - I)\Lambda P_{Y|X} P_{\hat{X}|Y} ) \nonumber \\
                     & = \mathsf{Tr}(\Lambda P_{Y|X} \1\1^\T) - \max_{P_{\hat{X}|Y} \in \R^{m\times n}_{\mathsf{sto}}} \mathsf{Tr}(\Lambda P_{Y|X} P_{\hat{X}|Y} ) \nonumber \\
                     & = 1 - \sum_{y \in \Y}\max\{\lambda_1 P_{Y|X}(y|x_1), \dots ,\lambda_n P_{Y|X}(y|x_n) \}.    
                \end{align}            
        Hence, we can define the quantity $\tilde{\tau}_{\max }(\lambda, P_{Y|X}) \triangleq R^*_\lambda - R^*_\lambda(P_{Y|X})$ as the \emph{max-DeGroot distance}:
            \begin{equation}
                 \begin{aligned}
                   \tilde{\tau}_{\max }(&\lambda, P_{Y|X}) \\ 
                   &  = \sum_{y \in \Y}\max\{\lambda_1  P_{Y|X}(y|x_1), \dots ,\lambda_n P_{Y|X}(y|x_n) \}  \\
                   &  \ \ \ \ \ \ \ \ \ \ \ \ - \max\{\lambda_1, \dots ,\lambda_n\}.            
                \end{aligned} 
            \end{equation}
            
        Observe that for the special case where $n = 2$, both min-DeGroot and max-DeGroot distances are equivalent and the same as classical DeGroot distance \cite{Raginsky2016}. Indeed, we have
            \begin{align}
                \nonumber \min\{& \lambda,\bar{\lambda}\} - \sum_{y \in \Y} \min \{\lambda P_{Y|X}(y|x_1), \bar{\lambda}P_{Y|X}(y|x_2) \} \\
                 & = \sum_{y \in \Y} \max \{\lambda P_{Y|X}(y|x_1), \bar{\lambda}P_{Y|X}(y|x_2) \} -  \max\{\lambda,\bar{\lambda} \}             \nonumber \\
                 & = \frac{1}{2}\|\lambda P_{Y|X= x_1} -  \bar{\lambda} P_{Y|X = x_2} \|_{1} - \frac{1}{2} |1-2\lambda|,
                \end{align}
        where $\bar{\lambda} = 1 - \lambda $ for $\lambda \in [0,1]$. 

\subsection{Simultaneously Maximal Coupling} \label{Sec: Simultaneously Maximal Coupling}
    From this subsection onward, we return to investigating Doeblin coefficients. In particular, in this subsection, we generalize Goldstein's simultaneously optimal coupling result, cf. \cite{Goldstein1979, Polyanskiy2017}, to the case where we have $n$ distributions rather than just two. Given a finite collection of (bivariate) probability distributions $P_{X_1, Y_{1}}, P_{X_{2}, Y_{2}}, \dots, P_{X_{n}, Y_{n}}$, the next theorem constructs a joint coupling which is simultaneously maximal with respect to the joint distributions of $(X_i, Y_i)$ for $i \in [n]$ and the marginal distributions of $X_{1}, \dots, X_{n}$. 
    \begin{theorem}[Simultaneously Maximal Coupling] \label{Simultaneous Coupling}
    Given $n$ probability distributions $P_{X_1, Y_{1}},P_{X_{2}, Y_{2}}, \dots, P_{X_{n}, Y_{n}}$, where the random variables $X_i \in \X$ and $Y_i \in \Y$ for $i\in [n]$ and finite alphabets $\X$ and $\Y$, there exists a coupling $P_{X_{1}, Y_{1}, X_{2}, Y_{2}, \dots , X_{n}, Y_{n}}$ of $P_{X_1, Y_{1}}, \dots, P_{X_{n}, Y_{n}}$ that is simultaneously maximal in the sense that 
        \begin{align*}
            \tau(P_{X_1, Y_1}, \dots, P_{X_{n}, Y_{n}})  &= \P(X_{1}=\dots=X_{n}, Y_{1}=\dots=Y_{n}) \\
            \tau(P_{X_{1}}, \dots , P_{X_{n}})  &=\P(X_{1}=\dots=X_{n}),
        \end{align*}  
        where $\P(\cdot)$ denotes the probability law under the coupling.
    \end{theorem}
    The proof is provided in \cref{Proof of Simultaneous Coupling}. It's worth mentioning that it is not always possible to construct a triply optimal coupling, one that is optimally coupled with respect to the joint distributions $P_{X_1,Y_1}, \dots, P_{X_n,Y_n}$ and both pairs of marginals $P_{X_1}, \dots, P_{X_n}$ and $P_{Y_1}, \dots, P_{Y_n}$ (even for $n=2$). An example illustrating this can be found in \cite[Remark 10]{Polyanskiy2017}. We will use \cref{Simultaneous Coupling} to establish the information contraction properties of Doeblin coefficients over Bayesian networks.
    
\subsection{Contraction over Bayesian Networks}\label{Sec: Contraction over Bayesian Networks}
    In this subsection, inspired by the ideas in \cite{Polyanskiy2017}, we derive the contraction properties of Doeblin coefficients over Bayesian networks. {Recall that a Bayesian network is a directed graphical model characterized by a directed acyclic graph with the vertices representing random variables taking values in fixed alphabet sets. The edges in the graph encode conditional dependencies between the vertex random variables. Specifically, the joint distribution of all vertex random variables (in topological ordering) can be factorized as a product of the conditional distributions of each vertex given its parents.  
    We refer readers to \cite[Section 4.2]{pml2Book} for detailed formal definitions. In this work, we will assume for simplicity that vertex random variables take values on finite sets. Examples of Bayesian networks with finite alphabet sets include Markov chains on finite state spaces \cite{LevinPeresWilmer2009}, broadcasting trees and Ising models \cite{Evansetal2000}, and noisy computation circuits \cite{evans1999signal}. Furthermore, we also refer readers to \cite[Section 4.2]{pml2Book} for other specific practically motivated examples of Bayesian networks with finite sets, e.g., models of student performance dependencies.} 
    \begin{figure}
        \centering
        \includegraphics[width = 0.43\textwidth]{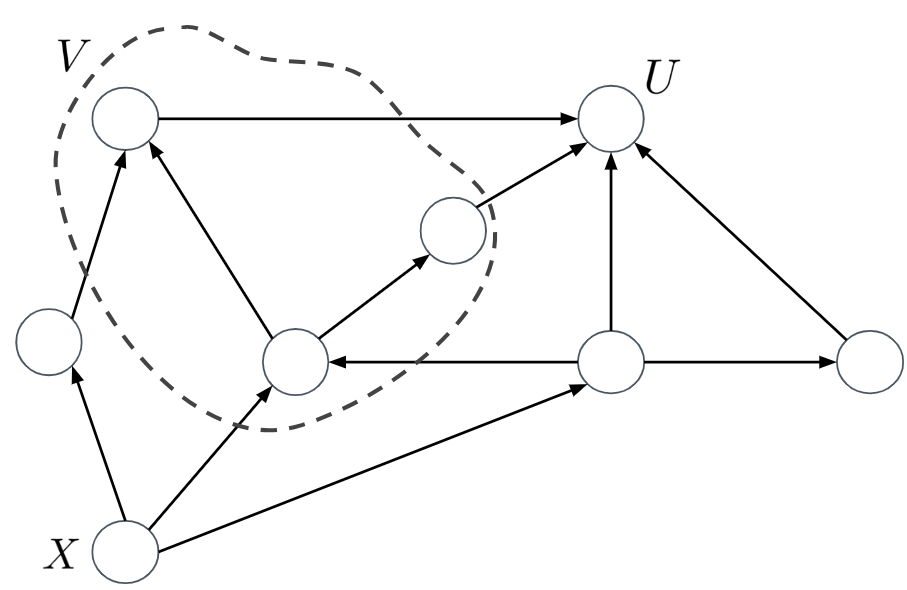} 
        \label{fig1}
        \caption{Illustration of a Bayesian network depicting a source node denoted by $X$, a sink node $U$, and a subset $V$ consisting of nodes enclosed within dotted lines. }
        \label{Figure 1}
    \end{figure} 

    Given the Doeblin coefficients for each of the channels comprising a Bayesian network, we are interested in obtaining a bound on the Doeblin coefficient of the composite channel from a single source to any sink of nodes. To this end, consider a Bayesian network represented by a finite directed acyclic graph with vertex set $\mathcal{V}$. As noted above, every vertex $U$ in $\mathcal{V}$ is a random variable that takes values in some finite alphabet set.  {(These alphabet sets can be different for different vertices; we do not define notation for them since they will not be required in the ensuing arguments.)} Let the source node be denoted by $X$ (which we assume is the unique source node with unspecified marginal), and suppose that every vertex $U$ other than the source node is associated with a conditional distribution $P_{U|\pa(U)}$, where $\pa(U)$ denotes the set of parents of $U$. These conditional distributions collectively define the Bayesian network \cite{pml2Book}. We assume that the vertices are topologically sorted, and for any node $U$ and \emph{subset} of nodes $V \subseteq \mathcal{V}$, we write $U > V$ when there is no directed path from $U$ to $V$. {We include an illustration of a Bayesian network in \cref{Figure 1} for the benefit of readers.}
    
    Moreover, in the spirit of \cite{Polyanskiy2017}, we define a \emph{site percolation} process on $\mathcal{V}$ so that each vertex $U$ in $\mathcal{V}$ is removed independently with probability $\tau_{U}$, where $\tau_{U} = \tau(P_{U|\pa(U)})$ denotes the Doeblin coefficient at $U$. Finally, let $\operatorname{perc}(V)$ be the probability (with respect to the percolation process) that there is an (open) path from $X$ to a subset of nodes $V \subseteq \mathcal{V}$.
    \begin{theorem}[Doeblin Coefficients in Bayesian Networks] \label{Doeblins Coefficient in Networks}
        For any node $U$ in $\mathcal{V}$ and any subset of nodes $V \subseteq \mathcal{V}$ such that $U>V$, we have 
        $$
        \tau(P_{V \cup\{U\} \mid X}) \geq \tau_U \tau(P_{V \mid X}) + (1- \tau_U) \tau(P_{V \cup \pa(U) \mid X}).
        $$    
    Furthermore, under the aforementioned site percolation process model, we have for every $V \subseteq \mathcal{V}$,
    $$
        \gamma(P_{V \mid X}) = 1 - \tau(P_{V \mid X}) \leq \operatorname{perc}(V).
    $$
    \end{theorem}

    The proof is provided in \cref{Proof of Doeblins Coefficient in Networks}. Note that, intuitively, $\tau(\cdot)$ is tending towards unity over the network, which is reasonable since a stochastic matrix typically transforms input probability distributions into ``more uniform'' distributions. Therefore, a Doeblin coefficient closer to one implies that the corresponding channel is \emph{losing information} contained in source distributions. In this sense, \cref{Doeblins Coefficient in Networks} captures information contraction over Bayesian networks as measured by Doeblin coefficients.
    Moreover, the percolation bound in \cref{Doeblins Coefficient in Networks} is a strengthening of that in \cite{Polyanskiy2017} as $1-\tau(P_{V|X}) \geq \eta_{\mathsf{TV}}(P_{V|X})$. 
 
    Next, consider a path $\pi = (X, \dots, U)$ that originates from the source node $X$ and terminates at a node $U$ in the set $V \subseteq \calV$. We say that such a path $\pi$ from $X$ to $V$ is ``shortcut-free,'' denoted as $X \stackrel{sf}{\rightarrow} V$, if there is no other path $\pi^\prime$ from $X$ to any node in $V$ that is a subset of $\pi$. 
    Furthermore, for any given path $\pi = (X, U_1, \ldots, U_n)$, we define $(1-\tau)^\pi$ as the product of $(1-\tau_{U_j})$ over all nodes $U_j$ along the path. Using this notation, we obtain the following corollary which states that for any subset $V \subset \calV$, the quantity $\tau(P_{V \mid X})$ is upper bounded by the sum of $(1-\tau)^\pi$ over all shortcut-free paths $\pi$ from $X$ to $V$.
\begin{corollary}[Bound on Doeblin Coefficient of Composite Channel]\label{Evans-Schulman}
    For any subset $V \subseteq \calV$, we have
    $$
        1 - \tau(P_{V \mid X}) \leq \sum_{\pi: X \stackrel{sf}{\rightarrow} V} (1-\tau)^\pi.
    $$
\end{corollary} 
    The proof of \cref{Evans-Schulman} is presented in \cref{Proof of Evans-Schulman}.
    One notable advantage of the aforementioned corollary is its computational simplicity, particularly in comparison to other measures such as the contraction coefficient for KL divergence. This simplicity may render it a convenient tool for downstream analysis. Indeed, results such as this play a crucial role in deriving impossibility results in various domains, such as reconstruction on graphs \cite{Evansetal2000,MakurMosselPolyanskiy2020,makur2022broadcasting} and reliable computation using noisy circuits \cite{evans1999signal}. Moreover, the corollary generalizes the results of \cite{Polyanskiy2017, evans1999signal} from TV distance to Doeblin coefficients. 

\subsection{Samorodnitsky's SDPI for Doeblin Coefficients}

    We next establish a generalization of Samorodnitsky’s SDPI for Doeblin coefficients in analogy with the results in \cite{Samorodnitsky2016, Polyanskiy2017, MakurZheng2020} for various $f$-divergences. To this end, consider discrete random variables $U, X_1, X_2, \dots, X_n$ and $Y_1, Y_2, \dots, Y_n$ on finite alphabets and let $X^n \triangleq (X_1, X_2, \dots, X_n)$ and $Y^n \triangleq (Y_1, Y_2, \dots, Y_n)$. Suppose $U \rightarrow X^n \rightarrow Y^n$ forms a Markov chain for a channel $P_{X^n|U}$ and a \emph{memoryless} channel $P_{Y^n|X^n}$, which is a tensor product of the channels $P_{Y_j|X_j}$ for $j \in [n]$:
    \begin{equation}
        P_{Y^n|X^n} = \prod_{j=1}^n P_{Y_j|X_j}.\vspace{-1mm}
    \end{equation} 
    Define the Doeblin coefficient of each individual channel as $\tau_j = \tau(P_{Y_j|X_j})$ for $j \in [n]$. 
    Using the tensorization and sub-multiplicativity properties in \cref{Properties of Doeblin coefficient}, we obtain the following bound on the Doeblin coefficient of $P_{Y^n|U}$:
        \begin{align}
            \tau(P_{Y^n|U}) & \nonumber \geq \tau(P_{Y^n|X^n}) + \tau(P_{X^n|U}) (1 - \tau(P_{Y^n|X^n}))  \\
            & \geq \prod_{j=1}^n \tau_j.        
        \end{align}     
    Alternatively, since $U \rightarrow X^n \rightarrow Y^n$ defines a Bayesian network, this bound can be obtained by applying \cref{Doeblins Coefficient in Networks} $n$ times as shown below:
    \begin{align}
         \tau (P_{Y^n|U}& )   \nonumber= \tau(P_{Y_1 \cup (Y_2, \dots, Y_n)|U} ) \\
         &  \geq \tau_{1} \tau(P_{(Y_2, \dots,Y_n)|U}) +(1 - \tau_1) \tau(P_{Y_1 \cup (X_2,\dots, X_n)|U}) \nonumber \\  
        &  \geq \tau_{1} \tau(P_{(Y_2, \dots,Y_n)|U}) \geq \prod_{j=1}^n \tau_j.    
    \end{align}
     This bound parallels the bound on contraction coefficients for operator convex $f$-divergences in \cite[Equation (61)]{MakurZheng2020}. In \cite[Section 6.2]{Polyanskiy2017} and \cite[Section 3.4]{MakurZheng2020}, the authors argue that stronger bounds can be obtained using more refined knowledge of the distribution $P_{U,X^n}$. 
    In this vein, the ensuing theorem presents tighter bounds on $\tau(P_{Y^n|U})$ in terms of the single-letter Doeblin coefficients $\{\tau_j : j \in [n]\}$ for the given Bayesian network structure.

    \begin{theorem}[Samorodnitsky’s SDPI for Doeblin Coefficients] \label{Samorodnitskys SDPI in terms of Doeblin Coefficient}
    For the Markov chain $U \rightarrow X^n \rightarrow Y^n$ with a memoryless channel $P_{Y^n|X^n} = \prod_{j=1}^n P_{Y_j|X_j}$, we have 
        $$ 
                 \tau(P_{Y^n|U}) \geq \sum_{T \subseteq [n]} P(T) \tau(P_{X_T|U}),   
        $$
    where $P(T)$ is the probability of generating a subset $T$ of $[n]$ by independently drawing each element $i \in [n]$ with probability $1-\tau_i$, and $X_T = \{X_i : i \in T\}$ for any subset $T \subseteq [n]$.  Specifically, when $\tau_j = \tau$ for all $j \in [n]$, the following bound holds: 
        \begin{equation*}
            \begin{aligned}
                \tau(P_{Y^n|U})  \geq \sum_{T \subseteq [n]} \tau^{n-|T|} (1-\tau)^{|T|} \tau(P_{X_T|U}).
            \end{aligned}
        \end{equation*}
     \end{theorem} 
     The proof is provided in \cref{Proof of Samorodnitskys SDPI in terms of Doeblin Coefficient}. 

\subsection{Application to PMF Fusion}
\label{Sec: Application to PMF Fusion}
    Finally, in this subsection, we will apply the coupling characterization of Doeblin coefficients developed earlier to the general problem of aggregation or \emph{fusion} of PMFs \cite{Koliander2022}. PMF fusion is a fundamental problem in statistics and machine learning where one tries to combine multiple sources of information or data into a single probability distribution, often for the purpose of estimating a probability distribution accurately. This problem is useful in many applications, such as in resource-constrained Bayesian inference, where multiple sources of data or prior beliefs need to be combined to produce a single posterior distribution \cite{GenestZidek1986, MitraRichardsShreyas2020, Jadbabaie2012, Jadbabaie2013, Lalitha2018, Nedic2017, Inan2022, Bordignon2021}. We next propose an optimization-based framework for PMF fusion, where the aggregated PMF is obtained as the solution to an optimization problem. This differs from previous approaches that relied on linear, log-linear, or Bayesian projection methods \cite{Nedic2017}. 
\subsubsection{PMF Fusion: Optimization-based Approach} 
    Suppose we have $n$ agents, each with their own PMFs $P_1,\dots, P_n \in \calP_m$, representing their beliefs about some phenomenon of interest. 
    The goal is to generate a combined belief by aggregating their individual PMFs. One approach to achieving this is to use an optimization method to find the aggregated PMF that maximizes or minimizes some measure of discrepancy between the individual PMFs. This approach has been studied using various measures, such as weighted KL divergence and weighted $\alpha$-divergences (see \cite{Koliander2022} for a detailed survey). 

    Inspired by Doeblin coefficients, we propose another approach that focuses on maximizing the probability of agreement {or \emph{``consensus''}} among the agents over all possible couplings of their beliefs. More precisely, to fuse the PMFs $P_1,\dots, P_n$, we first find the joint distribution or coupling that maximizes the likelihood that the agents' beliefs, represented as random variables $Y_1, \dots, Y_n$ with PMFs $P_1,\dots, P_n$, respectively, agree with each other over all possible couplings of their beliefs, i.e.,
        \begin{equation}
            \max_{\substack{ P_{Y_1, \dots, Y_n} : \\  P_{Y_1 } = P_1,  \dots ,P_{Y_n} =P_n} } \P( Y_1 =  \dots = Y_n), 
        \end{equation} 
    where $\P(\cdot)$ denotes the probability law corresponding to a coupling. From \cref{Maximal Coupling}, we know that the above optimization problem is precisely the Doeblin coefficient of the PMFs. Then, we define the optimal fused PMF $P_* \in \calP_m$ of $P_1,\dots, P_n$ as the conditional PMF of any single agent given that agreement occurs among all agents (with respect to the maximal coupling), i.e., $P_*(y) \propto \P(Y_1 = \dots = Y_n = y)$ for all $y \in \Y$. 
     
     Since \cref{Maximal Coupling} also tells us that $\P(Y_1= \dots = Y_n = y) = \min\{P_1(y), \dots, P_n(y) \}$ for the maximal coupling, we can write the optimal fused PMF as: 
        \begin{equation}
            \forall y \in \Y, \enspace P_*(y) = \frac{\min\{P_1(y), \dots, P_n(y) \}}{ \sum_{y\in \Y} \min\{P_1(y), \dots, P_n(y) \} } .
            \label{aggregation objective}
        \end{equation}
    We refer to the above fusion method as the \emph{min-rule}. It turns out that such a min-rule has been proposed in the literature, cf. \cite{MitraRichardsShreyas2020}. Indeed, the min-rule was used for distributed hypothesis testing in \cite{MitraRichardsShreyas2020} based on the intuition that if there is a true state of the world $y^* \in \Y$, and if there is an agent that can distinguish $y$ from $y^*$ for every false state $y \in \Y \backslash \{y^*\}$, then the min-rule can drive the beliefs of distributed agents to zero on each of the false states. For example, \cite{MitraRichardsShreyas2020} established that the min-rule enjoys faster asymptotic convergence than other belief averaging-based approaches, such as arithmetic pooling \cite{Jadbabaie2012, Jadbabaie2013} and geometric pooling \cite{Lalitha2018, Nedic2017, Inan2022, Bordignon2021}. However, the min-rule did not have a formal probabilistic interpretation. To remedy this, our development reveals the underlying probabilistic interpretation of the min-rule using the coupling for Doeblin coefficients.           

    It is worth mentioning that the min-rule in \eqref{aggregation objective} is only suitable under certain settings, such as group decision problems \cite{French2011}, where different experts are given equal importance and each expert is trying to construct optimal estimates based on their knowledge. If some of the agents are either adversarial or have imperfect knowledge about the environment, then the min-rule may not be appropriate. 

\section{Proofs of Properties and Coupling Constructions} \label{sec: Proofs}
    In this section, we will provide the proofs of various properties of the Doeblin coefficient discussed in \cref{Properties of Doeblin coefficient}. Additionally, we provide the extremal coupling characterization of Doeblin and max-Doeblin coefficients to prove \cref{Maximal Coupling,thm:MaxDoeblinCoupling,New Minimal Coupling for max-Doeblin Coefficient,Simultaneous Coupling for Doeblin and max-Doeblin}. Specifically, the proof of the properties of the Doeblin coefficient is presented in \Cref{Proof of Properties of Doeblin}. The coupling characterization of the Doeblin coefficient is detailed in \Cref{Proof of Maximal Coupling}. In \Cref{sec:Truelabel}, we delve into the coupling characterization. Initially, we establish this characterization for the case when $n=3$ and subsequently extend it to arbitrary $n$. Finally, \Cref{Proof of New Minimal Coupling for max-Doeblin Coefficient} and \Cref{Proof of Simultaneous Coupling for Doeblin and max-Doeblin} provide proofs of \Cref{New Minimal Coupling for max-Doeblin Coefficient,Simultaneous Coupling for Doeblin and max-Doeblin}.

\subsection{Proof of \cref{Properties of Doeblin coefficient}} \label{Proof of Properties of Doeblin}
    In this subsection, we will provide the proof of \cref{Properties of Doeblin coefficient}.
\begin{proof}[Proof of \cref{Properties of Doeblin coefficient}] \ \par
    \textbf{Part 1:} This follows from \cref{Definition of Doeblin}. We know that for any $i \in [n]$, 
    $$ 0 \leq\tau(W) \leq \sum_{y \in \Y} P_i(y) = 1. $$
    It is straightforward to verify that $\tau(W) = 1$ if $P_{1} = \dots = P_n$. To establish the converse, without loss of generality, assume that $P_1$ and $P_2$ differ at some $y_0 \in \Y$. Then, we have 
    $$ \tau(W) \leq \sum_{y \in \Y}\min\{P_1(y),P_2(y)\} =  1 - \|P_1-P_2\|_{\mathsf{TV}} < 1 , $$
    where the equality follows from the properties of TV distance \cite[Proposition 4.7]{LevinPeresWilmer2009}. Similarly, it is straightforward to verify that $\tau(W) = 0$ if and only if at least one of $P_{1}(y), \dots, P_{n}(y)$ is zero for all $y \in \Y$.

    \textbf{Part 2:} The first two properties, namely, total symmetry and positive definiteness follow immediately from the definition of $\tau(W)$ and the result in Part 1. Define the quantity $P_{\min}(y) \triangleq \min \{P_1(y), P_2(y), \dots, P_n(y)\}$ for $y \in \Y$. To prove the polyhedron inequality, we need to show that
        \begin{equation*}
            \begin{split}
                & (n -1)\bigg(1  -\sum_{y\in\Y}P_{\min}(y)  \bigg) \leq \sum_{i=1}^{n} \bigg( 1- \\
                & \ \ \ \ \   \sum_{y \in \Y} \min \{P_{1}(y), \ldots, P_{i-1}(y), P_{i+1}(y), \dots,P_{n+1}(y)\}\bigg).
            \end{split}
        \end{equation*}
    The above condition is equivalent to
            \begin{align}
                     \sum_{ y  \in \Y}  \bigg(  \sum_{i=1}^{n} \min \{P_{1}(y), \dots, &P_{i-1}(y), P_{i+1}(y), \dots, P_{n+1}(y)\}  \nonumber\\
                     &  -(n-1) P_{\min}(y) \vphantom{\sum_{x\in\Y}} \bigg) \leq 1.
                     \label{SimplifiedPolyhedron}
            \end{align}    
    Now define the event $\mathcal{A} \triangleq \{y \in \Y: P_{n+1}(y) \leq  P_{\min}(y) \}$. On the set $\mathcal{A}$, the terms inside the parentheses in \cref{SimplifiedPolyhedron} reduce to 
        $$
          n P_{n+1}(y)-(n-1) P_{\min}(y) \leq P_{n+1}(y).
        $$
    Now consider $y  \in \mathcal{A}^\complement $. 
    Without loss of generality, assume that $P_1(y)$ is the smallest among $\{ P_1(y),P_2(y), \dots,P_n(y)\}$. Then, the terms inside the parentheses in \cref{SimplifiedPolyhedron} reduces to 
            \begin{align}
                \nonumber (n-1) P_{1}(y)+\min  \{P_{2}(y), \dots, P_{n+1}&(y)\}   -(n-1) P_{1}(y) \\
                &  \leq P_{n+1}(y).             
            \end{align}
    Hence, on both $\mathcal{A}$ and $\mathcal{A}^\complement$, the terms inside the parentheses in \eqref{SimplifiedPolyhedron} are upper bounded by $P_{n+1}(y)$. Since the upper bound $P_{n+1}(y)$ sums to $1$ over all $y \in \Y$, this proves the polyhedron inequality.

    \textbf{Part 3:} For $V,W \in \R_{\mathsf{sto}}^{n\times m}$ such that $V = [V_1^\T,\ldots,V_n^\T]^\T$ and $W = [W_1^\T,\ldots,W_n^\T]^\T$ and 
    for some $\lambda \in [0,1]$, let $\bar{\lambda} \triangleq 1 - \lambda$. Then, we have 
    $$
        \begin{aligned}
             \tau( & \lambda V+\bar{\lambda} W)  \\
            & = \sum_{y\in \Y} \min \{\lambda V_{1}(y)+\bar{\lambda} W_{1}(y), \dots , \lambda V_{n}(y)+\bar{\lambda} W_{n}(y)\}\\
            &  \geq \lambda \sum_{y\in \Y}\min \{V_{1}(y), \ldots, V_{n}(y)\} \\ 
            & \ \ \ \ \ \ \ \ \ \ \ + \bar{\lambda} \sum_{y  \in \Y}  \min \{W_{1}(y), \ldots, W_{n}(y)\} \\
            & = \lambda \tau(V) + \bar{\lambda} \tau(W).
        \end{aligned}
    $$
    This proves the concavity of the Doeblin coefficient.

    \textbf{Part 4:} To prove sub-multiplicativity, we will utilize the following lemma from \cite{Chestnut2010, GohariGunluKramer2020}.
    \begin{lemma}[Doeblin Minorization and Degradation {\cite[Lemma 5]{GohariGunluKramer2020}}] \label{Doeblin Minorization and Degradation} 
        Let $\mathcal{A}$ and $\mathcal{R}$ be arbitrary discrete and finite sets. 
        Let $\mathcal{B} = \mathcal{A} \cup \{\mathtt e\}$ (here $\mathtt e$ is the erasure symbol). Let $A,B,R$ be random variables over $\calA,\mathcal{B},\mathcal{R}$ respectively such that $P_{B|A} = \textup{E}_\epsilon$ is an erasure channel with erasure probability $\epsilon$. Consider an arbitrary channel $P_{R|A}$, then there exists a conditional distribution $P_{R|B}$ such that
            \[P_{R|A}  = \textup{E}_\epsilon P_{R|B}, \]
        if and only if 
        $$ \tau(P_{R|A}) = \sum_{r \in \mathcal{R}} \min_a P_{R|A}(r|a) \geq \epsilon. $$
    \end{lemma}

    Let $P_{R|A} = VW \in \R_{\mathsf{sto}}^{k\times m}$ and let the elements of set $\mathcal{A}$ are indexed as $\{a_1, a_2, \dots, a_k\}$. Then the sub-multiplicativity property follows from \cref{Doeblin Minorization and Degradation} if we show a construction of $P_{R|B} \in \R^{(k+1) \times m}_{\mathsf{sto}}$ such that
        \begin{equation}
            VW = \text{E}_{1-(1-\tau(V))(1-\tau(W))} P_{R|B}. \label{Construction to prove}
        \end{equation}
    For simplicity, define some shorthand notations $\tau_v \triangleq \tau(V)$, $\tau_w \triangleq \tau(W)$, and row vectors $W_{\min}\in \R^m, V_{\min} \in \R^m $ such that their $j$-th entry $W_{\min}(j) = \min_{i \in [k]} W_{ij} $ for $j \in [m] $ and $V_{\min}(j) = \min_{i \in [n]} V_{ij} $ for $j \in [n]$. Let $V_j$ denote the $j$-th row of matrix $V$.
    For a fixed $b \in \mathcal{B}$, the row vector $P_{R | B=b}$ is constructed as follows:
        \begin{equation} \label{construction of PRgivenB}
          P_{R | B} = \begin{cases}
                \frac{(1-\tau_{v}) \tau_{w}}{\tau_{v}+\tau_{w}-\tau_{v} \tau_{w}} \frac{W_{\min}}{\tau_{w}}+\frac{\tau_{v} }{\tau_{v}+\tau_{w}-\tau_{v} \tau_{w} } \frac{V_{\min} W}{\tau_v} & \text{if } b=\mathtt e                 \vspace{1.5mm}\\           \frac{(V_{j}-V_{\min})}{(1-\tau_{v})} \frac{(W-\1 W_{\min})}{(1-\tau_{w})} & \text {if } b = a_j \end{cases}
        \end{equation}        
    We can verify that $P_{R|B}$ is a row-stochastic matrix as $W_{\min}/\tau_w \in \calP_m$ and $V_{\min}/\tau_v \in \calP_n.$ Additionally, $P_{R|B = \mathtt{e}}$ is a convex combination of probability vectors and $P_{R|B=a_j}$ is a probability vector obtained by passing $\frac{(V_{j}-V_{\min})}{(1-\tau_{v})} \in \calP_n$ through channel $\frac{(W-\1 W_{\min})}{(1-\tau_{w})} \in \R_{\mathsf{sto}}^{n\times m}$.
    Moreover, the construction of $P_{R|B}$ also satisfies \eqref{Construction to prove} as the $i$-th row of $\text{E}_{(1-(1-\tau(V))(1-\tau(W)))} P_{R|B}$ is
    $$
        \begin{aligned}
             & (\text{E}_{(1-(1-\tau(V))(1-\tau(W)))} P_{R|B})_i \\ 
             & = (1-\tau_{v})(1-\tau_{w}) P_{R \mid B=a_i} +(\tau_{v}+\tau_{w}-\tau_{v} \tau_{w}) P_{R \mid B=\mathtt e} \\
             & = V_{i}W - W_{\min}-V_{\min} W + \tau_{v} W_{\min}  + (1-\tau_{v}) W_{\min} \\ 
             & \ \ \ \ \ \ \ \  \ \ \ \ \ \ \ \ \ \ \ \ \ \ \ \ + V_{\min} W \\
             & = V_{i} W=(VW)_{i}.
        \end{aligned}
    $$

    \textbf{Part 5:} For $W \in \R^{n \times m}_{\mathsf{sto}},$ and $V \in \R^{l \times k}_{\mathsf{sto}},$ we have
    $$
        \begin{aligned}
            \tau(& W \otimes V)  =\sum_{x  \in [m], y \in [k]} \min_{i \in [n], j \in [l]} W_{ix} V_{jy} \\ 
            & =\sum_{x  \in [m], y \in [k]} \min \{W_{1x}, \ldots, W_{nx}\} \times \min \{V_{1y}, \ldots, V_{ly}\} \\
           &  =\bigg(\sum_{x  \in [m]} \min \{W_{1x}, \ldots, W_{nx}\} \bigg) \times \\
           & \ \ \ \ \ \ \ \ \ \ \ \ \ \ \ \ \ \  \bigg( \sum_{ y \in [k]} \min \{V_{1y}, \ldots, V_{ly}\} \bigg) \\
           & =\tau(W) \tau(V).
        \end{aligned}
    $$
    
    \textbf{Part 6:} This upper bound for $\tau(W)$ was proved in, e.g., \cite[Remark III.2]{Raginsky2016}. 

    \textbf{Part 7:} Observe that 
    $$\mathsf{Tr}(PW) = \sum_{j=1}^m \sum_{i=1}^n P_{ji} W_{ij} \geq \sum_{j=1}^m \min_{i \in [n]} W_{ij} \times \sum_{i=1}^n P_{ji}.$$ 
    Since for any $j \in [m]$, $\sum_{i=1}^n P_{ji} = 1$, we get $\mathsf{Tr}(PW) \geq \tau(W)$. 
    The equality holds when $P_{ji_0} = 1$ for $i_0 = \argmin_{k \in [n] } W_{kj}$ and $0$ otherwise. 
    
    \textbf{Part 8:} From the Bayesian decision theory, we know that for an observation model $P_{Y|X} \in \R^{n\times m}_{\mathsf{sto}} $, a prior PMF $P_X \in \calP_n$, a loss function $l: \X \times \X \rightarrow \R$, and any randomized estimator $P_{\hat{X}|Y} \in \R^{m \times n}_{\mathsf{sto}}$, the risk function is defined as
            $$ 
            \begin{aligned}
                R(& P_{Y |X} ,  P_{X}, l, P_{\hat{X}|Y} ) \triangleq \E[ l(X, \hat{X})]  \\
                &  = \sum_{x  \in \X}\sum_{y\in \Y}\sum_{\hat{x}\in \X} l(x, \hat{x}) P_X(x) P_{Y |X} (y|x) P_{\hat{X}|Y}(\hat{x}|y) \\
                & = \mathsf{Tr}(L^\T \diag(P_X) P_{Y|X} P_{\hat{X}|Y }),
            \end{aligned}
            $$
        where $L \in \R^{n \times n}$ is the matrix representation of the loss function $l : \X \times \X \rightarrow \R$. Specifically, the $(i,j)$th entry of the matrix $L$ is $l(x_i, \hat{x}_j)$ for $ i, j \in [n]$ and $x_i, \hat{x}_j \in \X$. By the choice of our loss function, we have $L = I$ and $\diag(P_{X}) = \frac{1}{n} I$. This gives the Bayes optimal risk $R^*_{\lambda}(P_{Y|X})$ as
            $$
            \begin{aligned}
               & R^*_{\lambda}(P_{Y|X})  = \min_{P_{\hat{X}|Y} \in \R^{m\times n}_{\mathsf{sto}}}  \P(\hat{X} (Y)  = X) \\ 
               & = \min_{P_{\hat{X}|Y} \in \R^{m\times n}_{\mathsf{sto}}} \mathsf{Tr}(I \frac{1}{n}I P_{Y \mid X} P_{\hat{X} \mid Y}) \\
                & = \frac{1}{n} \sum_{x \in \X} \min \{ P_{Y|X=x_1}(x), \ldots, P_{Y|X=x_n}(x)\} =\frac{\tau(P_{Y|X})}{n}. 
            \end{aligned}
            $$

\end{proof}
\subsection{Maximal Coupling Characterization of Doeblin Coefficient} \label{Proof of Maximal Coupling}
    In this subsection, we will provide the proof of \cref{Maximal Coupling}.
\begin{proof}[Proof of \cref{Maximal Coupling}]
    First, we will derive an upper bound on $\P(Y_1= \dots =Y_n)$ as follows. Define $\tilde{S}_i \triangleq \{ y\in \Y : P_{i}(y) \leq P_{j}(y)\  \forall \ j\neq i \}$ and $S_i \triangleq \tilde{S}_i \setminus \cup_{j<i} \tilde{S}_j $. Then $S_1,\ldots, S_n$ form a partition of $\Y$. Now, observe that the event $\{Y_1=\ldots=Y_n \}$ is contained in the event $\{\exists i: Y_i \in S_i \}$. Indeed, if $Y_1 = \ldots = Y_n = y$, then there exists $i$ such that $y \in S_i,$ which implies $Y_i \in S_i$.
    Hence, we have
    \begin{equation*}
    \begin{aligned}
        \P(Y_1=\ldots=Y_n) & \leq \P(\exists \ i, Y_i \in S_i) \stackrel{\zeta_1}{\leq} \sum_{i=1}^n \P(Y_i \in S_i) \\
        & \stackrel{\zeta_2}{=} \sum_{i=1}^n P_i(S_i), 
    \end{aligned}
    \end{equation*}
    where $\zeta_1$ follows from the union bound and $\zeta_2$ follows since coupling preserves the maginals. Hence,
    \begin{equation}
        \begin{aligned}
            \max_{\substack{ P_{Y_1, \dots, Y_n} : \\ P_{Y_1 } = P_1, \, \dots \, , \, P_{Y_n} =P_n} } & \P(Y_1=\ldots=Y_n)  \leq \sum_{i=1}^n \sum_{y \in S_i} P_i(y)  \\ 
           & = \sum_{y \in \Y}\min\{ P_1(y),\ldots,P_n(y)\},  \label{lower bound for Doeblin}
        \end{aligned}
    \end{equation}
    where the last step follows since the sets $S_i$ form a partition of $\Y$ and on each $S_i$ and for any $y \in S_i$ we have $P_i(y)  = \min\{P_1(y), \dots, P_n(y)\}.$
    Alternatively, \cref{lower bound for Doeblin} can also be obtained as 
    \begin{equation*}
        \begin{aligned}
        \P(Y_1  = \dots = Y_n) &  = \sum_{y  \in \Y} \P(Y_1 = \dots Y_n = y) \\    
             & \leq \sum_{y  \in \Y} \min\{ P_1(y) ,\P(Y_2 = \dots Y_n = y)\}\\
             & \leq \sum_{y  \in \Y} \min\{ P_1(y) ,P_2(y), \dots, P_n(y)\},
        \end{aligned} 
    \end{equation*}
    where the first inequality follows from the fact that $\P(A \cap B) \leq \min\{ \P(A), \P(B) \}$ and the next inequality follows by applying the same procedure recursively.
    Now we show a construction of a coupling that achieves the above upper bound.
    Let $c \triangleq \sum_{y  \in \Y} \min \{P_{1}(y), P_{2}(y), \ldots, P_{n}(y)\}$, and define distributions
    $$
        \begin{aligned}
            Q_0(y)& \triangleq \frac{\min \{P_{1}(y), P_{2}(y), \ldots, P_{n}(y)\}}{c} \\
            Q_{i}(y) & \triangleq \frac{P_{i}(y)-\min \{P_{1}(y), P_{2}(y), \ldots, P_{n}(y)\}}{1-c},
        \end{aligned}
    $$
    for $i\in [n]$. Define a random variable $Z \sim \text{Ber}(c)$ and independent of $Y_{1}, \ldots, Y_{n}$. We will now construct joint distribution $P_{Y_{1},\ldots, Y_{n}, Z}$ by defining the conditional distribution 
    \begin{equation} \label{conditionals}
        \begin{aligned}
            P_{Y_{1},\dots, Y_{n}|Z}(y_{1}, \dots, y_{n} | Z=1) & = Q_0(y) \I_{y_{1}=\ldots=y_{n} = y}, \\
            P_{Y_{1},\dots, Y_{n}|Z}(y_{1}, \dots, y_{n} | Z=0) & = \prod_{i=1}^{n} Q_{i}(y_{i}).
        \end{aligned}
    \end{equation}
    Moreover, observe that $P_{Y_{1},\dots, Y_{n}|Z}(y_{1}, \dots, y_{n} | Z=0)$ and $P_{Y_{1}, \dots, Y_{n}|Z}(y_{1}, \dots, y_{n} | Z=1)$ are mutually disjoint since $P_{Y_{1}, \dots, Y_{n}|Z}(y, y, \dots, y | Z=0) =0$ for all $y \in \Y$ as at least one of $P_1(y),\ldots,P_n(y) $ is equal to $\min\{P_1(y),\ldots,P_n(y)\}$.    
    Together with $Z \sim \text{Ber}(c)$ and \cref{conditionals}, we get the joint distribution $P_{Y_{1}, \ldots, Y_{n}, Z}$. Let $y^n \triangleq (y_1,\dots,y_n)$. Marginalizing $P_{Y_{1},\ldots, Y_{n}, Z},$ with respect to $Z,$ we get $P_{Y_{1}, \dots, Y_{n}}(y^n)$ as follows
    $$
        \begin{aligned}
            P_{Y_{1}, \ldots, Y_{n}}(y^n) & = P(Z=1) P_{Y_{1}, \dots, Y_{n} \mid Z}(y^{n} \mid Z=1)+ \\
            & \ \ \ \ \ \ \ \ \ \ \ \ P(Z=0) P_{Y_{1}, \dots, Y_{n} \mid Z}(y^{n} \mid Z=0) \\
             & =  c Q_0(y) \I_{y_{1}=\dots =y_{n} = y} + ( 1-c) \prod_{i=1}^{n} Q_{i}(y_{i}).
        \end{aligned}
    $$
    We will now show that the coupling $P_{Y_{1},\ldots, Y_{n}}$ has the right marginals.
    In this vein, define $y^{(i)} \triangleq (y_1,\dots,y_{i-1},y_{i+1},\dots,y_n)$. Observe that for all $i \in [n]$ and all $y  \in \Y$
    \[
    \begin{aligned}
        P_{Y_{i}}(y)  &  =c \sum_{y^{(i)}} Q_0(y) \I_{y_{1}=\ldots=y_{n}=y}  \\ 
        & \ \ \ \ \ \ \ \ \ \ \ \ \ \ \ + (1-c)Q_{i}(y) \sum_{y^{(i)}}  \prod_{j \neq i} Q_{j}(y_{j})\\ 
        & =  \min \{P_{1}(y), \dots, P_{n}(y)\}  +P_{i}(y) \\ 
        & \ \ \ \ \ \ \ \ \ \ \  \ \ \ \ \ \ \ \ \ - \min \{P_{1}(y), \dots, P_{n}(y)\}\\
        & = P_{i}(y).
    \end{aligned}
    \]
    Hence, $P_{Y_{1}, \ldots , Y_{n}}$ has the right marginals. Finally, since the measure of the set $\{Y_{1}=\dots=Y_{n}\}$ is same as the probability $\{Z = 0\}$ (as $P_{Y_1,\dots,Y_n|Z}(y, y,\dots,y|Z=1)= 0$), hence we have $\P(Y_{1}=Y_{2}=\dots =Y_{n})= c = \sum_{y  \in \Y} \min \{P_{1}(y), \ldots, P_{n}(y)\}$. 
\end{proof}

\subsection{Minimal Coupling for Max-Doeblin Coefficient}\label{sec:Truelabel}
In this subsection, we will provide the proof of \cref{thm:MaxDoeblinCoupling}.  Firstly, we would like to note that the coupling construction for $n=2$ is identical to the standard Dobrushin's maximal coupling (or equivalently, the one depicted in \cref{Proof of Maximal Coupling} for $n=2$). To demonstrate that this coupling achieves $\tau_{\max}(P_1,P_2)$ for $n=2$, we observe the following: 
 \[
    \begin{aligned}
      \sum_{y\in \Y}&\P(\cup_{i=1}^2\{ Y_i =y\} )\\
      & = \sum_{y\in \Y} (\P(Y_1 =y) + \P( Y_2 =y)  - \P(Y_1=Y_2 =y)) \\
     &  = \sum_{y\in \Y} (P_1(y) + P_2(y) - \min\{P_1(y),P_2(y)\}) \\
    & = \sum_{y\in \Y} \max\{P_1(y), P_2(y)\} = \tau_{\max}(P_1,P_2).
    \end{aligned}
\]
 Next, to better illustrate the key ideas, we will provide the proof of \cref{thm:MaxDoeblinCoupling} for the specific scenario of $n=3$ and provide the corresponding coupling construction. Following this, we will provide a general proof for arbitrary $n$.

\subsubsection{Proof of \cref{thm:MaxDoeblinCoupling} for $n=3$}\label{sec: Proof of MaxDoeblinCoupling} \ \\
    Observe that, for any coupling $\P(\cdot)$ we have
        \begin{align}
           \nonumber \sum\limits_{y  \in \Y} \P(\cup_{i=1}^{3}\{Y_{i} =y\}) &  \geq \sum\limits_{y  \in \Y} \max \{P_{1}(y), \P(\cup_{i=2}^{3}\{Y_{i}=y\})\}  \\
            \nonumber & \geq \sum_{y  \in \Y} \max \{P_1(y), P_{2}(y), P_{3}(y)\} \\
            & = \tau_{\max}(P_1,P_2,P_3),\label{lower bound n3}
        \end{align}
    where the first inequality follows from the fact that $\mathbb{\P}(A \cup B) \geq \max \{\P(A), \P(B)\}$ combined with the coupling constraint $\P(Y_{1}=y)=P_{1}(y)$. The second inequality follows similarly by using the same step. 

    Assuming $\tau_{\max_2}(P_1, P_2, P_3) \leq 1$, we will now demonstrate a construction of coupling that satisfies the lower bound in \cref{lower bound n3}. Later, in \cref{Proof of New Minimal Coupling for max-Doeblin Coefficient}, we will slightly modify the coupling construction to consider the case when $\tau_{\max_2}(W) > 1$. 

    We begin by defining the following quantities:
    $$
    \begin{aligned}
        \tau_{12} & \triangleq \sum_{y\in \Y}\min\{P_1(y), P_2(y)\},\\
        \tau_{23} & \triangleq \sum_{y\in \Y}\min\{P_2(y), P_3(y)\}, \\ 
        \tau_{13} & \triangleq \sum_{y\in \Y}\min\{P_1(y), P_3(y)\}, \\ 
        \tau & \triangleq \tau(P_1,P_2,P_3).
    \end{aligned}
    $$
    We will utilize a useful relation between $\tau_{\max}(P_1,P_2,P_3), \tau_{12},\tau_{23},\tau_{13},\tau$ using the maximum-minimums identity (for $n=3$). For any $y \in \Y$, we have
        \begin{align}
      \nonumber      &\max \{P_{1}(y), P_{2}(y), P_{3}(y)\}  = P_{1}(y)+P_{2}(y)+P_{3}(y) \\ 
\nonumber            &  \ \ \ \ -\min \{P_1(y), P_{2}(y)\}-\min \{P_{2}(y), P_{3}(y)\}  \\
             &  \ \ \  - \min \{P_{1}(y), P_{3}(y)\}+\min \{P_{1}(y), P_{2}(y), P_{3}(y)\}.\label{minimum maximum principle n3}
        \end{align}
    Summing over both sides for all $y \in \Y,$ we obtain
    \begin{equation}
         \tau_{\max }(P_1,P_2,P_3) =3-(\tau_{12}+\tau_{23}+\tau_{13})+\tau. \label{relation between weights}
    \end{equation}
    Furthermore, we define $P_{\min}(y) \triangleq \min \{P_{1}(y), P_{2}(y), P_{3}(y)\}$ as before. Additionally, we introduce the following distributions:
    \begin{equation}        \label{bunch of eqns}
    \begin{aligned}
     & R_1(y)  \triangleq   \frac{1}{1+\tau-\tau_{12}-\tau_{13}} \times \bigg(
     P_{1}(y)+P_{\min}(y) \\ 
     & \ \ \ \ \ \ \ \ \ \ \ \ \ -\min \{P_{1}(y),P_{2}(y)\}-\min \{P_{1}(y), P_{3}(y)\} \bigg), \\
     & R_2(y)  \triangleq \frac{1}{1+\tau-\tau_{12}-\tau_{23}}\bigg( P_2(y)+P_{\min}(y) \\
     & \ \ \ \ \ \ \ \ \ \ \ \ \ -\min \{P_{1}(y),P_{2}(y)\}-\min \{P_{2}(y),P_{3}(y)\}\bigg), \\
     & R_3(y)  \triangleq  \frac{1}{1+\tau-\tau_{23}-\tau_{13}}\bigg( P_3(y)+P_{\min}(y) \\
     & \ \ \ \ \ \ \ \ \ \ \ \ \ - \min\{P_{2}(y),P_{3}(y)\} - \min\{P_{1}(y), P_{3}(y) \}\bigg), \\
     & Q_{0}(y)  \triangleq \frac{P_{\min}(y) }{\tau}, \\
     &Q_{1,1}(y, y^{\prime}) \triangleq R_1(y) \times \frac{\min \{P_{2}(y^{\prime}), P_{3}(y^{\prime})\}- P_{\min}(y^\prime) }{\tau_{23}-\tau},\\
     &Q_{1,2}(y, y^{\prime}) \triangleq R_2(y) \times \frac{\min \{P_{1}(y^{\prime}), P_{3}(y^{\prime})\}- P_{\min}(y^\prime)}{\tau_{13}-\tau}, \\
     &Q_{1,3}(y, y^{\prime}) \triangleq R_3(y) \times \frac{\min \{P_{1}(y^{\prime}), P_{2}(y^{\prime})\} - P_{\min}(y^{\prime})}{\tau_{12}-\tau}, \\
     & Q_{2}(y_{1}, y_{2}, y_{3})  \triangleq R_1(y_1) \times R_2(y_2) \times R_3(y_3).
    \end{aligned}
    \end{equation}      
    One can verify that each of $R_i (y), Q_{1, i} $ for $i\in [3]$ and $Q_0, Q_2$ are indeed distributions, i.e., they are always non-negative and sum to $1$. Specifically, we can show the non-negativity of the following expression as follows:
    $$
        \begin{aligned}
            & P_{1}(y) +  P_{\min}(y)-\min \{P_{1}(y),P_{2}(y)\}-\min \{P_{1}(y), P_{3}(y)\} \\
            & = \max\{P_1(y),P_2(y) ,P_3(y) \} -\max\{ P_2(y), P_3(y) \}.
        \end{aligned}
    $$
    Similarly, we can demonstrate $Q_{1, i}$ for $i \in [3]$ are also non-negative. Additionally, we have normalized $Q_0, Q_{1, i}$ for $i\in [3]$, and $Q_2$ such that they sum to $1$. 
    Next, we define the coupling distribution as a convex combination of $Q_0$, $Q_{1, i}$ for $i\in [3]$, and $Q_2$, as follows:  
        \begin{align}
\nonumber             \P&(y_{1} , y_{2}, y_{3})=\tau Q_{0}(y) \I_{y_{1}=y_{2}=y_{3}=y}+(\tau_{23}-\tau) \times \\
         \nonumber   & Q_{1,1}(y_{1}, y^{\prime}) \I_{y_{2}=y_{3}=y^{\prime}} + (\tau_{13}-\tau) Q_{1,2}(y_{2}, y^{\prime}) \I_{y_{1}=y_{3}=y^{\prime}}  \\
          \nonumber  & \ \ \ \ \ \  + (\tau_{12}-\tau) Q_{1,3}(y_{3}, y^{\prime}) \I_{y_{1}=y_{2}=y^{\prime}} \\
            & \ \ \ \ \ \  +(1+2\tau-\tau_{12}-\tau_{13} -\tau_{23}) Q_{2}(y_{1}, y_{2}, y_{3}).\label{convex combination n3}
        \end{align}
    To verify that the expression in \cref{convex combination n3} is indeed a convex combination, observe that the weights $\tau, \tau_{23} -\tau, \tau_{13} -\tau, \tau_{12} - \tau$ are all non-negative and using \cref{relation between weights}, we can simplify the weight corresponding to $Q_2(\cdot)$ as follows:
    \begin{align}
        \nonumber 1+ & 2\tau-\tau_{12}-\tau_{13}-\tau_{23}  = \tau_{\max}(P_1,P_2,P_3) + \tau -2 \\
        & = \sum_{y\in \Y} (\max\{ P_1(y), P_2(y), P_3(y)\} + P_{\min}(y) ) -2 \nonumber\\
         & \stackrel{\zeta}{=} 1- \sum_{y\in \Y} \max\nolimits_{2}\{ P_1(y), P_2(y), P_3(y)\}\nonumber\\
         & = 1 - \tau_{\max_2}(P_1,P_2,P_3). \label{somelabel}
    \end{align}
    where $\zeta$ follows since \[
        \begin{aligned}
        &\sum_{y \in \Y} (\max\{ P_1(y),  P_2(y), P_3(y)\} + \\
        & \max\nolimits_2 \{P_1(y), P_2(y), P_3(y) \} + \min\{P_1(y), P_2(y), P_3(y)\}) \\
        & =  \sum_{y \in \Y}  (P_1(y)+ P_2(y)+ P_3(y)) = 3.
        \end{aligned}
    \]
    Hence, the weight corresponding to $Q_2(\cdot)$ in \cref{convex combination n3} is non-negative by our assumption that $\tau_{\max_2}(P_1,P_2,P_3) \leq 1$. Moreover, note that the weights in \cref{convex combination n3} sum to 1. Furthermore, note that each of the distributions used in the convex combination in \cref{convex combination n3} are mutually disjoint to each other. This is because $Q_{1,i}(y,y) = 0$ for $i \in \{1,2,3\}$ and $Q_2(y_1,y_2,y_3)= 0$ whenever any two of the variables $y_1,y_2,y_3$ are equal. Thus, $Q_0(y)$ is supported on the set $\{Y_1= \dots = Y_n\}$, $Q_{1,i} $ is supported on the set $Y_i\neq Y_j=Y_k \neq Y_i$ for $j,k \in [3] \setminus \{i\}$ and $j \neq k,$ and $Q_2(y)$ is supported on the set where all three $Y_1,Y_2,Y_3$ are distinct.

    As our next step, we will verify that the coupling has the correct marginals. Without loss of generality, we will verify the marginals of $Y_1$. Marginalizing \cref{convex combination n3} with respect to $Y_2, Y_3$ gives,
    $$
    \begin{aligned}
       &  \sum_{y_{2}, y_{3} \in \Y}  \mathbb{P}(y_{1}, y_{2}, y_{3})   = P_{\min}(y_1)  + \frac{\tau_{23}-\tau}{1+\tau-\tau_{12}-\tau_{13}}\times \\     
       & \bigg(P_{1}(y_{1}) +P_{\min}(y_1)-\min \{P_{1}(y_{1}), P_{2}(y_{1})\} - \\ 
       & \min \{P_{1}(y_{1}), P_{3}(y_{1})\}\bigg)   +\min \{P_{1}(y_{1}), P_{2}(y_{1})\}-P_{\min}(y_1) \\ 
       & \ \ \ \ \ \  +\min \{P_{1}(y_{1}), P_{3}(y_{1})\}-P_{\min}(y_1) \\ 
       & \ \ \ \ \ \   + \frac{(1+ 2\tau-\tau_{12}-\tau_{13}-\tau_{23})}{1+\tau-\tau_{12}-\tau_{13}}  \times \bigg(P_{1}(y_{1})+ \\
       &  P_{\min}(y_{1})-\min\{ P_{1}(y_{1}), P_{2}(y_{1})\}-\min \{P_{1}(y_{1}),P_{3}(y_{1})\} \bigg).
    \end{aligned}
    $$
    Simplifying the above expression, we obtain
    $$
    \begin{aligned}
     &\sum_{y_{2}, y_{3} \in \Y} \P(y_{1}, y_{2}, y_{3})  = \min \{ P_{1}(y_{1}), P_{2}(y_{1})\} + \\
     &   \min\{P_{1}(y_{1}), P_{3}(y_{1})\} -P_{\min} (y_1) + (1+\tau-\tau_{12}-\tau_{13}) \times \\
    &  \ \ \ \ \ \ \ \ \ \bigg(\frac{P_{1}(y_{1})+P_{\min}(y_1)-\min \{P_{1}(y_{1}), P_{2}(y_{1})\} }{1+\tau-\tau_{12}-\tau_{13}}\\
    & \ \ \ \ \ \ \ -\frac{\min\{P_{1}(y_1),P_{3}(y_1)\} }{1+\tau-\tau_{12}-\tau_{13}} \bigg) \\
    & =P_{1}(y).
    \end{aligned}
    $$

    Finally, to demonstrate that the coupling satisfies the lower bound in \cref{lower bound n3}, we will use the inclusion-exclusion principle to simplify $\mathbb{P}(\bigcup_{i=1}^{3}\{Y_{i}=y\})$ as
    $$
    \begin{aligned}
         & \mathbb{P}\bigg(\bigcup_{i=1}^{3}\{Y_{i}=y\}\bigg) =\mathbb{P}(Y_{1}=y)+\mathbb{P}(Y_{2}=y)+\mathbb{P}(Y_{3}=y) \\
         & \ \ \ \ \ \ \   -(\mathbb{P}(Y_{1}=y, Y_{2}=y)+\mathbb{P}(Y_{2}=y, Y_{3}=y) \\
        & \ \ \ \ \ \  \ \ \ \  +\mathbb{P}(Y_{1}=y, Y_{3}=y)) +\mathbb{P}(Y_{1}=Y_{2}=Y_{3}=y)\\
        & \ \ \ \ \ \  = P_1(y)+P_{2}(y)+P_{3}(y) - (\mathbb{P}(Y_{1}=y, Y_{2}=y) \\
        & \ \ \  +\mathbb{P}(Y_{2}=y, Y_{3}=y) +\mathbb{P}(Y_{1}=y, Y_{3}=y)) + P_{\min}(y),
    \end{aligned}
    $$
    where the second equality follows since the coupling preserves marginals, we get $ \P(Y_{i}=y)=P_{i}(y)$. Furthermore, by the construction of coupling $\mathbb{P}(Y_{1}=Y_{2}=Y_{3}=y) = \min\{P_{1}(y), P_{2}(y), P_{3}(y)\}$. This is because $Q_{1,i}(y,y)$ and $Q_2(y,y,y) = 0$ for all $y \in \Y$.
    To show that the lower bound in \cref{lower bound n3} is achievable, we will prove that for $i\neq j$ and $i,j \in \{1,2,3\}$, we have
    \begin{equation}
        \mathbb{P}(Y_{i}=y, Y_{j}=y)=\min \{P_{i}(y), P_{j}(y)\}. \label{coupledminimum}    
    \end{equation}
    Then achievability of \cref{lower bound n3} follows from \cref{coupledminimum} by the maximum-minimums principle.
    To prove \cref{coupledminimum}, we begin by letting $k = \{1,2,3\} \setminus \{i,j\}$ and therefore, we can write 
    $$
        \begin{aligned}
         \mathbb{P}(Y_{i}=Y_{j}=y) & =\sum_{y^{\prime} \in \Y} \mathbb{P}(Y_{i}=Y_{j}=y, Y_{k}=y^{\prime}) \\
         & =\sum_{y^{\prime} \in \Y }(\tau_{ij}-\tau) Q_{1,k}(y^{\prime},y)+ P_{\min}(y) \\
         & =\min \{P_{i}(y), P_{j}(y)\}-P_{\min}(y) +P_{\min}(y) \\
         & =\min \{P_{i}(y), P_{j}(y)\}.
        \end{aligned}
    $$
    Thus, we have shown that $ \mathbb{P}(Y_{i}=y, Y_{j}=y)=\min\{P_{i}(y), P_{j}(y)\}$, which completes the proof of \cref{thm:MaxDoeblinCoupling} for $n=3$. \qed

\subsubsection{Proof of \cref{thm:MaxDoeblinCoupling} for general $n$} \label{Proof of MaxDoeblinCoupling for general n}
    In this subsection, we will generalize the construction presented in \cref{sec: Proof of MaxDoeblinCoupling} for general $n$. The coupling is formed by taking the convex combination of disjoint distributions $Q_{k,\calA}, Q_{n-1}$ defined (later) for all possible $k \in \{0,1,\ldots, n-2\}$ and $\calA \subseteq [n]$ with $|\calA| = k$ such that $Q_{k,\calA}$ is supported on the set where random variables $\{Y_j: j \in \calA^\complement \}$ are equal and all $\{Y_j: j \in \calA\}$ are different from one another. Similar to construction for $n=3$ in \cref{sec: Proof of MaxDoeblinCoupling}, the construction for general $n$ utilizes the maximum-minimums principle (for $n$ elements). 
\begin{proof}[Proof of \cref{thm:MaxDoeblinCoupling}] \ \\
    Observe that, for any coupling $\P(\cdot)$ we have
        \begin{align}
            \nonumber \sum_{y  \in \Y}  \P(\cup_{i=1}^{n}\{Y_{i} &=y\})  \geq \sum_{y  \in \Y} \max \{P_{1}(y), \mathbb{P}(\cup_{i=2}^{n}\{Y_{i}=y\})\}, \\
             & \geq \sum_{y  \in \Y} \max \{P_1(y), P_{2}(y), \ldots, P_{n}(y)\}. \label{lower bound for general n}
        \end{align}
    where the first inequality follows from the fact that $\mathbb{\P}(A \cup B) \geq \max \{\P(A), \P(B)\}$ and the coupling constraint $\P(Y_{1}=y)=P_{1}(y)$, and the second inequality follows by repeating the same steps recursively. 

    Before presenting the coupling construction achieving the lower bound in \cref{lower bound for general n}, we first introduce some additional notation. For any set $\calI \subseteq [n]$, let
    \begin{equation}
    \begin{aligned}\label{min notation}
        P_{\min,\calI}(y)  &\triangleq \min_{i \in \calI} P_i(y), \\
        P_{\max,\calI}(y)  &\triangleq \max_{i \in \calI} P_{i}(y), \\
        \tau_{\calI}  &\triangleq \sum_{y  \in \Y } P_{\min,\calI}(y).
    \end{aligned}
    \end{equation}
    Furthermore, we will use the shorthand notation
    \begin{equation}
    \begin{aligned}
        P_{\min,(i)}(y)  &\triangleq P_{\min,[n]\setminus \{i\}}(y) \text{, } \\ 
        \tau_{\min,(i)}(y) & \triangleq \sum_{y\in \Y}P_{\min,(i)}(y),\\
        P_{\max,(i)}(y)  & \triangleq P_{\max,[n]\setminus \{i\}}(y) \text{, } \\
         \tau_{\max,(i)}(y) & \triangleq \sum_{y\in \Y}P_{\max,(i)}(y).
    \end{aligned}
    \end{equation}
    Also, let $\tau_{\max} \triangleq \tau_{\max}(P_1, \ldots,P_n)$ and $\tau_{\max\nolimits_2} \triangleq \tau_{\max_2}(P_1, \ldots,P_n)$.
    We will frequently utilize the maximum-minimums identity stated below (using the notation just defined).
    \begin{lemma}[Maximum-Minimums Identity {\cite[Section 7.2]{ross2010first}}]\label{lem: max min inclusion rule}
        For any $n \geq 2,$ let $P_1,P_2,\ldots,P_n \in \calP_m$, and for any $y  \in \Y$, we have
        \begin{equation} 
            P_{\max,[n]}(y) = \sum_{k=1}^n (-1)^{k-1} \sum_{\calI \subseteq [n], |\calI| = k} P_{\min, \calI}(y).
        \end{equation}
    \end{lemma}

    We will use the following distributions in our construction of the minimal coupling. For any $k \in [n]$, let
    \begin{equation}\label{coupling distributions1}
    \begin{aligned}
        R_k(y)  \triangleq \frac{\sum_{l=1}^n (-1)^{l-1} \sum_{\calI \subseteq [n], |\calI| = l: k \in \calI } P_{\min, \calI}(y) }{ \sum_{l=1}^n (-1)^{l-1} \sum_{\calI \subseteq [n], |\calI| = l: k \in \calI} \tau_{\calI}(y)}.\\
    \end{aligned}
    \end{equation}
    Now for any set $\calA = \{a_1,\ldots, a_k\} \subseteq [n]$ with cardinality $k$ such that $k \in \{0,1, \ldots, n-2\}$, define a distribution     \begin{equation} \label{coupling distributions2}
    \begin{aligned}
        & Q_{k,\calA}(y, y_{a_1}, \ldots,y_{a_{k}})  \triangleq \bigg( \prod_{i=1}^k R_{a_i} (y_{a_i}) \bigg) \times \\
        & \ \ \ \ \ \ \ \ \ \ \ \ \frac{\sum_{l=0}^{k} (-1)^{l} \sum_{ \calI \subseteq \calA : |\calI| = l} P_{\min, \calA^\complement \cup \calI}(y) }{ \sum_{l=0}^{k} (-1)^{l} \sum_{ \calI \subseteq \calA: |\calI| = l} \tau_{\calA^\complement \cup \calI} }.
    \end{aligned}
    \end{equation}
    Also, define a product distribution $Q_{n-1}(y_1,\dots,y_n)$ as
    \begin{equation} \label{coupling distributions3}
        Q_{n-1}(y_1,\dots,y_n) \triangleq \prod_{i=1}^n R_{i} (y_{i}).
    \end{equation}
    \sloppy Note that $Q_{n-1}(y_1,\dots,y_n)$ is the same as $Q_{n-1,[n]\setminus \{1\}}(y_1, y_{2}, \ldots,y_{n})$. The following lemma verifies that $R_k(y)$, $Q_{k,\calA}(y, y_{a_1}, \ldots,y_{a_{k}}),$ and $Q_{n-1}(y_1,\dots,y_n)$ are indeed PMFs over their respective domains.

    \begin{lemma}[PMF Property]\label{lem: these are PMFs}
        $R_k$ is a PMF over $\Y$ for any $k \in [n]$. Similarly, the distributions $Q_{k,\calA}(y, y_{a_1}, \dots ,y_{a_{k}})$ are PMFs over the product space $\Y^{k+1}$ for any $k \in [n-2] \cup \{0\}$ and $Q_{n-1}(y_1,\ldots,y_n)$ is a PMF over the product space $\Y^n$.  
    \end{lemma}
    The proof is provided in \cref{Proof of these are PMFs}. \\
    Now, define the coupling distribution to be the convex combination of the PMFs with the following weights
    \begin{equation}
    \begin{aligned} \label{coupling for general n}
     &\P(y_{1},\ldots, y_{n})  =  
    \sum_{k=0}^{n-2} \sum_{\substack{\calA \subseteq [n] : \\
    |\calA|=k}}\bigg(\sum_{l=0}^{k}(-1)^{l} \sum_{\substack{I \subseteq \calA :\\
    |\calI|=l}} \tau_{\calA^{\complement} \cup \calI} \bigg)  \times \\ 
    & \ \ \ \ \ Q_{k, \calA}(y, y_{a_1}, \ldots, y_{a_{k}})\bigg(\prod_{u \in \calA^{\complement}} \I_{y_{u}=y}\bigg) + \\
    & \bigg(1-\sum_{k=0}^{n-2} \sum_{\substack{\calA \subseteq [n]: \\
    |\calA|=k}}\bigg(\sum_{l=0}^{k}(-1)^{l} \sum_{\substack{\calI \subseteq \calA \\
    |\calI|=l}} \tau_{\calA^{\complement} \cup \calI}\bigg)\bigg) Q_{n-1}(y_{1}, \ldots, y_{n}).
    \end{aligned}
    \end{equation}
    Now we will show that the weights of the convex combination in \cref{coupling for general n} are indeed non-negative and sum to one. 
    First of all, note that the weight corresponding to $Q_{k,\calA}$ is the same as its normalizing constant, i.e., the denominator term in \cref{coupling distributions2} and as shown in \cref{lem: these are PMFs} these weights are non-negative (as they are normalizing constants for the PMF). Moreover, they sum to one as the weight assigned to $Q_{n-1}$ is one minus the sum of weights assigned to other PMFs $Q_{k,\calA}$. Now, in the following lemma, we will show that the weight assigned to $Q_{n-1}(y_{1}, \ldots, y_{n})$ is $1 - \sum_{y  \in \Y} \max\nolimits_{2}\{P_{1}(y), \ldots, P_{n}(y)\}$ which is non-negative by the assumption of the theorem. 

    \begin{lemma}[Simplifying the Sum of Weights] \label{lem: the three equivalences}
        The sum of all the weights assigned to $Q_{k,\calA}$ for $k\in[n-2]\cup\{0\}$ and $\calA \subseteq [n]$ and $|\calA| \leq n-2$ in \cref{coupling for general n} can be simplified as follows:
        \begin{align}
             &\sum_{y  \in \Y} \sum_{k=0}^{n-2} \sum_{\substack{\calA \subseteq [n] \\ |\calA|=k}}\bigg(\sum_{l=0}^{k}(-1)^{l} \sum_{\substack{\calI \subseteq \calA \\        |\calI|=l}} P_{\min , \calA^{\complement} \cup \calI }(y)\bigg) \nonumber \\  \label{one of three}
            & = \sum_{y\in\Y} \sum_{k=2}^n (-1)^{k}(k-1) \sum_{\substack{ \calI \subseteq [n]\\|\calI| = k } } P_{\min,\calI}(y)  \\
            & =\sum_{y \in \Y} \max\nolimits_{2}\{P_{1}(y), \ldots, P_{n}(y)\} = \tau_{\max_2}.\label{three of three}
        \end{align}
    \end{lemma}
    The proof of \cref{lem: the three equivalences} is provided in \cref{Proof of the three equivalences}. Hence, the above lemma establishes that if $\sum_{y  \in \Y} \max _{2}\{P_{1}(y), \ldots, P_{n}(y)\} \leq 1$, then \cref{coupling for general n} is indeed a valid convex combination. Next, we will demonstrate that the coupling distribution defined in \cref{coupling for general n} has the right marginals. For $i\in [n]$, let $y^n_{(i)} \triangleq (y_1,\ldots,y_{i-1},y_{i+1},\ldots, y_n)$ and the marginal with respect to $Y_i$ can be written as
    \begin{equation} \label{incomplete right marginals}
        \begin{aligned}
             &\mathbb{P}(Y_{i}=y_i)  =\sum_{y_{(i)}^{n} \in \Y^{n-1}} \mathbb{P}(y_{1}, \ldots, y_{n}) \\
             & = \sum_{y_{(i)}^{n} \in \Y^{n-1}} \sum_{k=0}^{n-2} \sum_{ \substack{\calA \subseteq [n]: \\|\calA| =k} } \bigg( \bigg( \sum_{l=0}^{k} (-1)^{l} \sum_{ \substack{\calI \subseteq \calA:\\ |\calI| = l} } P_{\min, \calA^\complement \cup \calI}(y) \bigg)   \times \\ 
             & \ \ \ \ \ \ \ \ \ \ \ \ \ \ \ \ \ \bigg( \prod_{u \in \calA} R_{u} (y_{u}) \bigg)\bigg(\prod_{u \in \calA^{\complement}} \I_{y_{u}=y} \bigg) \bigg)  \\
             & \ \ \ \ \ \ \ \ \ \ \ \ \ \ \ \ \  + \sum_{y_{(i)}^{n} \in \Y^{n-1}} ( 1 - \tau_{\max_2} )\bigg( \prod_{j=1}^n R_j(y_j) \bigg)\\
            & = \sum_{k=0}^{n-2} \calT_k + ( 1 - \tau_{\max_2} )R_i (y_i),
        \end{aligned}
    \end{equation}
    where $\calT_k$ for $k \in \{ 0,\ldots, n-2\}$ is defined as
    \begin{equation}
    \begin{aligned}
      \calT_k  \triangleq & \sum_{y_{(i)}^{n} \in \Y^{n-1}}\sum_{ \substack{\calA \subseteq [n]: \\|\calA| =k} }  \bigg( \bigg( \sum_{l=0}^{k} (-1)^{l} \sum_{ \substack{\calI \subseteq \calA:\\ |\calI| = l} } P_{\min, \calA^\complement \cup \calI}(y) \bigg)  \\ 
     & \times \bigg( \prod_{u \in \calA} R_{u} (y_{u}) \bigg)\bigg(\prod_{u \in \calA^{\complement}} \I_{y_{u}=y} \bigg) \bigg).
    \end{aligned}
    \end{equation}
We decompose $\calT_k \triangleq \calT_{k,1} + \calT_{k,2} $ as
\begin{align}
    \calT_{k,1} &\triangleq  \sum_{y_{(i)}^{n} \in \Y^{n-1}}\sum_{ \substack{\calA \subseteq [n]: \\|\calA| =k,\\ i \notin \calA} } \bigg( \bigg( \sum_{l=0}^{k} (-1)^{l} \sum_{\substack{ \calI \subseteq \calA :\\ |\calI| = l}} P_{\min, \calA^\complement \cup \calI}(y) \bigg)  \vphantom{\sum_{ \substack{\calA \subseteq [n]: \\|\calA| =k} }^k} \nonumber\\
    & \ \ \ \  \times \bigg( \prod_{u \in \calA} R_{u} (y_{u}) \bigg)\bigg(\prod_{u \in \calA^{\complement}} \I_{y_{u}=y} \bigg) \bigg), \label{definition of Tk1} \\
    \calT_{k,2} &\triangleq  \sum_{y_{(i)}^{n} \in \Y^{n-1}}\sum_{ \substack{\calA \subseteq [n]: \\|\calA| =k,\\ i \in \calA} } \bigg( \bigg( \sum_{l=0}^{k} (-1)^{l} \sum_{\substack{ \calI \subseteq \calA :\\ |\calI| = l}} P_{\min, \calA^\complement \cup \calI}(y) \bigg) \nonumber\\ 
    & \ \ \ \ \times \vphantom{\sum_{ \substack{\calA \subseteq [n]: \\|\calA| =k} }^k}  \bigg( \prod_{u \in \calA} R_{u} (y_{u}) \bigg)\bigg(\prod_{u \in \calA^{\complement}} \I_{y_{u}=y} \bigg) \bigg). \label{definition of Tk2} 
\end{align}
Note that in the definition of $\calT_{k,1}$, since $i \notin \calA$, hence the marginalization reduces the product component $\prod_{u \in \calA} R_u(y_u)$ to one. Thus, one obtains
\begin{equation}
\begin{aligned}
    \calT_{k,1} = \sum_{ \substack{\calA \subseteq [n]: \\|\calA| =k,\\ i \notin \calA} }  \sum_{l=0}^{k} (-1)^{l} \sum_{ \substack{ \calI \subseteq \calA :\\ |\calI| = l}} P_{\min, \calA^\complement \cup \calI}(y_i). 
\end{aligned}
\end{equation}
Similarly in the definition of $\calT_{k,2}$, since $i \in \calA$, thus the marginalization reduces the sum over $P_{\min,\calA^{\complement} \cup \calI }$ to a constant (in form of $\tau_{\calA^\complement}$). Moreover, for any set $\calA$, the terms in product of $\prod_{u \in \calA\setminus \{ i\}} R_u(y_u)$ gets marginalized to one. Thus, one obtains
\begin{equation}
    \calT_{k,2} = \bigg(\sum_{ \substack{\calA \subseteq [n]: \\|\calA| =k,\\ i \in \calA} }  \sum_{l=0}^{k} (-1)^{l} \sum_{ \substack{ \calI \subseteq \calA :\\ |\calI| = l}} \tau_{\calA^\complement \cup \calI}\bigg) R_i(y_i).  
\end{equation}

Now we will simplify the terms $\sum_{k=0}^{n-2} \calT_{k,1}$ and $\sum_{k=0}^{n-2}\calT_{k,2}$. First, we will show that 
\begin{equation}
\begin{aligned}
    \sum_{k=0}^{n-2} \calT_{k,1} & = \sum_{k=0}^{n-2}  \sum_{ \substack{\calA \subseteq [n]\setminus \{i\}: \\|\calA| =k }}   \sum_{l=0}^{k} (-1)^{l} \sum_{ \substack{ \calI \subseteq \calA :\\ |\calI| = l}} P_{\min, \calA^\complement \cup \calI}(y_i) 
    \\
    & = \sum_{l=2}^n (-1)^l \sum_{\substack{\calI \subseteq [n]: \\|\calI| = l,\\ i \in \calI } } P_{\min,\calI}(y_i). \label{simplification of Tk1}
\end{aligned}
\end{equation}  
This follows since $\calT_{k, 1}$ is a sum of terms  of the form $P_{\min,\calV}(y_{i})$ where $\calV \subseteq [n]$ and clearly $\calV$ contains $\{i\}$. Clearly, from \cref{simplification of Tk1}, for any term corresponding to the set $\calV$ in the sum $\sum_{k=0}^{n-2} \calT_{k,1}$, we have $|\calV| \geq 2$. We focus on identifying the coefficient of $P_{\min,\calV}(y_i)$ in the sum $\sum_{k=0}^{n-2} \calT_{k,1}$ for any $\calV \subseteq [n]$ such that $i \in \calV$ and $|\calV| = u $ for some $u \geq 2$. Now, corresponding to $k=n-u$,  the coefficient of $P_{\min,\calV}(y_i)$ is ${u\choose 0}$ when $\calA = \calV^\complement $ and $l=0$. Similarly, corresponding to $k = n-u+1$, we get coefficient of $P_{\min,\calV}(y_i)$ as $-{u-1 \choose 1}$, when $\calA = \calV^{\complement} \cup \{t\}$ where $t \in \calV\setminus \{i\}$ and $l=1$. Continuing in a similar fashion, corresponding to $|\calA|=n-u+v$, we get coefficient as $(-1)^{v} {u-1 \choose v}$ for $v \in \{0, \ldots, u-2\}$. Hence, the coefficient of $P_{\min,\calV}(y)$ is
$$
    \sum_{v=0}^{u-2}(-1)^{v} {u-1 \choose v} =(-1)^{u}.
$$
Since $\sum_{k=0}^{n-2} \calT_{k,1}$ contains all sets $\calV \subseteq [n]$ with cardinality at least two such that $i \in \calV $, $\sum_{k=0}^{n-2} \calT_{k,1} $ simplifies to 
\begin{equation}
    \sum_{k=0}^{n-2} \calT_{k,1} = \sum_{l=2}^n (-1)^l \sum_{\substack{\calI \subseteq [n]: \\|\calI| = l,\\ i \in \calI } } P_{\min,\calI}(y_i).
\end{equation}
Now we will simplify $\sum_{k=0}^{n-2}\calT_{k,2}$ as 

\begin{align}
   \nonumber  &\sum_{k=0}^{n-2} \calT_{k, 2} = \\
   \nonumber  &   \sum_{y  \in \mathcal{Y}} \sum_{k=0}^{n-2} \sum_{\substack{\calA \subseteq [n] \\ |\calA|=k \\
    i \in \calA}} \sum_{l=0}^{k}(-1)^{l} \sum_{\substack{\calI \subseteq \calA ; \\
    |\calI|=l}} P_{\min , \calA^{\complement} \cup \calI }(y) R_i(y_i)\\
   \nonumber  & =  \sum_{y  \in \mathcal{Y}} \sum_{k=0}^{n-2} \sum_{\substack{\calA \subseteq [n] \\ |\calA|=k}}^{n-2} \sum_{l=0}^{k}(-1)^{l} \sum_{\substack{\calI \subseteq \calA ; \\     |\calI|=l}} P_{\min , \calA^{\complement} \cup \calI }(y)R_i(y_i)  \\  
     \nonumber & \ \ \  - \sum_{y  \in \mathcal{Y}} \sum_{k=0}^{n-2} \sum_{\substack{\calA \subseteq [n]\setminus \{i\} \\ |\calA|=k}}^{n-2} \sum_{l=0}^{k}(-1)^{l} \sum_{\substack{\calI \subseteq \calA ; \\     |\calI|=l}} P_{\min , \calA^{\complement} \cup \calI }(y)R_i(y_i) \\ 
    \nonumber & \stackrel{\zeta_1}{=} \sum_{y \in \Y} \max\nolimits_{2}\{P_{1}(y), \ldots, P_{n}(y)\} R_i(y_i) \\ 
     \nonumber& \ \ \ \  -  \sum_{y  \in \mathcal{Y}} \sum_{l=2}^n (-1)^l \sum_{\substack{\calI \subseteq [n]: \\|\calI| = l,\\ i \in \calI } } P_{\min,\calI}(y)R_i(y_i)  \\
     & \stackrel{\zeta_2}{=} ( \tau_{\max\nolimits_2} -1 + (\tau_{\max} - \tau_{\max,(i) })  )R_i(y_i),\label{simplifaction of Tk2}
\end{align}
where $\zeta_1$ follows from \cref{lem: the three equivalences} and \cref{simplification of Tk1} and  $\zeta_2$ follows by re-arrangement of \cref{lem: max min inclusion rule} as follows  
\[
\begin{aligned}
     P_{\max,[n]}(y) =  \sum_{l=1}^n (-1)^{l-1} \sum_{\substack{\calI \subseteq [n],\\ |\calI| = l\\i\in \calI} } P_{\min, \calI}(y) + P_{\max,(i)}(y). \end{aligned}
\]
Now substituting the simplified expression of $\sum_{k=0}^{n-2}\calT_{k,1}$ and $\sum_{k=0}^{n-2}\calT_{k,2}$ in \cref{simplification of Tk1} and \cref{simplifaction of Tk2} into \cref{incomplete right marginals}, we get \[
    \begin{aligned}
         & \mathbb{P}(Y_{i} =y_i) = (1-\tau_{\max\nolimits_2})R_i(y_i) \\ 
         & \ \ \ \ \ \ \ \ \ \ \  \ \ \ \ + \ \sum_{l=2}^n (-1)^l \sum_{\substack{\calI \subseteq [n]: \\|\calI| = l,\\ i \in \calI } } P_{\min,\calI}(y_i) \\ 
         &\ \ \ \ \ \ \ \ \ \ \ \ \ \ \  +  (\tau_{\max\nolimits_2} -1 + (\tau_{\max} - \tau_{\max,(i) })  )R_i(y_i) \\
          & = \sum_{l=2}^n (-1)^l \sum_{\substack{\calI \subseteq [n]: \\|\calI| = l,\\ i \in \calI } } P_{\min,\calI}(y_i) + \sum_{l=1}^n (-1)^{l-1} \sum_{\substack{\calI \subseteq [n]: \\|\calI| = l,\\ i \in \calI } } P_{\min,\calI}(y_i) \\
         & = P_{i}(y_i).
    \end{aligned}
\]
    Now, it remains to show that the coupling construction satisfies the lower bound \cref{lower bound for general n}. To prove this, we will show that for any $\calA \subseteq [n],$ we have
    \begin{equation} \label{measure of intersections is min}
        \mathbb{P}(\cap_{l \in \calA}\{Y_{l}=y\})=P_{\min, {\calA} }(y).
    \end{equation}
    If this is true, then the achievability of the lower bound follows by using \cref{lem: max min inclusion rule} and inclusion-exclusion rule for probability.
    \begin{align}
        &\sum_{y\in\Y} \P(  \cup_{i=1}^n \{Y_i =y\})  
         \nonumber \\ 
        \nonumber & = \sum_{y \in \Y} \sum_{k=1}^n  (-1)^{k-1} \sum_{ \substack{\calI \subseteq [n]: \\ |\calI|=k} } \P(\cap_{i\in \calI} \{Y_i = y\})  \\
         & = \sum_{y \in \Y} \sum_{k=1}^n  (-1)^{k-1} \sum_{ \substack{\calI \subseteq [n]: \\ |\calI|=k} } P_{\min,\calI } (y) 
         = \sum_{y \in \Y}P_{\max,[n]}(y).
    \end{align}

    To prove \cref{measure of intersections is min}, similar to $n=3$, in the following lemma we will show that the various components used in \cref{convex combination n3} are mutually disjoint. 
    \begin{lemma}[Orthogonality of Components] \label{lem: Orthogonality of components} The following statements hold:
    \begin{enumerate}
        \item For any $i \neq j$, and for any $y  \in \Y$ we have 
            $$ R_{i}(y) R_{j}(y)=0,  \ \forall\ y  \in \Y. $$
        \item $Q_{k, \calA}(y, y_{a_{1}}, \ldots, y_{a_{k}})=0$ whenever any two of its coordinates are equal.
        \item The distributions used in convex combination \cref{coupling for general n} are mutually disjoint to one another. 
    \end{enumerate}
    \end{lemma}
    The proof of \cref{lem: Orthogonality of components} is provided in \cref{Proof of Orthogonality of components}. 
    
    Now to show \cref{measure of intersections is min}, for any $\calV \subseteq [n]$ such that $|\calV^{\complement}| = u$ and $\calV^{\complement} = \{v_1,\ldots,v_u\}$, we have
    \begin{align}
    &\nonumber \mathbb{P}(\cap_{l \in \calV } \{Y_{l}=y\}) \\
    & \nonumber = \sum_{y_{v_1}, \ldots, y_{v_u} \in \Y^u}\P(\cap_{l \in \calV} \{Y_{l}=y\} \cap (\cap_{q =1}^u \{Y_{v_q} = y_q\}) \\
    \nonumber & \stackrel{\zeta_1}{=} \sum_{y_{v_1}, \ldots, y_{v_u}\in \Y^u} \sum_{k=0}^{u} \sum_{ \substack{\calA \subseteq \calV^\complement\\ |\calA| = k} }  \bigg( \sum_{l=0}^{k}(-1)^{l} \sum_{\substack{I \subseteq \calA :\\
    |\calI|=l}} \tau_{\calA^{\complement} \cup \calI} \bigg) \times
       \\ 
     \nonumber & \ Q_{k,\calA}(y,y_{a_1},\dots,y_{a_k}) \times \bigg( \prod_{q \in \calV^\complement \setminus \calA} \I_{y_{q} = y}\bigg) \times \prod_{q =1 }^k R_{a_q}(y_{a_q}) \\
     \nonumber & \stackrel{\zeta_2}{=} \sum_{k=0}^{u} \sum_{ \substack{\calA \subseteq \calV^\complement\\ |\calA| = k} } \sum_{l=0}^{k} (-1)^{l} \sum_{ \calI \subseteq \calA : |\calI| = l} P_{\min, \calA^\complement \cup \calI}(y)  \\
     & \stackrel{\zeta_3}{=} P_{\min, \calV}(y),  \label{last step showing min is achievabile}
     \end{align}  
    where $\zeta_1$ follows from \cref{coupling for general n} and  \cref{lem: Orthogonality of components}. Hence, only those `components' of coupling $Q_{k,\calA}$ are taken into consideration where $\calA \subseteq \calV^\complement$.  $\zeta_2$ follows from definition of $Q_{k,\calA}(\cdot)$ in \cref{coupling distributions2} and since the product distribution corresponding to  $R_{a_q}(v_{a_q})$ gets marginalized. The explanation for $\zeta_3$ is as follows: Note that $\zeta_2$ is the sum of terms of the form $P_{\min,\calZ}(y)$ where $\calZ \supset \calV$. Also, $\calV$ is the smallest set in various terms of $\zeta_2$ and has coefficient $1$ (corresponding to $k=u, \calA = \calV^\complement, l= 0, \calI = \phi$). Next, we will show that any other term has a coefficient $0$. 
    Consider any term $P_{\min,\calZ}(y)$ where $|\calZ|> n-u$  and $\calV \subseteq \calZ$ (i.e., $\calZ $ has at least one element of $\calV^\complement$). Define $\calZt = \calZ \setminus \calV$. Observe that the term $P_{\min,\calZ}(y)$ in $\zeta_2$ is observed when $\calI = \calZt \cap \calA$. Therefore, the coefficient of $P_{\min,\calZ}(y)$ when $|\calZt \cap \calA| = q$ is $(-1)^q { |\calZt| \choose q}$. Hence, the overall coefficient is given by  
    $$\sum_{q=0}^{|\calZt|} (-1)^q { |\calZt| \choose q} = 0.$$
    Thus, proving the theorem.
\end{proof}

\subsubsection{Proof of \cref{lem: these are PMFs}} \label{Proof of these are PMFs}
    First, we will show that $R_k$ is indeed a PMF. Note that the numerator term of $R_k(y)$ in \cref{coupling distributions1} is
    \begin{align}
    \nonumber   & \sum_{l=1}^n (-1)^{l-1} \sum_{\substack{\calI \subseteq [n]\\|\calI| = l,\\ k\in \calI} } P_{\min,\calI}(y) = 
         \sum_{l=1}^n (-1)^{l-1}\sum_{\substack{\calI \subseteq [n]\\|\calI| = l}  } P_{\min,\calI}(y) \\ 
  \nonumber   & \ \ \ \ \ \ \ \ \ \ \ \ \ \ \ \ \  - \sum_{l=1}^n (-1)^{l-1}\sum_{\substack{\calI \subseteq [n]\\|\calI| = l, k\notin \calI} } P_{\min,\calI}(y) \\
    \nonumber    & =P_{\max,[n]}(y) \quad-\sum_{l=1}^{n}(-1)^{l-1} \sum_{\substack{\calI \subseteq[n]\setminus \{k\} \\
        |\calI|=l}} P_{\min, \calI}(y) \\
        &  =P_{\max ,[n]}(y)-P_{\max ,(k)}(y).\label{numerator of Rk}
    \end{align}
    which is always non-negative. Moreover, the denominator is also non-negative since it is obtained by summing non-negative terms over all $y \in \Y$. Since $R_k(y)$ is normalized to have a sum of $1$, hence it is a PMF over $\Y$.

    Now we will show $Q_{k,\calA}(y, y_{a_1}, \dots ,y_{a_{k}})$ is a PMF over the product space $\Y^{k+1}$. To prove this, we need to show that for any $\calA \subseteq [n]$ with $|\calA| = k \leq n-2$, $S_\calA(y)$ defined as
    \begin{equation}    
        S_\calA(y) \triangleq    \frac{\sum_{l=0}^{k} (-1)^{l} \sum_{ \calI \subseteq \calA : |\calI| = l} P_{\min, \calA^\complement \cup \calI}(y) }{ \sum_{l=0}^{k} (-1)^{l} \sum_{ \calI \subseteq \calA: |\calI| = l} \tau_{\calA^\complement \cup \calI} }, \label{definition of Sy}
    \end{equation}
    is a PMF over $\Y$. Using the same technique as \cref{numerator of Rk}, we can simplify the numerator of $S_\calA(y)$ as
    \begin{equation}
    \begin{aligned}\label{numerator of QkA}
         &\sum_{l=0}^{k} (-1)^{l} \sum_{ \substack{\calI \subseteq \calA :\\ |\calI| = l}} P_{\min, \calA^\complement \cup \calI}(y)  \stackrel{\zeta_1}{=} \sum_{l=0}^{k} (-1)^{l} \sum_{ \substack{ \calI \subseteq \calA \cup \mathfrak{a}^\complement:\\ |\calI| = l+1 \\ \mathfrak{a}^\complement \in \calI }} P_{\min, \calI}(y)  \\
        & \stackrel{\zeta_2}{=} \max \{P_{\mathfrak{a}^\complement }(y), P_{a_1}(y), \ldots, P_{a_k}(y)  \}-P_{\max , \calA}(y)\\
        & = \max \{P_{\min, \calA^\complement }(y), P_{a_1}(y), \ldots, P_{a_k}(y)  \}-P_{\max , \calA}(y)\\
        & = \max \{P_{\min, \calA^\complement }(y), P_{\max, \calA} (y)\}-P_{\max , \calA}(y) \geq 0,
    \end{aligned}
    \end{equation}
    where in $\zeta_1$ we treat the set $\mathcal{A}^\complement$ as a single element $\mathfrak{a}^\complement$ and use the same techniques as in \cref{numerator of Rk} for $\zeta_2$. Similarly, the denominator of $S_\calA(y)$ is non-negative as it is obtained by summing non-negative terms over all $y \in \Y$. Hence, $S_{\calA}(y)$ is a PMF. Finally, since $R_i(y)$ is a probability mass function (PMF), the fact that $Q_{n-1}(y_1,\ldots,y_n)$ has a product structure demonstrates that it is also a PMF.  \qed

\subsubsection{Proof of \cref{lem: the three equivalences}}\label{Proof of the three equivalences}
    First, we will prove \cref{one of three}. For any fixed $y$, we need to show that the term $\tilde{\calT}$
    \begin{equation}
    \begin{aligned}
         \tilde{\calT} & \triangleq  \sum_{k=0}^{n-2} \sum_{\substack{\calA \subseteq [n]:\\ |\calA|=k} } \bigg( \sum_{l=0}^k (-1)^l \sum_{\substack{\calI \subseteq \calA\\|\calI|=l}} P_{\min,\calA^{\complement} \cup \calI } (y)\bigg) \\
        & = \sum_{k=2}^n (-1)^{k}(k-1) \sum_{\substack{ \calI \subseteq [n]\\|\calI| = k } } P_{\min,\calI}(y). 
    \end{aligned}
    \end{equation}
    To see this, consider any set $\calV \subseteq [n]$ such that $|\calV| = u$. Clearly, the cardinality of any set $\calV$ on the left in \cref{one of three} is at least two. We focus on identifying the coefficient of term $P_{\min,\calV}(y)$ in $\tilde{\calT}$. For $u \geq 2$, corresponding to $|\calA|=n-u$ we get coefficient of $P_{\min,\calV}(y)$ as $u\choose 0$ when $l=0$. Similarly, corresponding to $|\calA|=n-u+1$, we get coefficient as $-{u \choose 1}$, when $l=1$. Similarly corresponding to $|\calA|=n-u+v$, we get coefficient as $(-1)^{v} {u \choose v}$ for $v \in \{0, \ldots, u-2\}$. Hence, the coefficient of $P_{\min,\calV}(y)$ is
    $$
        \sum_{v=0}^{u-2}(-1)^{v} {u \choose v} =(-1)^{u}(u-1).
    $$
    {Since $\tilde{\calT}$ contains all sets $\calV \subseteq [n]$ and with cardinality at least two such that $i \in \calV $, hence we get
    \begin{equation}
         \tilde{\calT} = \sum_{k=2}^n (-1)^{k} (k-1) \sum_{\substack{\calI \subseteq [n]: \\|\calI| = k } } P_{\min,\calI}(y).
    \end{equation}
    The equivalence of the right-hand side of \cref{one of three} and \cref{three of three} follows from the maximum-minimums principle for the $k$-th ranked element \cite[Proposition 2]{maximumminimumidentity}, which is stated below specifically for the second largest element.
    \begin{lemma}[Maximum-Minimums Principle for Second Largest Element {\cite[Proposition 2]{maximumminimumidentity}}]
    For any $x_1,x_2,\ldots,x_n \in \R$,
    $$
        \begin{aligned}
        & \max\nolimits_2\{x_1 , \ldots,x_n \} \\
        &  \  \ \ \ \ \ \ \ \ \ = \sum_{r=2}^n (-1)^{r}{r-1 \choose 1} \sum_{i_1 < \ldots<i_r } \min\{ x_{i_1}, \ldots,x_{i_r} \}.
        \end{aligned}
    $$
    \end{lemma}
    This proves the equivalence of \cref{one of three} and \cref{three of three}. 
\qed

\subsubsection{Proof of \cref{lem: Orthogonality of components}} \label{Proof of Orthogonality of components} \ \par
    \textbf{Part 1:} Note that from \cref{numerator of Rk}, we have
     $$\begin{aligned}
        R_{i}(y) & R_{j}(y) \propto \\
& (P_{\max }(y)-P_{\max,(i)}(y))(P_{\max }(y)-P_{\max, (j)}(y)).    
     \end{aligned}
     $$ 
     Hence, $R_{i}(y) \neq 0$ if and only if $P_{i}(y)>P_{\max, (i) }(y)$ and similarly $R_{j}(y) \neq 0$ if and only if $P_{j}(y)>P_{\max,(j)}(y)$. Since both the conditions cannot be true simultaneously for $i\neq j$, hence $R_{i}(y) R_{j}(y)=0$.

    \textbf{Part 2:} For $i\neq j$ and $a_i, a_j \in \calA$, whenever $y_{a_{i}}=y_{a_{j}}$, $Q_{k,\calA}(\cdot) =0$ (by definition in \cref{coupling distributions2} and by Part 1). Now we will show that whenever $y=y_{a_{i}}$ for some $a_i \in \calA$, then we need to show $Q_{k,\calA}(y,y_{a_1}, \ldots,y_{a_{i-1}}, y,y_{a_{i+1}},\ldots, y_{a_k}) =0$ for all $y, y_{a_{1}}, \ldots,y_{a_{i-1}}, y_{a_{i+1}}, \ldots, y_{a_{n}} \in \Y$. Note that by \cref{numerator of Rk} and \cref{numerator of QkA}, we have  
    $$ 
        \begin{aligned}
             & Q_{k,\calA}(y,y_{a_1}, \ldots,y_{a_{i-1}}, y,y_{a_{i+1}},\ldots, y_{a_k} )  \propto  \\
             & ( \max\{  P_{\min, \calA^{\complement}}(y), P_{\max,\calA }(y)\} - P_{\max, \calA}(y)) \times \\
            & \ \ \ \ \ \ \ \ \ \ \ \  (P_{\max}(y) - P_{\max,(a_i)}(y) ) \times\prod_{\substack{j=1\\ j\neq i}}^k R_j(y_{a_j}).
        \end{aligned}
    $$
    Clearly, the above term is zero because $P_{\max}(y) - P_{\max,(a_i)}(y)  \neq 0$ if and only if $P_{a_i}(y) > P_{\max ,(i)}(y)$ which implies $\max \{ P_{\min,\calA^{\complement} } (y), P_{ \max, \calA}(y)\}=P_{ \max,\calA }(y)$ as $a_i \in \calA$. 

    \textbf{Part 3:} Follows, directly from Part 1 and Part 2. 
    This is because by Part 1, $Q_{n-1}(\cdot)$ is only supported on the set where all the co-ordinates are different from each other. Moreover, for any set $\calA, \calA^\prime \subseteq [n]$ such that $|\calA| = k,  |\calA^\prime| = k^\prime$ and $k\neq k^\prime$, $Q_{k,\calA}(\cdot)$ and $Q_{k^\prime,\calA}(\cdot)$  are clearly disjoint from one another by Part 2. 
    Similar reasoning applies when $|\calA|=|\calA^\prime| = k$, (but $\calA \neq \calA^\prime$).
    \qed

\subsection{Proof of \cref{New Minimal Coupling for max-Doeblin Coefficient}}\label{Proof of New Minimal Coupling for max-Doeblin Coefficient}
    The coupling construction for $\tau_{\max_2}(P_1,P_2,P_3) \leq 1$ is the same as that described in \cref{sec: Proof of MaxDoeblinCoupling}. 
    For the case where $\tau_{\max_2}(P_1,P_2,P_3) > 1$, the lower bound $\sum\limits_{y  \in \Y} \P(\cup_{i=1}^{3}\{Y_{i} =y\})$ in \cref{lower bound n3} can be strengthened as presented in the following lemma.
    \begin{lemma}[Strengthened Lower Bound for any Coupling] \label{lem: strengthened lower bound}
    For any random variables $Y_1, Y_2, Y_3$ with PMFs $P_1, P_2, P_3$, respectively, and for any coupling $\P$ of $P_1, P_2, P_3$, the following statement holds:
    $$ 
    \begin{aligned}
    \sum\limits_{y  \in \Y} \P(\cup_{i=1}^{3}&\{Y_{i} =y\}) \geq \\
    &  \tau_{\max}(P_1,P_2,P_3) + (\tau_{\max_2}(P_1,P_2,P_3) -1)_+.
    \end{aligned}
    $$
    \end{lemma}
    The proof is provided in \cref{Proof of strengthened lower bound}. 
    Next, we will present a coupling construction specifically designed for the case when $\tau_{\max_2}(P_1, P_2, P_3) > 1$. This construction aims to achieve the lower bound stated in \cref{lem: strengthened lower bound} by modifying the previous coupling construction used for the case $\tau_{\max_2}(P_1, P_2, P_3) \leq 1$. Let $\tau_2 \triangleq \tau_{\max_2}(P_1,P_2,P_3)$ and define the following distributions:
 \begin{align*}
     & R_1(y)  \triangleq  \frac{1}{1+\tau-\tau_{12}-\tau_{13} + \frac{2(\tau_2 -1)}{3} } \times \bigg( P_{1}(y)+P_{\min}(y) \\
     &  - \min \{P_{1}(y),P_{2}(y)\}-\min \{P_{1}(y), P_{3}(y)\}  + \frac{\tau_2 -1}{3}\\ 
     &  \ \ \ \ \ \ \times\bigg( \frac{\min\{P_1(y),P_2(y)\}-P_{\min}(y)}{\tau_{12} -\tau} + \\
     &  \ \ \ \ \ \      \frac{\min\{P_1(y),P_3(y)\}-P_{\min}(y)}{\tau_{13} -\tau}\bigg) \bigg), \\
     & R_2(y)  \triangleq  \frac{1}{1+\tau-\tau_{12}-\tau_{23} + \frac{2(\tau_2 -1)}{3} } \times \bigg( P_2(y)+ P_{\min}(y)  \\
     & -\min \{P_{1}(y),P_{2}(y)\}-\min \{P_{2}(y),P_{3}(y)\}  + \frac{\tau_2 -1}{3}\\ 
     &  \ \ \ \ \ \ \times  \bigg( \frac{\min\{P_1(y),P_2(y)\}-P_{\min}(y)}{\tau_{12} -\tau} + \\ 
     &  \ \ \ \ \ \  \frac{\min\{P_2(y),P_3(y)\}-P_{\min}(y)}{\tau_{23} -\tau}\bigg) \bigg), \\
     & R_3(y)  \triangleq \frac{1}{1+\tau-\tau_{13}-\tau_{23} + \frac{2(\tau_2 -1)}{3} } \times \bigg( P_3(y)+ P_{\min}(y)\\
     &  -\min \{P_{1}(y),P_{3}(y)\}-\min \{P_{2}(y),P_{3}(y)\}     + \frac{\tau_2 -1}{3}  \\ 
     &  \ \ \ \ \ \ \times \bigg( \frac{\min\{P_1(y),P_3(y)\}-P_{\min}(y)}{\tau_{13} -\tau} + \\
     &  \ \ \ \ \ \ \frac{\min\{P_2(y),P_3(y)\}-P_{\min}(y)}{\tau_{23} -\tau}\bigg) \bigg). 
    \end{align*}

    Now, we substitute the above defined $R_i(y)$ into the definition of $Q_{1,i}(y,y^\prime)$ in \cref{bunch of eqns} for $i \in [3]$. Compared with \cref{bunch of eqns}, the definition of $Q_{0}(y)$ remains the same as before, and we will no longer utilize the product distribution $Q_2(\cdot)$ in the construction of the modified coupling. Now, we define the coupling distribution as a convex combination, as described below:
        \begin{align}
        \nonumber    & \P(y_{1}, y_{2}, y_{3})  = \tau Q_{0}(y) \I_{y_{1}=y_{2}=y_{3}=y} + \\ 
         \nonumber   & \bigg(1 + \tau -\tau_{12} - \tau_{13} + \frac{2(\tau_2 -1)}{3} \bigg)  Q_{1,1}(y_{1}, y^{\prime}) \I_{y_{2}=y_{3}=y^{\prime}} \\
         \nonumber   & + \bigg(1 + \tau -\tau_{12} - \tau_{23} + \frac{2(\tau_2 -1)}{3} \bigg) Q_{1,2}(y_{2}, y^{\prime}) \I_{y_{1}=y_{3}=y^{\prime}} \\
            & + \bigg(1 + \tau -\tau_{23} - \tau_{13} + \frac{2(\tau_2 -1)}{3}  \bigg) Q_{1,3}(y_{3}, y^{\prime}) \I_{y_{1}=y_{2}=y^{\prime}}.\label{convex combination new3}
        \end{align}
    One can verify that each of the weights used in the above convex combination are non-negative and they sum to $1$ as
    $$
    \begin{aligned}
         & \tau + 3 + 3\tau - 2(\tau_{12}+\tau_{13}+\tau_{23}) + 2(\tau_2 -1)  \\ 
         & = 1 - 2(\tau_{12}+\tau_{13}+\tau_{23} - 2\tau) + 2\tau_2  = 1,
    \end{aligned}
    $$
    where the last equality follows using \cref{somelabel}.
    
    As our next step, we will verify that the coupling has the correct marginals. Without loss of generality, we will check the marginals of the coupling $\P$ with respect to $Y_1$.
    $$
    \begin{aligned}
         & \sum_{y_{2}, y_{3} \in \Y} \mathbb{P}(y_{1}, y_{2}, y_{3})   = P_{\min}(y_1) + (P_{1}(y_{1})+P_{\min}(y_1) - \\ 
         &  \min \{P_{1}(y_{1}), P_{2}(y_{1})\}-\min \{P_{1}(y_{1}), P_{3}(y_{1})\}) + \frac{\tau_2 -1}{3} \times  \\
         &\ \  \ \ \ \ \ \ \ \ \ \ \   \bigg( \frac{\min\{P_1(y_1),P_2(y_1)\}-P_{\min}(y_1)}{\tau_{12} -\tau} + \\
         & \ \  \ \ \ \ \ \ \ \ \ \ \ \ \ \ \ \ \ \  \frac{\min\{P_1(y_1),P_3(y_1)\}-P_{\min}(y_1)}{\tau_{13} -\tau}\bigg) \\
         &\ \  \ \ \ \ \ \ \ \ \ \ \   +  \bigg(1 + \tau -\tau_{12} - \tau_{23} + \frac{2(\tau_2 -1)}{3}  \bigg)\\ 
         &\ \  \ \ \ \ \ \ \ \ \ \ \ \ \ \ \ \ \ \ \ \ \ \times \frac{\min\{P_{1}(y_{1}), P_{3}(y_{1})\}-P_{\min}(y_1)}{\tau_{13} - \tau} \\
         &\ \  \ \ \ \ \ \ \ \ \ \ \   + \bigg(1 + \tau -\tau_{23} - \tau_{13} + \frac{2(\tau_2 -1)}{3}  \bigg)  \\
         &\ \  \ \ \ \ \ \ \ \ \ \ \  \ \ \ \ \ \ \ \ \ \ \times   \frac{\min\{P_{1}(y_{1}), P_{2}(y_{1})\}-P_{\min}(y_1)}{\tau_{12} - \tau} \\
         & = P_1(y_1).
    \end{aligned}
    $$

    Finally, to demonstrate that the coupling satisfies the lower bound in \cref{lower bound n3}, we will use the inclusion-exclusion principle to simplify $\mathbb{P}(\bigcup_{i=1}^{3}\{Y_{i}=y\})$ as
    \begin{align*}
         & \sum_{y \in \Y} \mathbb{P}\bigg(\bigcup_{i=1}^{3}\{Y_{i}=y\}\bigg) \\ 
         & = \sum_{y \in \Y}\bigg( \mathbb{P}(Y_{1}=y)+\mathbb{P}(Y_{2}=y)  +\mathbb{P}(Y_{3}=y) \\
         & \ \ \ \ \ \ \ \ \ \ \ -(\mathbb{P}(Y_{1}=y, Y_{2}=y)+\mathbb{P}(Y_{2}=y, Y_{3}=y) \\
         & \ \ \ \ \ \ \ \ \ \ \  +\mathbb{P}(Y_{1}=y, Y_{3}=y)) +\mathbb{P}(Y_{1}=Y_{2}=Y_{3}=y) \bigg)\\
         &= \sum_{y \in \Y}\bigg( P_1(y) + P_{2}(y) + P_{3}(y) - (\mathbb{P}(Y_{1}=y, Y_{2}=y) \\
         & \ \  +\mathbb{P}(Y_{2}=y, Y_{3}=y) +\mathbb{P}(Y_{1}=y, Y_{3}=y)) + P_{\min}(y) \bigg)\\
         & = 3 + \tau - \sum_{y \in \Y} (\mathbb{P}(Y_{1}=y, Y_{2}=y)+\mathbb{P}(Y_{2}=y, Y_{3}=y) \\ 
         & \ \ \ \ \ \ \ \ \ \ \ \ +\mathbb{P}(Y_{1}=y, Y_{3}=y)) \\ 
         & = 3 + \tau - (\tau + 1+ \tau -\tau_{12}-\tau_{13} + \frac{2(\tau_2 -1)}{3} ) - \\
         & \ \ \ \ \ \ \ \  (\tau + 1+ \tau -\tau_{12}-\tau_{23} + \frac{2(\tau_2 -1)}{3} ) - \\
         &\ \ \ \ \ \ \ \ \ \   (\tau + 1+ \tau -\tau_{13}-\tau_{23} + \frac{2(\tau_2 -1)}{3} ) \\
         & =  2 - \tau  = \tau_{\max}(P_1,P_2,P_3) + \tau_2 -1, 
            \end{align*}
    where the second equality follows since the coupling preserves marginals, we get $ \P(Y_{i}=y)=P_{i}(y)$. Furthermore, by the construction of coupling $\mathbb{P}(Y_{1}=Y_{2}=Y_{3}=y) = \min\{P_{1}(y), P_{2}(y), P_{3}(y)\}$. \qed 

\subsubsection{Proof of \cref{lem: strengthened lower bound}}
\label{Proof of strengthened lower bound} 
When $\tau_{\max_2}(P_1,P_2,P_3) \leq 1,$ then the lower bound derived in \cref{lower bound n3} suffices to prove the claim. Now we will focus on the case when $\tau_{\max_2}(P_1, P_2, P_3) > 1$.
For distributions $P_1, P_2, P_3$ and $y,y^\prime \in \Y$, define $A_{yy^\prime} \triangleq \{Y_{a_y} = y^\prime\} $ where $a_y$ is the smallest index $i \in [3]$ such that $a_y = \argmax_{i} P_i(y)$. Let $C_{yy^\prime} \triangleq \{Y_{c_y} = y^\prime\} $ where $c_y$ is the largest index $i \in [3]$ such that  $c_y =\argmin_{i} P_i(y)$. Finally, let $B_{yy^\prime} \triangleq \{Y_{b_y} = y^\prime\} $, where $b_y =\argmax_{2} P_i(y) $ or equivalently $b_y = [3] \setminus \{a_y,c_y\}$ is the index such that $P_{b_y}(y)$ is neither the maximum nor the minimum among $P_i(y)$.
Using the inclusion-exclusion rule, we get 
\begin{equation*}
\begin{aligned}
    &\sum_{y \in \Y}\P(\cup_{i=1}^3 \{Y_i = y\} ) \\
    & = \sum_{y \in \Y} ( \P(A_{yy} ) + \P(B_{yy} \cup C_{yy}) - \P(A_{yy} \cap (B_{yy} \cup C_{yy})) ) \\
    & \stackrel{\zeta}{\geq} \sum_{y \in \Y} ( \max\{P_1(y), P_2(y), P_3(y)\} + \\
    & \ \ \ \ \ \   \max\nolimits_2\{P_1(y), P_2(y), P_3(y)\}   -  \P(A_{yy} \cap (B_{yy} \cup C_{yy})) ) \\ 
    &  = \tau_{\max}(P_1,P_2,P_3) + \tau_{\max_{2}} (P_1,P_2,P_3) \\
    & \ \ \ \ \ \ - \sum_{y \in \Y} \P(A_{yy} \cap (B_{yy} \cup C_{yy}))\\
    &  \geq  \tau_{\max}(P_1,P_2,P_3) + \tau_{\max_{2}} (P_1,P_2
    ,P_3)  -1,
\end{aligned}
\end{equation*}
where $\zeta$ follows since $\P(A_{yy}) =\P(\{Y_{a_y} = y\}) = \max\{P_1(y),P_2(y),P_3(y)\}$, by definition of $A_{yy}$ and the fact that the coupling preserves the marginals, and in addition, $\P(B_{yy} \cup C_{yy}) \geq \max\{\P(B_{yy}),\P(C_{yy})\} = \max\nolimits_2 \{P_1(y), P_2(y), P_3(y)\} $ by the same reasoning.
Now, we claim that the sum $\sum_{y \in \Y} \P(A_{yy} \cap (B_{yy} \cup C_{yy})) \leq \P(\Y^3) = 1$. Notice that 

\begin{align*}
 &\sum_{y \in \Y} \P(A_{yy} \cap (B_{yy} \cup C_{yy}))  = \sum_{y \in \Y} \P(A_{yy} \cap B_{yy} \cap C_{yy}) \\ 
 & \ \ \ \ \ \ \ \ \ \  + \sum_{y \in \Y}\sum_{\substack{y^\prime \in \Y: \\ y^\prime \neq y}}  \P(A_{yy} \cap B_{yy^\prime} \cap C_{yy}) \\
 & \ \ \ \ \ \ \ \ \ \ +  \sum_{y \in \Y}\sum_{\substack{y^\prime \in \Y: \\ y^\prime \neq y}} \P(A_{yy} \cap B_{yy} \cap C_{yy^\prime}) \\
 & \ \ = \underbrace{ \sum_{y \in \Y} \P(Y_1 = Y_2 = Y_3 = y) }_{\zeta_1}  \\
  & \ \ \ \ + \underbrace{\sum_{y \in \Y} \sum_{\substack{y^\prime \in \Y: \\ y^\prime \neq y}}  \P(\{ Y_{a_y} = Y_{c_y} = y\} \cap \{ Y_{b_y} = y^\prime \}  ) }_{\zeta_2} \\
  & \ \ \ \ \ \ \ \ + \underbrace{\sum_{y \in \Y} \sum_{\substack{y^\prime \in \Y: \\ y^\prime \neq y}}  \P(\{ Y_{a_y}  = Y_{b_y} = y\} \cap \{ Y_{c_y} = y^\prime \} ) }_{\zeta_3} \\
 & \stackrel{\zeta}{\leq} \P(\Y^3) = 1,
\end{align*}
where $\zeta$ follows since the above decomposition is the measure of disjoint sets. This bookkeeping may be somewhat non-trivial to see. The terms $\zeta_1,\zeta_2, \zeta_3$ are sums of measures of various sets. Notice that any set in $\zeta_1$ is clearly disjoint from any set in either $\zeta_2$ or $\zeta_3$. Moreover, any set in $\zeta_2$ is disjoint from any set in $\zeta_3$ (as for any $y$, the indices  $a_y,  b_y,c_y$ are unique). Now we show that the various sets in $\zeta_2$ (or $\zeta_3$) are disjoint from one another. This follows because for fixed $y^\prime$ but different $y$ we get disjoint sets and for different $y^\prime$ but fixed $y$ we get another disjoint set and by varying both $y, y^\prime$, no overlapping sets can be obtained, which completes the proof. \qed 

\subsection{Proof of \cref{Simultaneous Coupling for Doeblin and max-Doeblin}}\label{Proof of Simultaneous Coupling for Doeblin and max-Doeblin}

        The coupling construction presented in the proof of \cref{thm:MaxDoeblinCoupling} (cf. \cref{Proof of MaxDoeblinCoupling for general n}) achieves both bounds simultaneously. Specifically, in \cref{last step showing min is achievabile}, substituting $\calV = [n]$ implies that $\P(\cap_{i=1}^n \{Y_i = y\}) = \min\{P_1(y),\ldots,P_n(y)\}$. Finally, summing over all $y \in \Y$ on both sides proves the proposition. 
\qed

\section{Simultaneous Coupling Construction and Application to Bayesian Networks} \label{sec: proofs2}
In this section, we will first provide the construction of our simultaneously optimal coupling in \Cref{Proof of Simultaneous Coupling} so as to prove \cref{Simultaneous Coupling}. Leveraging this construction, we then provide the proof of \Cref{Doeblins Coefficient in Networks} in \Cref{Proof of Doeblins Coefficient in Networks}. Then as an application of \cref{Doeblins Coefficient in Networks}, we provide the proof of \Cref{Evans-Schulman} in \cref{Proof of Evans-Schulman}. 

\subsection{Proof of \cref{Simultaneous Coupling}}\label{Proof of Simultaneous Coupling}
    First, similar to \cref{Maximal Coupling} we will derive an upper bound on $\P(X_1=\dots=X_n,Y_1=\dots=Y_n)$ and $\P(X_1=\dots=X_n)$ 
    $$
        \begin{aligned}
        & \P(X_1=\dots=X_n,Y_1=\dots=Y_n)  \\
        &= \sum_{x \in \X,y\in \Y} \P(X_1 = \dots = X_n = x,Y_1 = \dots =Y_n = y )     \\
             & \stackrel{\zeta}{\leq} \sum_{x \in \X,y\in \Y} \min\{ P_{X_1,Y_1}(x,y), \\
             & \ \ \ \ \ \ \ \ \ \ \ \ \ \ \  \P(X_2 = \dots = X_n = x,Y_2 = \dots = Y_n = y)\}\\
             & \leq \sum_{x \in \X,y\in \Y} \min\{ P_{X_1,Y_1}(x,y), \dots, P_{X_n,Y_n}(x,y)\},
        \end{aligned}
    $$
    where $\zeta$ follows from the fact that $\P(A \cap B) \leq \min\{\P(A),\P(B)\}$ and substituting $\A = \{X_1=x,Y_1=y\}$ and $B = \cap_{i=2}^n\{X_i =x, Y_i=y\}$. The last inequality follows by applying the same inequality recursively.
    And, similarly, we have the following upper bound 
    $$
        \begin{aligned}
        \P(X_1  = \dots = X_n) \leq \sum_{x \in \X} \min\{ P_{X_1}(x),\dots, P_{X_n}(x)\}.
        \end{aligned}
    $$   
    Now, before presenting the construction of the coupling achieving the above bounds, we define constants $c_{XY}, c_X$ as
    \[\begin{aligned}
        c_{XY} & \triangleq \sum_{x \in \X,y\in\Y} \min \{P_{X_1,Y_1}(x,y), \dots, P_{X_n,Y_n}(x,y)\} \\
        c_X & \triangleq   \sum_{x \in\X} \min\{P_{X_1}(x),\dots,P_{X_n}(x)\}. 
    \end{aligned}
    \] 
    Observe that $c_X \geq c_{XY}$. If $c_{X}=c_{XY}$, then we have
    \begin{equation*}
        \begin{aligned}
         & \sum_{x \in \mathcal{X}} \min \{P_{X_{1}}(x), \dots , P_{X_{n}}(x)\} \\ 
         & \ \ \ \ \ \ \ \ =  \sum_{x \in \mathcal{X}} \sum_{y \in \Y} \min \{P_{X_1, Y_{1}}(x, y), \ldots , \ P_{X_{n}, Y_{n}}(x, y)\}, 
        \end{aligned}
    \end{equation*}
        which gives
        $$ 
        \begin{aligned}
            &\sum_{x \in \X}\bigg(\min \bigg\{\sum_{y \in \Y}  P_{X_{1},Y_{1}}(x, y)-P_{\min}(x, y), \ldots, \\
             & \ \ \ \ \ \ \ \ \ \ \ \ \ \ \ \ \ \ \ \ \ \     \sum_{y \in \Y} P_{X_{n},Y_{n}}(x, y)-P_{\min }(x, y)\bigg\}\bigg)=0,
        \end{aligned}
        $$
    where $P_{\min}(x, y) = \min \{P_{X_{1},Y_1}(x, y), \ldots . ., P_{X_{n}, Y_{n}}(x, y)\}$. Note that $P_{X_{i}, Y_i}(x, y) \geq P_{\min }(x, y)$ for all $x \in \mathcal{X}, y \in \mathcal{Y}$. Hence, for the above equation to hold, for every $x$ there should exist an $i_x$ such that
    \begin{equation*}
        P_{X_{i_x}, Y_{i_x}}(x, y)=P_{\min }(x, y) \text { for all }  y \in \Y,
    \end{equation*}
    which is only possible if
    \begin{equation*}
        P_{X_{i_x}, Y_{i_x}}(x, y)=P_{X_1,Y_1}(x, y)=\dots =P_{X_{n},Y_{n}}(x,y).
    \end{equation*}
    This implies that all joint distributions $P_{X_1, Y_{1}} , \ldots , P_{X_{n}, Y_{n}}$ are identical. In this case, the theorem holds trivially. 
    
    Now, consider the case $c_X > c_{XY}$. Let $x^n \triangleq (x_1,x_2,\dots, x_n)$ and $y^n \triangleq (y_1,y_2,\dots,y_n)$, also define $P_{X_{\min} }(x) \triangleq \min\{P_{X_1}(x), \dots,P_{X_n}(x) \}$.  Now, define the following distributions over the product space $\X^n \times \Y^n$
    $$
        \begin{aligned}
            R_1(x,y) & \triangleq  \frac{P_{\min}(x,y)}{c_{XY}}, \\
             R_{2}(x,y^n) &\triangleq  \frac{P_{X_{\min} }(x)  - \sum_{y \in \Y} P_{\min}(x,y)}{c_X -c_{XY}}    \\
             & \ \ \ \ \ \  \times \prod_{i=1}^n \frac{P_{X_i,Y_i}(x,y_i) -P_{\min}(x,y_i) }{P_{X_i}(x) - \sum_{y\in\Y}P_{\min}(x,y) }, \\
             R_3(x^n,y^n) & \triangleq \prod_{i=1}^n \frac{ P_{X_i,Y_i}(x_i,y_i) - P_{\min}(x_i,y_i) }{P_{X_i}(x_i) - \sum_{y \in \Y}P_{\min}(x_i,y) }  \\ 
             & \ \ \ \ \ \ \times\prod_{i=1}^n \frac{P_{X_i}(x_i) -P_{X_{\min}}(x_i) }{1 - c_{X}    }.
        \end{aligned}
    $$
    
     Define a random variable $Z$ which is independent of $X_1, Y_1,X_2, Y_2,\dots, X_n,Y_n$ as   
    $$ Z = \begin{cases}
    0  \text{ with prob } c_{XY}\\
    1  \text{ with prob } c_X - c_{XY}\\
    2  \text{ with prob } 1 - c_X.
    \end{cases} $$
    We will now construct a joint distribution $P_{X_{1},Y_1,\ldots X_{n},Y_n, Z}$. First, we construct the conditional distribution as follows
    $$
    \begin{aligned}
    P_{X_{1},Y_{1}, \dots, X_{n},Y_{n}|Z}(x^{n}, y^n | Z=0) & = R_1(x,y) \I_{x^n =x,y^n = y} \\
     P_{X_{1},Y_{1}, \dots, X_{n},Y_{n}|Z}(x^{n}, y^{n} | Z=1) & = R_2(x,y^n )\I_{x^n = x}\\
     P_{X_{1},Y_{1}, \dots, X_{n},Y_{n}|Z}(x^{n}, y^{n} | Z=2) & = R_3(x^n,y^n ).
    \end{aligned}
    $$

    The above equations give us the joint distribution $P_{X_{1}, Y_{1}, \ldots, X_{n},Y_{n}, Z}$.  Marginalizing with respect to $Z$ we get $P_{X_{1}, Y_1, \dots, X_{n},Y_n}$ as follows
    $$
    \begin{aligned}
     & P_{X_{1},Y_1,  \ldots, X_{n},Y_n}(x^n,y^n) \\
     &  \ \ \ \ \ \ \ = \sum_{i=0}^2 P(Z=i) P_{X_{1},Y_1, \dots, X_{n},Y_n \mid Z}(x^{n},y^n \mid Z=i).
    \end{aligned}
    $$
    Now we will show that the above joint distribution has the right marginals jointly with respect to $X$ and $Y$. Define $x^{(i)} \triangleq (x_{1}, \ldots, x_{i-1}, x_{i+1}, \ldots, x_{n})$ and $y^{(i)} \triangleq (y_{1}, \ldots, y_{i-1}, y_{i+1}, \ldots, y_{n})$. Observe that for all $x, y$ such that $x_{i}=x $ and $y_{i}=y$
    \begin{align*}
         & \sum_{x^{(i)},y^{(i)}} P_{X_1,Y_1, \ldots, X_{n}, Y_{n}}(x^{n},  y^{n}) \\
         &  =\sum_{x^{(i)}, y^{(i)}}  c_{X Y} R_{1}(x, y) \mathbbm{1}_{x^{n}=x, y^{n}=y} \\ 
         & \ \ \ \ \ \ \ \ \ \ \ \ \ \ \ \  +\sum_{x^{(i)}} \sum_{y(i)}(c_{X}-c_{X Y}) R_{2}(x, y^{n}) \mathbbm{1}_{x^{n}=x} \\
         & \ \ \ \ \ \ \ \ \ \ \ \ \ \ \ \  +\sum_{x^{(i)}} \sum_{y^{(i)}}(1-c_{X}) R_{3}(x^{n}, y^{n}) \\
         & =P_{\min }(x, y)+\bigg(P_{X_{\min} }(x) -\sum_{y \in y} P_{\min }(x, y)\bigg) \\ 
         & \ \ \ \ \ \ \ \ \ \ \ \ \ \ \ \ \times \bigg( \frac{P_{X_{i}, Y_{i}}(x, y)-P_{\min }(x, y)}{P_{X_{i}}(x)-\sum_{y \in \Y} P_{\min }(x, y)} \bigg)  \\ 
         & \   \times \sum_{y^{(i)}} \prod_{j \neq i} \frac{P_{X_{j},Y_{j}}(x, y_{j})-P_{\min }(x, y_{j})}{P_{X_{j}}(x)-\sum_{y \in \Y} P_{\min }(x, y)}  \\ 
         & \ \ \ \ +\frac{(P_{X_{i}, Y_{i}}(x, y)-P_{\min }(x, y))(P_{X_{i}}(x)-P_{X_{\min} }(x))}{P_{X_{i}}(x)-\sum_{y \in y} P_{\min }(x, y)} \\ 
         & \ \ \ \ \ \times \sum_{x^{(i)},y^{(i)}} \prod_{j\neq i} \bigg( \frac{P_{X_j,Y_j}(x_j,y_j) - P_{\min}(x_j,y_j) }{P_{X_j}(x_j) - \sum_{y \in \Y} P_{\min}(x_j,y) } \\ 
         & \ \ \ \ \ \ \ \ \ \ \times  \frac{P_{X_j}(x_j) - P_{\min}(x_j) }{1 - c_X}  \bigg)\\
         & =P_{\min }(x, y)+\bigg(P_{X_{\min}}(x)-\sum_{y \in y} P_{\min }(x, y)\bigg) \\ 
         & \ \ \ \ \ \ \ \ \ \ \ \ \ \ \ \ \times  \frac{P_{X_{i}, Y_{i}}(x, y)-P_{\min }(x, y)}{P_{X_{i}}(x)-\sum_{y \in \Y} P_{\min }(x, y)} \\
         & \ \ \ \ \  +\frac{(P_{X_{i}, Y_{i}}(x, y)-P_{\min }(x, y))(P_{X_{i}}(x)-P_{X_{\min }}(x))}{P_{X_{i}}(x)-\sum_{y \in \Y} P_{\min }(x, y)} \\
         & =P_{\min }(x, y)+\frac{(P_{X_{i}, Y_{i}}(x, y)-P_{\min}(x, y)) }{P_{X_{i}}(x)-\sum_{y \in y} P_{\min }(x, y)}  \\ 
         & \ \ \ \ \ \ \ \ \ \ \ \ \ \ \ \ \times  (P_{X_i}(x)-\sum_{y \in \Y} P_{\min}(x,y)) \\
         & =P_{\min }(x, y)+P_{X_{i}, Y_{i}}(x, y)-P_{\min }(x, y) \\
         & =P_{X_{i}, Y_{i}}(x, y).
        \end{align*}

    Now, it remains to show that the above coupling satisfies the following two conditions
    \begin{equation*}
    \begin{aligned}
         \P(X_{1}=\ldots =X_{n}, Y_{1}=\ldots = Y_{n}) & =\tau(P_{X_{1}, Y_{1}}, \ldots, P_{X_{n}, Y_{n}}) \\
         \mathbb{P}(X_{1}=X_{2}=\ldots =X_{n})& =\tau(P_{X_{1}}, \ldots , P_{X_{n}}).
        \end{aligned}
    \end{equation*}

    Define events $A=\{X_{1}=X_{2}=\ldots=X_{n}\},  B=\{Y_{1}=Y_{2}=\ldots=Y_{n}\}$. Observe that under both $R_{2}$ and $R_{3}$ the measure of event $A \cap B$ is zero. This is because for any $y \in \Y$ and $x \in \X,$ if $\{X_{1}=X_{2}=\ldots=X_{n}=x\}$ and $\{Y_{1}=Y_{2}=\ldots=Y_{n}=y\}$ then the following term is zero for all $x \in X$
    $$ \prod_{i=1}^n (P_{X_i,Y_i}(x,y) - P_{\min}(x,y)) = 0. $$
    Hence,
    \begin{equation*}
    \begin{aligned}
      P(A \cap B) =\mathbb{P}(Z=0) =c_{XY} =\tau(P_{X_1, Y_1}, \dots , P_{X_n,Y_{n}}).
    \end{aligned}
    \end{equation*}    
    Next we will show $\mathbb{P}(A)=\tau(P_{X_{1}}, \dots , P_{X_{n}})$. Note that the measure of set $A$ under  $R_{3}(\cdot)$ is $0$. Hence, 
    \begin{equation*}
    \begin{aligned}
      \mathbb{P}(X_{1} & =X_{2}=\ldots =X_{n}=x) \\
     & =c_{X Y} \sum_{y^{n}} R_{1}(x, y)+(c_{X}-c_{X Y}) \sum_{y^{n}} R_{2}(x, y) \\
     & =\sum_{y} P_{\min }(x, y)+\bigg(P_{X_{\min }}(x)-\sum_{y} P_{\min }(x, y)\bigg) \\ 
     & \ \ \ \ \ \ \ \ \ \ \ \  \times \bigg(\sum_{y^{n}} \prod_{i=1}^{n} \frac{P_{X_{i}, Y_{i}} (x, y_i)-P_{\min }(x, y_{i})}{P_{X_{i}}(x)-\sum_{y \in \Y} P_{\min }(x, y)}\bigg) \\
     & =\sum_{y} P_{\min }(x, y)+P_{X_{\min }}(x)-\sum_{y} P_{\min }(x, y) \\
     & =P_{X_{\min }}(x).
    \end{aligned}
    \end{equation*}
    Thus, we have
    \begin{equation*}
    \begin{aligned}
    \mathbb{P}(A)  &=\sum_{x \in \X} \mathbb{P}(X_{1}=X_{2}=\dots=X_{n}=x) \\
    & =\sum_{x \in \X} P_{X_{\min }}(x)=\tau(P_{X_{1}}, P_{X_{2}}, \ldots, P_{X_{n}}).
    \end{aligned}
    \end{equation*}
    This completes the proof. \hfill $\square$

\subsection{Proof of \cref{Doeblins Coefficient in Networks}} \label{Proof of Doeblins Coefficient in Networks}
    Let $Z= \pa(U)$ and assume that it takes values in the set $\mathcal{Z}$. Let $Q_{i}$ denote the conditional distribution of $V ,Z$ given $X = x_{i}$, i.e., $Q_{i}=P_{V,Z \mid X=x_{i}} \text{ for } i \in [n]$. Moreover, define the random variables $(V_{i}, Z_{i}) \sim Q_{i}$ for $i \in [n]$. By \cref{Simultaneous Coupling}, there exists a simultaneously maximal coupling $ P_{V_1, Z_{1}, \dots ,V_{n},Z_{n}}$ of $Q_{1}, \dots, Q_{n}$, with corresponding probability operator denoted by $\P(\cdot)$, such that
    \begin{equation*}
        \begin{aligned}
             \P(V_{1}= \dots =V_{n}, Z_{1}=\dots=Z_{n} )  =\tau(Q_{1}, Q_{2}, \ldots, Q_{n}) , \\
             \P(V_{1}=\dots=V_{n})  =\tau(P_{V \mid X=x_{1}}, \ldots, P_{V \mid X=x_{n}}).
        \end{aligned}
    \end{equation*}
    Conditioned on $Z_1 = z_{1}, \ldots, Z_n = z_{n}$ for $z_1,\dots,z_n \in \Z$, define the random variables $U_{i} \sim P_{U \mid Z_{i}=z_{i}}$. Construct a maximal coupling of $P_{U \mid Z=z_{1}}, \dots , P_{U \mid Z=z_{n}}$, and let $(U_{1}, U_{2}, \ldots ,U_{n})$ be distributed according to this coupling (conditioned on $Z_1 = z_{1}, \ldots, Z_n = z_{n}$). Then, the simultaneously maximal coupling of $V_{1},Z_{1}, V_{2},Z_{2}, \ldots, V_{n},Z_n$ (earlier) and the maximal coupling of $U_{1}, U_{2}, \ldots, U_{n}$ define the joint distribution $P_{V_1, U_{1}, Z_{1}, \ldots ,V_{n}, Z_{n}, U_{n}}$, which satisfies the Markov relation
    \begin{equation*}
        (V_{1}, V_{2}, \dots ,V_{n}) \longrightarrow (Z_{1}, Z_{2}, \ldots ,Z_{n}) \longrightarrow (U_{1}, U_{2}, \ldots ,U_{n}).
    \end{equation*}

    Therefore, we have
    \begin{equation*}
    \begin{aligned}
          1-\mathbb{P}(&U_{1}=\dots=U_{n} \mid Z_{1} =z_1, \ldots, Z_{n} = z_n) \\
         & = 1-\tau(P_{U \mid Z_1=z_1}, \ldots, P_{U \mid Z_n=z_{n}})\\
        & \leq(1-\tau(P_{U \mid Z})) \I_{\{Z_1=\dots= Z_n \}^\complement},
    \end{aligned}
    \end{equation*}
    where the first equality follows since operator $\P(\cdot)$ is based on the maximal coupling of $P_{U \mid Z=z_{1}}, \dots , P_{U \mid Z=z_{n}}$, and let $(U_{1}, U_{2}, \ldots ,U_{n})$ and the inequality holds since the state space of $Z$ may be larger than $\{z_1, \ldots, z_{n}\}$ and the two bounds are equal on the set $\{Z_1=\dots= Z_n \}$. Now, taking expectation conditioned on $V_1=\dots = V_n$ on both sides, gives
    $$
    \begin{aligned}
        & 1 - \E[\mathbb{P}(U_{1} =\cdot\cdot =U_{n} | Z_{1} =z_1,\dots, Z_{n} = z_n)| V_1=\cdot\cdot = V_n]  \\
         & \leq (1-\tau(P_{U \mid Z})) \P(\{Z_1=\dots= Z_n \}^\complement | V_1=\dots = V_n), 
    \end{aligned}
    $$
    which simplifies to
    $$
    \begin{aligned}
      & 1 - \mathbb{P}(U_{1}=\ldots=U_{n}| V_1=\dots = V_n) \\
      &\leq (1-\tau(P_{U \mid Z})) (1- \P(\{Z_1=\dots= Z_n \} | V_1=\dots = V_n)) .  
    \end{aligned}
    $$
    Now, multiplying both sides by $\mathbb{P}(V_{1}=\ldots=V_{n})$ gives
    $$ 
        \begin{aligned}
           &\P(V_{1}=\ldots=V_{n}) -\P(U_{1}=\ldots=U_{n}, V_{1}=\ldots=V_{n}) \\ 
         & \leq (1-\tau(P_{U | Z})) (\P(V_1=\dots = V_n) \\
         & \ \ \ \ \ \ \ \ \ \ \ \ \ \ \ \ \  - \P(Z_1=\dots = Z_n, V_1=\dots= V_n)).
        \end{aligned}
    $$
    By construction of simultaneous maximal coupling, we have 
    $$ 
    \begin{aligned}
       & \tau(P_{V|X})-(1-\tau(P_{U \mid Z}))(\tau(P_{V| X})-\tau(P_{Z, V \mid X}))  \\
       & \leq  \P(U_{1}=\ldots=U_{n}, V_{1}=\ldots=V_{n}) \leq \tau(P_{V, U| X}), 
    \end{aligned}
    $$
    which on simplification gives
    \begin{equation}
        \tau(P_{V, U \mid X}) \geq \tau(P_{V \mid X}) \tau(P_{U \mid Z}) + (1- \tau(P_{U|Z})) \tau(P_{Z,V|X}) . \label{firstpart}
    \end{equation}

    We will now prove the percolation bound. Using the argument in \cite[Theorem 5]{Polyanskiy2017}, the set-function $\operatorname{perc}$ satisfies the following recursion: 
    $$
    \begin{aligned}
     &\operatorname{perc}(V \cup\{U\})  \triangleq \P(\exists \pi: X \rightarrow V \cup\{U\}]) \\
    &=\tau_{U} \operatorname{perc}(V)+(1-\tau_{U})\operatorname{perc}(V \cup \operatorname{pa}(U)).
    \end{aligned}
    $$
    The recursion holds trivially for $V=\{X\}$, since both sides are equal to 1. 
    Moreover, we have the following recursive relation from \eqref{firstpart}:  
    \begin{equation*}
       1- \tau(P_{V, U \mid X}) \leq  \tau_U (1-\tau(P_{V \mid X})) + (1- \tau_U) (1 - \tau(P_{Z,V|X})). 
    \end{equation*}
    Then, by induction on the maximal element of $V$ and applying \eqref{firstpart}, we get the following for all $V$:
    \begin{equation*}
        1 - \tau(P_{V \mid X}) \leq \operatorname{perc}(V).        
     \end{equation*} 
    This yields the desired result. \hfill $\square$

\subsection{Proof of \cref{Evans-Schulman}} \label{Proof of Evans-Schulman}

Similar to the analyses in \cite{evans1999signal,Polyanskiy2017}, note that if $\calA$ and $\calB$ are disjoint sets of nodes, then
\begin{equation}
\begin{aligned}
     \sum_{\pi: X \stackrel{sf}{\rightarrow} \calA \cup \calB} (1- \tau)^\pi & = \sum_{\pi: X \stackrel{sf}{\rightarrow} \calA \text { avoid } \calB} (1-\tau)^\pi \\ 
     & \ \ \  \ \ \ \ + \sum_{\pi: X \stackrel{sf}{\rightarrow} {\calB \text { avoid } \calA}} (1-\tau)^\pi . \label{Firstpart}
\end{aligned}
\end{equation}
Let $\pi: X \rightarrow V$ and $\pi_1$ be another path same as $\pi$ but without the last node, then
\begin{equation}\label{secondpart}
\pi: X \stackrel{s f}{\rightarrow} V \quad \Longleftrightarrow \quad \pi_1: X \stackrel{s f}{\rightarrow} \mathrm{pa}(V) \backslash \{V\} .
\end{equation}
Now we can represent $V=(V^{\prime}, W)$ with $W>V^{\prime}$, denote $Q=\mathrm{pa}(W) \backslash V $ and by induction we assume the following two relations 
\begin{equation} \label{gatheredeqn}
\begin{gathered}
 1 - \tau(P_{V^{\prime} \mid X})  \leq \sum_{\pi: X \stackrel{sf}{\rightarrow} V^\prime}   (1- \tau)^\pi, \\
1 - \tau(P_{V^{\prime}, Q \mid X})  \leq \sum_{\pi: X \stackrel{sf}{\rightarrow} \{V^{\prime}, Q\}} (1 - \tau)^\pi.
\end{gathered}
\end{equation}
By \cref{Firstpart} and \cref{secondpart} we have the following relation
\begin{align}
 & \sum_{\pi: X \stackrel{sf}{\rightarrow} V} (1 - \tau)^\pi   =\sum_{\pi: X \stackrel{sf}{\rightarrow} V^{\prime}} (1 - \tau)^\pi+\sum_{\pi: X \stackrel{s f}{\rightarrow} W \text{avoid} V^{\prime}} (1 - \tau)^\pi \nonumber \\
 & = \sum_{\pi: X \stackrel{s f}{\rightarrow}   V^{\prime}} (1-\tau)^\pi + (1-\tau_W) \sum_{\pi: X \stackrel{s f}{\rightarrow} Q \text { avoid } V^{\prime}} (1 - \tau)^\pi. \label{thirdpart}
\end{align}
Next, \cref{Doeblins Coefficient in Networks} and induction hypotheses \cref{gatheredeqn} gives
\begin{align}
& 1 -\tau(P_{V|X}) \nonumber \\ 
& \stackrel{\zeta_1}{\leq}  \tau_W (1-\tau(P_{V^\prime|X})) + (1-\tau_W)(1-\tau(P_{V^\prime ,Q})) \nonumber \\ 
 & \stackrel{\zeta_2}{\leq}   \tau_W  \sum_{\pi: X\stackrel{sf}{\rightarrow } V^{\prime}} (1 - \tau)^\pi + (1- \tau_W) \sum_{\pi: X \stackrel{sf}{\rightarrow} V^{\prime}, Q}(1 - \tau)^\pi \nonumber\\
&  =  \tau_W \sum_{\pi: X\stackrel{sf}{\rightarrow} V^{\prime}} (1 - \tau)^\pi + (1- \tau_W) \sum_{\pi: X \stackrel{sf}{\rightarrow}Q \text{ avoid } V^{\prime}}(1 - \tau)^\pi\nonumber \\
& \ \ \ \ \ \ \ \ \ \ \ \ \ \ + (1- \tau_W) \sum_{\pi: X \stackrel{sf}{\rightarrow} V^{\prime} \text { avoid } Q }(1 - \tau)^\pi\nonumber\\
 & \stackrel{\zeta_3}{=}  \sum_{\pi: X\stackrel{sf}{\rightarrow} V^{\prime}} (1 - \tau)^\pi + (1- \tau_W) \sum_{\pi: X \stackrel{sf}{\rightarrow} Q \text { avoid } V^{\prime}}(1 - \tau)^\pi \nonumber\\ 
& \ \ \ \ \ \ \ \ \ \ \ \ \ \ - (1- \tau_W) \sum_{\pi: X \stackrel{sf}{\rightarrow} Q \text { pass } V^{\prime}}(1 - \tau)^\pi \nonumber\\
 & \leq \sum_{\pi: X \stackrel{sf}{\rightarrow} V^{\prime}} (1 -\tau)^\pi + (1-\tau_W) \sum_{\pi: X \stackrel{sf}{\rightarrow} Q \text { avoid } V^{\prime}} \tau^\pi \nonumber\\ 
& = \sum_{\pi: X \stackrel{sf}{\rightarrow} V} (1 - \tau)^\pi. \label{fifthpart}
\end{align}
where $\zeta_1$ follows from \cref{Doeblins Coefficient in Networks} and $\zeta_2$ from induction hypotheses \cref{gatheredeqn} and in $\zeta_3$, we split the summation over $\pi: X \stackrel{sf}{\rightarrow} V^\prime$ into paths that avoid and pass nodes in $Q$. Finally, \cref{fifthpart} follows from \cref{thirdpart}.
\qed

\section{Proof of \cref{Samorodnitskys SDPI in terms of Doeblin Coefficient}} \label{Proof of Samorodnitskys SDPI in terms of Doeblin Coefficient}
    We begin by presenting a generalization of \cref{Doeblin Minorization and Degradation} for the channel $P_{Y^n|X^n} = \prod_{j=1}^n P_{Y_j|X_j}$. We will show that if $X^n$ is first passed through an erasure channel $P_{{E}^n|X^n}$ given by the tensor product $P_{{E}^n|X^n} = \prod_{i=1}^n \text{E}_{{E}_i|X_i}(\tau_i) $, where $\text{E}_{{E}_i|X_i}(\tau_i)$ is an erasure channel with erasure probability $\tau_i$, then there exist a conditional distribution $P_{Y^n|{E}^n}$ such that
        \begin{equation}
            P_{Y^n|X^n} = P_{{E}^n|X^n} P_{Y^n|{E}^n}.
        \end{equation}
    To see this, we refer back to \cref{Doeblin Minorization and Degradation} which assures us that for each $P_{Y_j|X_j}$ there exist a conditional distribution $P_{Y_j|{E}_j}$ such that for all $j \in [n]$
        $$ P_{Y_j|{X}_j} = \text{E}_{{E}_j|X_j}(\tau_j) P_{Y_j|{E}_j}.$$ 
    Hence, taking the tensor product of all the $n$ equations on both sides, we get
        \begin{equation}
            \begin{aligned}
                P_{Y^n|{X}^n} & = \prod_{j=1}^n P_{Y_j|{E}_j} = \prod_{j=1}^n (\text{E}_{{E}_j|X_j}(\tau_j) P_{Y_j|{E}_j}) \\ 
                & = \bigg(\prod_{j=1}^n \text{E}_{{E}_j|X_j}(\tau_j) \bigg) \bigg(\prod_{j=1}^n P_{Y_j|{E}_j} \bigg) ,
            \end{aligned}
        \end{equation}   
    where the last step follows from the mixed product property. 
    Thus, the Markov chain $U \rightarrow X^n \rightarrow Y^n$ can be equivalently written as $U \rightarrow X^n \rightarrow \text{E}^n \rightarrow Y^n$.
    By sub-multiplicativity of the Doeblin coefficient, we get
    $$
    \begin{aligned}
        \tau(P_{Y^n|U})  & \stackrel{\zeta_1}{\geq} \tau(P_{{E}^n|U}) + \tau(P_{Y^n|{E}^n})(1- \tau(P_{{E}^n|U}))  \\&
         \geq \tau(P_{{E}^n|U})
           \stackrel{\zeta_2}{=} \sum_{T \subseteq [n]} P(T) \tau(P_{X_T|U}),   
    \end{aligned}
    $$
    where $\zeta_1$ follows since $P_{{E}^n|U} = P_{{X}^n|U} P_{{E}^n|{X}^n}$ and $\zeta_2$ follows since $P_{{E}_i|{X}_i}$ is an erasure channel with erasure probability $\tau_i$. 
    Finally, when $\tau_j = \tau$ for all $j \in [n]$, then the next part follows since $P(T) = \tau^{n -|T|}(1-\tau)^{|\tau|}$, 
    which completes the proof of \cref{Samorodnitskys SDPI in terms of Doeblin Coefficient}. \hfill $\square$

\section{Conclusion} \label{Sec: Conclusion}
This paper presents new insights into the structural and geometric properties of Doeblin coefficients. For instance, we demonstrate that Doeblin coefficients possess desirable properties that make them suitable for defining a notion of distance between multiple probability distributions, and they possess a maximal coupling characterization that generalizes the coupling characterization for TV distance. We also introduce a new coefficient, namely the max-Doeblin coefficient, which serves as another generalization of TV distance, and we provide its coupling characterization as well (subject to a certain condition). Additionally, we develop the contraction properties of Doeblin coefficients over Bayesian networks. Our results highlight that the Doeblin coefficient exhibits contraction properties similar to other canonical contraction coefficients of SDPIs. We also show an approach for fusing PMFs based on the coupling characterization for Doeblin coefficients. These results open several new directions for future research. For example, determining (for $n \geq 4$) the optimal value in \cref{thm:MaxDoeblinCoupling} under the condition $\tau_{\max_2}(W) > 1$ is an open problem. In the context of PMF fusion, future directions include identifying the axioms that uniquely characterize the min-rule and developing more robust versions of it that consider agent reliability. 

\appendices

\section{Proof of Properties of Max-Doeblin Coefficient} \label[appendix]{Proof of Properties of Max-Doeblin coefficient}
In this section, we will provide the proof of \cref{Properties of max-Doeblin coefficient}. Much of the proof follows analogously to the proof of \cref{Properties of Doeblin coefficient}.
\begin{proof}[Proof of \cref{Properties of max-Doeblin coefficient}] \ \par
\textbf{Part 1:} We begin by recalling the definition of $\gamma_{\max}(\cdot)$:
    $$ 
    \begin{aligned}
         \gamma_{\max }(P_{1}, &\ldots, P_{n}) = \\
         & \frac{1}{n-1}\bigg(\sum_{y  \in \Y} \max \{P_{1}(y), \ldots, P_{n}(y)\}-1\bigg).
    \end{aligned}
    $$
    If $\gamma_{\max }(P_{1}, \ldots, P_{n})=0$, then we have
    $$
        \sum_{y  \in \Y} \max \{P_{1}(y), \ldots, P_{n}(y)\}=1.
    $$
    Note that for any $i \in [n]$, we have
    $$
        \max \{P_{1}(y), \ldots, P_{n}(y)\} \geq P_{i}(y).
    $$
    Summing over all $y \in \Y$ on both sides, we obtain
    $$
        \sum_{y  \in \Y} \max \{P_{1}(y), \ldots, P_{n}(y)\} \geq \sum_{y  \in \mathcal{Y}} P_{i}(y)=1.
    $$
    Therefore, the lower bound is achieved if and only if
    $$
        \max \{P_{1}(y), \ldots, P_{n}(y)\}=P_{i}(y), \ \forall y \in \Y, i \in [n].
    $$
    This implies that $P_{1}(y)=\ldots=P_{n}(y)$ for all $y\in\mathcal{Y}$. 

    Next, let us consider the case where $\gamma_{\max}(P_1, \ldots, P_{n})=1$. This implies 
    $$
        \sum_{y  \in \Y} \max \{P_{1}(y), \ldots, P_{n}(y)\}=n.
    $$
    Since $P_{i}(y) \geq 0$ for any $y \in \Y$, we have
    $$
        \max \{P_{1}(y), \ldots, P_{n}(y)\} \leq P_{1}(y)+P_{2}(y)+\ldots+P_{n}(y).
    $$
    Since the sum of probabilities over all outcomes is equal to $1$, we have $\sum_{y \in \mathcal{Y}} P_{i}(y) = 1$. Thus, we get 
    $$
        \sum_{y  \in \Y} \max \{P_{1}(y), \ldots, P_{n}(y)\} \leq n.
    $$
    Since the above upper bound holds with equality, it is only possible when $P_{1}, P_{2}, \ldots, P_{n}$ are all pairwise disjoint. Conversely, mutually disjoint PMFs have $\sum_{y \in \Y}\max \{P_{1}(y), \ldots, P_{n}(y)\} = n$. 

    \textbf{Part 2:} Clearly $\gamma_{\max}(P_{1}, \ldots, P_{n})$ is symmetric by definition and positive definiteness follows from Part 1. Now, for any $P_{n+1} \in \calP_m$, we will prove the polyhedron inequality stated as 
     \begin{equation*}
                    \begin{aligned}
                        (n-1) \gamma_{\max}(P_{1}, &\dots, P_{n} ) \leq  \\
                        & \sum_{i=1}^{n} \gamma_{\max}(P_{1}, \dots, P_{i-1}, P_{i+1}, \dots, P_{n+1}).
                    \end{aligned}
            \end{equation*}
    From the definition of $\gamma_{max}(\cdot)$, we get
    $$
        \begin{aligned}
         &\sum_{y  \in \Y} \max \{P_{1}, \ldots, P_{n}(y)\}-1 \leq \frac{1}{n-1} \times   \\ 
         &\sum_{i=1}^{n} \sum_{y\in \Y} \max \{P_{1}, \ldots, P_{i-1}, P_{i+1}, \ldots, P_{n+1}(y)\} - \frac{n}{n-1}.
        \end{aligned}
    $$
    This simplifies to
    \begin{equation}
    \begin{aligned}\label{Simplified max Polyhedron}
     \sum_{y  \in \Y}\bigg(\sum_{i=1}^{n} \max \{&P_{1}(y), \ldots, P_{i-1},  P_{i+1}, \ldots, P_{n+1}(y)\} \\ 
     &  -(n-1) \max \{P_{1}, \ldots, P_{n}(y)\} \bigg)  \geq 1.
    \end{aligned}
    \end{equation}
    Analogous to proof of \cref{Properties of Doeblin coefficient}, define event $\calA \triangleq \{y: P_{n+1}(y) \geq \max \{P_{1}(y),  \ldots, P_{n}(y)\}\}$. On the set $\calA$ the term inside parentheses of \cref{Simplified max Polyhedron} reduces to
    $$
        n P_{n+1}(y)-(n-1) \max \{P_{1}(y), \ldots, P_{n}(y)\} \geq P_{n+1}(y).
    $$
\sloppy    Now consider any $y  \in \calA^{\complement}$. Without loss of generality, for any fixed $y \in \Y$ assume $P_1(y)$ is largest among $P_1(y),P_2(y), \dots,P_n(y)$, the term the term inside parentheses simplifies to     
    $$
    \begin{aligned}
        (n-1) P_{1}(y)+\max  \{P_{2}(y), \ldots, P_{n+1}(y)\}&-(n-1) P_{1}(y) \\
         &\geq P_{n+1}(y).
    \end{aligned}
    $$
    Hence, on both $\mathcal{A}$ and $\mathcal{A}^\complement$, the terms inside the parentheses in \cref{Simplified max Polyhedron} is lower bounded by $P_{n+1}(y)$ for all $y  \in \Y$. Since the lower bound $P_{n+1}(y)$ sums to $1$, this proves the polyhedron inequality.

\textbf{Part 3:} For $V,W \in \R_{\mathsf{sto}}^{n\times m}$ such that $V = [V_1^\T,\ldots,V_n^\T]^\T$ and $W = [W_1^\T,\ldots,W_n^\T]^\T$ and 
    for some $\lambda \in [0,1]$, let $\bar{\lambda} \triangleq 1 - \lambda$. Then, we have 
    $$
        \begin{aligned}
             & \tau_{\max}(\lambda V+\bar{\lambda} W) \\
             &=  \sum_{y\in \Y} \max \{\lambda V_{1}(y)+\bar{\lambda} W_{1}(y), \dots , \lambda V_{n}(y)+\bar{\lambda} W_{n}(y)\}\\
             & \leq \lambda \sum_{y\in \Y}\max \{V_{1}(y), \ldots, V_{n}(y)\} \\
             & \ \ \ \ \ \ \ \ + \bar{\lambda} \sum_{y  \in \Y}  \max \{W_{1}(y), \ldots, W_{n}(y)\} \\
            & = \lambda \tau_{\max}(V) + \bar{\lambda} \tau_{\max}(W).
        \end{aligned}
    $$

\textbf{Part 4:} To prove sub-multiplicativity, we will decompose $W$ as a (weighted) difference of two row-stochastic matrices as follows 
    \[
        W = \tau_{\max}(W)\1 w^\T - (\tau_{\max}(W) -1) S_1,
    \]
    where $w \in \R^n $ such that $w_j \triangleq \max_{i} \frac{W_{ij}}{\tau_{\max}(W)}$ and $S_1 \triangleq \frac{\tau_{\max}(W)\1 w^\T - W}{\tau_{\max}(W) -1}$. Note that $S_1$ is a stochastic matrix. Hence, 
    \begin{equation}
    \begin{aligned}
        VW & = \tau_{\max}(W)V\1 w^\T  -  (\tau_{\max}(W) -1)V S_1\\ 
           & = \tau_{\max}(W)\1 w^\T  -  (\tau_{\max}(W) -1)V S_1.
    \end{aligned}
    \end{equation}
    Hence, we have
    $$
    \begin{aligned}
         & \tau_{\max}( VW) \\ 
         & \ \ \ = \sum_j \max_i  (\tau_{\max}(W)\1 w^\T -  (\tau_{\max}(W) -1)V S_1)_{ij} \\
                         &\ \ \ \leq \tau_{\max}(W) \sum_j \max_i (\1 w^\T)_{ij}   = \tau_{\max}(W).
    \end{aligned}
    $$
    Similarly, we can decompose $V$ as \[
        V = \tau_{\max}(V)\1 v^\T - (\tau_{\max}(V) -1) S_2,
    \]
    where $v \in \R^n $ such that $v_j \triangleq \max_{i} \frac{V_{ij}}{\tau_{\max}(W)}$ and $S_2$ defined in a similar fashion as $S_1$. Hence, 
        $$VW = \tau_{\max}(V) \1 v^\T W  -   (\tau_{\max}(V) -1)  S_2 W. $$
    This gives
    \begin{equation}
    \begin{aligned}
         &\tau_{\max}(VW) \\ 
          & \ \ = \sum_j \max_i  (\tau_{\max}(V)\1 v^\T W -  (\tau_{\max}(V) -1)  S_2 W )_{ij} \\
                         & \ \ \leq \tau_{\max}(V) \sum_j \max_i (\1 v^\T W)_{ij}   = \tau_{\max}(V).
    \end{aligned}
    \end{equation}
    Thus, we obtain $\tau_{\max}(VW) \leq \min\{ \tau_{\max}(V), \tau_{\max}(W) \}$.

    \textbf{Part 5:} For $W \in \R^{n \times m}_{\mathsf{sto}},$ and $V \in \R^{l \times k}_{\mathsf{sto}},$ we have
    $$
        \begin{aligned}
           &  \tau_{\max}(W \otimes V)  = \sum_{x  \in [m], y \in [k]} \max_{i \in [n], j \in [l]} W_{ix} V_{jy} \\
           & =\sum_{x  \in [m], y \in [k]} \max \{W_{1x}, \ldots, W_{nx}\} \times \max \{V_{1y}, \ldots, V_{ly}\} \\
             & =\bigg(\sum_{x  \in [m]} \max \{W_{1x}, \ldots, W_{nx}\} \bigg) \\ 
             & \ \ \ \ \ \ \ \  \times  \bigg( \sum_{ y \in [k]} \max \{V_{1y}, \ldots, V_{ly}\} \bigg)  \\
           & =\tau_{\max}(W) \tau_{\max}(V).
        \end{aligned}
    $$
    
    \textbf{Part 6:} This lower bound for $\tau_{\max}(W )$ was proved in  \cite[Lemma 2]{Issa2020}.
    
    \textbf{Part 7:}  It is easy to see that 
    $$\mathsf{Tr}(PW) = \sum_{j=1}^m \sum_{i=1}^n P_{ji} W_{ij} \leq \sum_{j=1}^m \max_{i \in [n]} W_{ij} \times \sum_{i=1}^n P_{ji}.$$ 
    Since for any $j \in [m]$, $\sum_{i=1}^n P_{ji} = 1$, we get $\mathsf{Tr}(PW) \leq \tau_{\max}(W)$.         
    The equality holds when for any $j \in [m]$, $P_{ji_0} = 1$ when $i_0 = \argmax_{k \in [n] } W_{kj}$ and $0$ otherwise.

\textbf{Part 8:} From the Bayesian decision theory, we know that for an observation model $P_{Y|X} \in \R^{n\times m}_{\mathsf{sto}} $, 
    and a prior PMF $P_X \in \calP_n$, and a utility function $u: \X \times \X \rightarrow \R$, and any randomized estimator $P_{\hat{X}|Y} \in \R^{m\times n}_{\mathsf{sto}}$, the average utility $R$ is defined as
            $$ 
            \begin{aligned}
                R( &P_{Y |X} ,  P_{\X}, u, P_{\hat{X}|Y} )  \triangleq \E[ u(X, \hat{X})] \\
                 & = \sum_{x \in \X}\sum_{y\in \Y}\sum_{\hat{x}\in \X} u(x, \hat{x}) P_X(x) P_{Y |X} (y|x) P_{\hat{X}|Y}(\hat{x}|y) \\
                 & = \mathsf{Tr}(U^\T \diag(P_X) P_{Y|X} P_{\hat{X}|Y }),
            \end{aligned}
            $$
        where $U \in \R^{n \times n}$ is the matrix representation of the utility function $u : \X \times \X \rightarrow \R$. Specifically, the $(i,j)$th entry of the matrix $U$ is $u(x_i, \hat{x}_j)$ for $ i, j \in [n]$ and $x_i, \hat{x}_j \in \X$. By the choice of our utility function, we have $U = I$ and $\diag(P_{X}) = \frac{1}{n} I$. This gives the optimal average utility $R^*_{\lambda}(P_{Y|X})$ as
            $$
            \begin{aligned}
               R^*_{\lambda}(P_{Y|X}) & = \max_{P_{\hat{X}|Y} \in \R^{m\times n}_{\mathsf{sto}}}  \P(\hat{X} (Y)  = X) \\
              & = \max_{P_{\hat{X}|Y} \in \R^{m\times n}_{\mathsf{sto}}} \mathsf{Tr}\left(I \frac{1}{n}I P_{Y \mid X} P_{\hat{X} \mid Y}\right) \\
              & = \frac{1}{n} \sum_{y \in \Y} \max \{ P_{Y|X=x_1}(y), \ldots, P_{Y|X=x_n}(y)\} \\    
              & =\frac{\tau_{\max}(P_{Y|X})}{n}.
            \end{aligned}
            $$
\end{proof}

\section*{Acknowledgment} The authors would like to thank William Lu and Shreyas Sundaram for discussions.

%%%%%%
%% References:
\bibliographystyle{IEEEtran}
\bibliography{references}

\end{document}